\pdfoutput=1
\documentclass[usegraphicx,usenatbib]{mn2e}
\usepackage{color}
\usepackage{amssymb}
\usepackage{epsf,multirow}
\usepackage{graphicx}
\usepackage{txfonts}
\usepackage{textcomp}
\usepackage[usenames,dvipsnames]{xcolor}
\usepackage{soul,color} 

\def\nar{New Astronomy Review}

\def\aap{A\&A}
\def\apj{ApJ}
\def\apjs{ApJS}
\def\apjl{ApJL}

\def\mnras{MNRAS}
\def\aj{AJ}

\def\aj{AJ}

\def\apj{ApJ}
\def\apjl{ApJ}
\def\apjs{ApJS}

\def\aap{A\&A}

\def\mnras{MNRAS}

\def\pasp{PASP}
\def\pasj{PASJ}

\def\aj{AJ}

\def\apj{ApJ}
\def\apjl{ApJ}
\def\apjs{ApJS}

\def\aap{A\&A}

\def\mnras{MNRAS}

\def\pasp{PASP}
\def\pasj{PASJ}

\defcitealias{Mazzalay2013a}{Paper\,I}
\newcommand{\citeta}{\citetalias}
\newcommand{\citepa}{\citepalias}

\def\kms{~km~s$^{-1}$}
\def\micron{~$\mu$m}

\def\deg{$^{\circ}$}

\def\Ms{~M$_{\sun}$}

\def\n{NGC~}
\def\S{SINFONI}

\def\H2{H$_2$}
\def\Ha{H$\alpha$}
\def\Hb{H$\beta$}
\def\HeI{He\,{\sc i}}
\def\OIII{[O\,{\sc iii}]}

\def\Brg{Br$\gamma$}

\title[Molecular gas in the centre of galaxies -- II.]{Molecular gas in the centre of nearby galaxies from VLT/SINFONI integral field spectroscopy -- II. Kinematics\thanks{Based on observations at the European Southern Observatory (ESO) Very
Large Telescope [083.B-0126(A) and 083.B-0126(B)].}}

\author[X. Mazzalay et al.]{X. Mazzalay,$^{1}$\thanks{E-mail: ximena@mpe.mpg.de} W. Maciejewski,$^2$ P. Erwin,$^{1}$ R. P. Saglia,$^{1,3}$  R. Bender,$^{1,3}$ M. H. Fabricius,$^{1,3}$ \newauthor N. Nowak,$^{4}$ S. P. Rusli,$^{1,3}$ and J. Thomas$^{1,3}$\\
$^1$ Max-Planck-Institut f\"ur extraterrestrische Physik, Postfach 1312, 85741 Garching, Germany\\
$^2$ Astrophysics Research Institute, Liverpool John Moores University, Twelve Quays House, Egerton Wharf, Birkenhead CH41 1LD, UK\\
$^3$ Universit\"atssternwarte, Scheinerstrasse 1, 81679 M\"unchen, Germany\\
$^4$ Stockholm University, Department of Astronomy, Oskar Klein Centre, AlbaNova, SE-10691 Stockholm, Sweden\\
}

\begin{document}

\date{Accepted for publication in MNRAS}


\maketitle

\begin{abstract}
We present an analysis of the \H2\ emission-line gas kinematics in the inner $\lesssim 4$~arcsec radius of six nearby spiral galaxies, based on AO-assisted integral-field observations obtained in the $K$-band with \S/VLT. 
Four of the six galaxies in our sample display ordered \H2\ velocity fields, consistent with gas moving in the plane of the galaxy and rotating in the same direction as the stars. However, the gas kinematics is typically far from simple circular motion. We can classify the observed velocity fields into four different types of flows, ordered by increasing complexity: (1) circular motion in a disc (\n3351); (2) oval motion in the galaxy plane (\n3627 and \n4536); (3) streaming motion superimposed on circular rotation (\n4501); and (4) disordered streaming motions (\n4569 and \n4579). The \H2\ velocity dispersion in the galaxies is usually higher than 50\kms\ in the inner 1--2~arcsec radii. The four galaxies with ordered kinematics have $v/\sigma < 1$ at radii less than 40--80~pc. The radius at which $v/\sigma = 1$ is independent of the type of nuclear activity. While the low values of $v/\sigma$ could be taken as an indication of a thick disc in the innermost regions of the galaxies, other lines of evidence (e.g. \H2\ morphologies and velocity fields) argue for a thin disc interpretation in the case of \n3351 and \n4536. 
We discuss the implications of the high values of velocity dispersion for the dynamics of the gaseous disc and suggest caution when interpreting the velocity dispersion of ionized and warm tracers as being entirely dynamical. 
Understanding the nature and role of the velocity dispersion in the gas dynamics, together with the full 2D information of the gas, is essential for obtaining accurate black hole masses from gas kinematics.

\end{abstract}

\begin{keywords}
galaxies: nuclei -- infrared: galaxies -- galaxies: ISM -- ISM: molecules -- galaxies: kinematics and dynamics\\
\end{keywords}


\section{Introduction}\label{s_intro}

With the advent of integral field spectrographs on large telescopes, it is now possible to obtain high-spatial-resolution two-dimensional (2D) kinematics of the gas and stars, allowing a more complete picture of the nuclear kinematics and avoiding the misinterpretations inherent to the one-dimensional information of the velocity fields \citep[e.g.][]{McDermid2007,Fu2012,Emsellem2013}. Analysis of the full 2D gas and stellar kinematics obtained with the SAURON integral field unit \citep[IFU,][]{Bacon2001} of a relatively large sample of galaxies have shown that the nuclear ionized-gas kinematics is rarely consistent with simple coplanar circular motions \citep[e.g.][]{Sarzi2006,Ganda2006}. Moreover, from the analysis of 16 galaxies, \citet{Dumas2007} found that deviations from axisymmetric rotation in the gas velocity fields are more frequent in active than in inactive galaxies, suggesting a link between the nuclear gas streaming and the nuclear activity. However, due to the relatively low spatial resolution provided by SAURON ($\gtrsim1$~arcsec), these studies were based on global integrated parameters without analysing the kinematics in detail.

Despite the observed complexity, the 2D gas flows in the centres of galaxies are still often analysed in simplified ways: departures from coplanar circular motions are commonly interpreted in terms of tilted rings \citep[e.g.][]{Neumayer2007,Seth2010} or radial flows \citep[e.g.][]{Wong2004a}. The assumption of tilted rings often lacks physical justification, and radial flows lead to continuity problem, as sink or source is needed in the galaxy centre. Clearly, a better understanding of the gas flows in the centre of galaxies is urgently needed. \citet{Spekkens2007} showed that oval flow can fit the data equally well as the radial flow, thus avoiding the continuity problem. Recently, Maciejewski, Emsellem \& Krajnovic (2012)\nocite{Maciejewski2012} developed a method of recovering the radial and tangential velocity components for an oval flow from integral-field data. Another approach is to deduce from the integral-field data the mechanism that drives the flow, and then fit the data with an appropriate model. This has been done in the case of shocks in galactic bars \citep[e.g.][]{Mundell1999} or nuclear spiral shocks \citep[e.g.][]{Davies2009a}.

Correct interpretation of integral-field data can have implications for radial redistribution of matter in galactic nuclei and therefore for understanding their evolution. Large departures from circular motion also imply that the observed velocity is not a good approximation for the circular velocity, and therefore cannot be used directly to estimate the enclosed mass, like in the case of mass estimates of supermassive black holes (SMBHs) in centres of galaxies. The measurement of SMBH masses using circumnuclear emission-line gas kinematics is often seen as a relatively simple method. Studies using high-spatial-resolution \textit{HST} single-aperture and long-slit spectroscopy \citep[e.g.][]{Ferrarese1996,Macchetto1997,Bower1998,vanderMarel1998,Ferrarese1999,Verdoes2000,Verdoes2002,Barth2001b,Marconi2003b,Marconi2006,Capetti2005,Atkinson2005,deFrancesco2006,deFrancesco2008,Walsh2010,Walsh2013} have usually assumed that the gas is in a dynamically cold, thin disc (rotating in circular orbits) in the plane of the galaxy. However, non-axisymmetric perturbations in the gravitational potential (e.g. bars) or non-gravitational processes, such as AGN-driven outflows \citep[e.g.][]{Rodr'iguez-Ardila2006b,Mazzalay2010,Mazzalay2013b,Muller-S'anchez2011}, can easily lead to strong non-circular gas motions. Moreover, a substantial amount of intrinsic velocity dispersion in gas has been observed in many galaxies; since its origin is not well understood, it is usually ignored or treated in a very simplified manner \citep[e.g.][]{vanderMarel1998,Verdoes2000,Barth2001a}.

This is the second of two companion papers presenting new integral field spectroscopic (IFS) data of six galaxies that display emission lines in their spectra taken from the observational spectroscopic survey of 33 nearby galaxies using the near-infrared (NIR) spectrograph \S\ \citep{Eisenhauer2003,Bonnet2004} at the Very Large Telescope (VLT). While the majority of the programme is focussed on the stellar and dark matter components of the sample \nocite{Nowak2007,Nowak2008,Nowak2010,Rusli2011,Rusli2013a,Thomas2013a} (Nowak et al. 2007, 2008, 2010; Rusli et al. 2011, 2013a, 2013b; Thomas et al. 2013; Bender et al. in preparation; Erwin et al. in preparation; Saglia et al. in preparation), here and in \citet[][hereafter Paper\,I]{Mazzalay2013a} we focus on their gaseous component. The \S\ data provide us the opportunity to study the gas in the innermost regions of galaxies at high spatial resolutions using AO, with the advantages inherent to the NIR spectral region: e.g. low dust extinction and a number emission lines, in particular those of \H2, which are usually less affected by non-gravitational forces than ionized gas \citep[e.g.][]{Riffel2011c}. In \citeta[][]{Mazzalay2013a} we analyse the morphology and physical conditions of the warm molecular (traced by \H2) and ionized (traced by \Brg\ and/or \HeI) emission-line gas in the centres of the galaxies, as well as star formation associated with it.

Here, we present a detailed study of the kinematics in each galaxy of our sample, placing it in the context of the large-scale structure of galaxies. We complement our gas analysis with the stellar kinematics derived from the same dataset. In Section~\ref{s_obs} we describe the \S\ observations and the analysis of the data. The 2D kinematic maps of the gas and stars of each galaxy are presented in Section~\ref{s_results}, together with the relevant large-scale context. A detailed description of the gas kinematics and its interpretation can also be found in this Section. In Section~\ref{s_discussion} we discuss our results and its implications on the gas dynamics, the structure of the gaseous disc and the estimation of SMBH masses from kinematic gas measurements. A summary and our main conclusions can be found in Section~\ref{s_summary}.

\section{Observations and analysis of the data}\label{s_obs}

In this section we give a description of the observations and the procedures used to derive the kinematic information for the six galaxies in our sample. Our sample galaxies, together with some of their main properties, are listed in Table~\ref{t_prop}.

\subsection{IFS observations and data reduction}

The observations and data reduction process were described in detail in \citeta{Mazzalay2013a}. In short, the observations were carried out with the \S\ integral field spectrograph at the VLT in the $K$-band (1.95--2.45\micron\ range). Data for all the galaxies were obtained using the high-spatial resolution configuration (hereafter HR data). These observations were AO-assisted (using either the galaxy nucleus as a natural guide star or a laser guide star) and cover a field of view (FOV) of $3\times 3$~arcsec, with a final spatial sampling of $0.05 \times 0.05$~arcsec. The resulting spatial resolution varies between $\sim 0.1$--0.2~arcsec in full width at half maximum (FWHM). Additionally, low-spatial resolution data (hereafter LR data) were obtained for some of the galaxies in the sample, covering a FOV of at least $8\times 8$~arcsec, with a final spatial sampling of $0.125 \times 0.125$~arcsec. The spatial resolution achieved with this configuration varies between 0.45 and 0.75~arcsec.
Both high- and low-spatial-resolution configurations give a spectral resolution of $\sigma_{\rm inst} \sim 35$\kms\ at 2.12\micron.

The data were reduced using a custom pipeline incorporating {\sc esorex} \citep{Modigliani2007} and {\sc spred} \citep{Schreiber2004,Abuter2006}. The reduction steps include bias subtraction, flat-fielding, bad pixel removal, detector distortion and wavelength calibrations, sky subtraction \citep{Davies2007a}, reconstruction of the object data cubes and, finally, telluric and flux calibrations.

\subsection{Gas kinematics}\label{s_gaskin}

In order to extract the kinematics of the molecular and atomic gas from the \S\ data we used a Markov chain Monte Carlo (MCMC) technique to fit a Gaussian function to the  \H2~2.12\micron, \Brg\ and \HeI\ emission lines observed in the continuum-subtracted spectra of the galaxies of the sample. 
From the central wavelength and width of the Gaussian function we derived the velocity and velocity dispersion ($\sigma$) of the gas in each spatial bin. For the description and subtraction of the continuum we applied the penalized pixel fitting method (pPXF) of \citet{Cappellari2004}, using as the stellar template a combination of the spectra of six late-type stars observed with the same configuration as the galaxy data. Spectral regions containing emission lines or spurious lines due to bad pixels and/or bad sky-lines subtraction were masked to ensure the best representation of the stellar continuum. The adaptive spatial Voronoi binning method of \citet{Cappellari2003} was applied to all our maps, as described in \citeta{Mazzalay2013a}. For this, we chose a signal-to-noise ratio (SNR) threshold for each measured line in each galaxy so as not to compromise the high-spatial resolution in high-SNR regions and, at the same time, be able to extract information in regions of low SNR. 

The emission lines of all the galaxies except \n4579 were well described by a single Gaussian component. The emission-line profiles of \n4579 display strong asymmetries and double components in some regions. Therefore, a single-Gaussian fit is insufficient to properly describe the emission lines, and since a multi-Gaussian approach would lead to large uncertainties in the kinematic parameters (due to the large number of degrees of freedom to constrain such fits), no kinematic maps were constructed for this galaxy. Note that in a multi-Gaussian approach, although the centre and width of the Gaussian-profiles are very uncertain, the total integrated area gives a good estimate of the total emission-line flux \citepa[see][]{Mazzalay2013a}. An analysis of the line profiles observed in different regions of \n4579, along with channel maps, are presented in Sect.~\ref{s_4579}. 

The parameters derived from the Gaussian fitting were used to construct the velocity and velocity dispersion 2D maps. The velocities in each map are given relative to the mean stellar velocity measured in the corresponding dataset (see Section~\ref{s_stellkin} for details on the stellar kinematics). The gas kinematic maps are shown and analysed in the next section. These are centred at the location of the maximum in the 2.1--2.3\micron\ continuum flux, indicated in the figures by a plus sign. The velocity dispersion was corrected for instrumental broadening, subtracting $\sigma_{\rm inst} = 35$\kms\ in quadrature. Regions with very low emission-line fluxes were masked out in the kinematic maps due to the high uncertainty in the line properties. These rejected regions (characterized by an emission-line amplitude smaller than five times the rms scatter of the continuum) are shown in grey in the 2D maps. 

In general, the uncertainties of the kinematic parameters given by their variance within the MCMC samples are $\sim 5$\kms. Since the Gaussian fitting is performed with the continuum-subtracted spectra, this uncertainty does not take into account the errors in the determination of the continuum, which will particularly affect the values derived for the velocity dispersion. In order to derive more realistic uncertainties for the kinematic parameters, we repeated our analysis by subtracting continua equal to the original continuum value $\pm 1$ times the continuum noise level.  As anticipated, this does not significantly change  the values derived for the central velocity, but it does have an impact in the determination of the velocity dispersion, with typical differences between 10--15 \kms. Therefore, we estimate an accuracy of the measurements of $\sim 5$\kms\ in the case of the velocity of the gas and of $\sim 15$\kms\ for the velocity dispersion.

\subsection{Stellar kinematics}\label{s_stellkin}

The stellar kinematics of the galaxies was derived from the \S\ data as part of our efforts to measure the central SMBH masses in these objects via stellar dynamical modelling. A detailed description of the methodology and results can be found in Erwin et al. (in preparation). Here, we use the stellar kinematics as a reference for the analysis of the emission-line gas.

The stellar kinematic analysis followed the basic approach of \citet{Nowak2007, Nowak2008}, with some minor alterations. We derived non parametric line-of-sight velocity distributions (LOSVDs) using the maximum penalized likelihood technique of \citet{Gebhardt2000a}. This involves convolving a set of template stellar spectra with the LOSVD in order to match the observed spectra, focusing on the spectral region of the first two CO bandheads [$^{12}$CO(2--0) and $^{12}$CO(3--1)]. The observed spectra were first spatially binned using the Voronoi method \citep{Cappellari2003}, which ensures that the SNR is high (at least 50) and homogeneous enough to obtain reliable kinematics. As templates we used the spectra of 4--6 late-type stars, chosen from a larger set of 11 stars (all observed using the same instrumental configuration as was used for the galaxies) based on how well their individual CO line-strength indices \citep{M'armol-Queralt'o2008} agreed with the range of indices measured for binned spectra of a particular galaxy. 
Finally, a parametrization of the LOSVD by Gauss--Hermite moments gives the first four moments, $v$, $\sigma$, $h_3$ and $h_4$, from which we constructed the velocity (relative to the mean $v$ measured for each galaxy) and velocity dispersion maps shown in Section~\ref{s_results}.

\section{Individual galaxies}\label{s_results}

In this section we describe and analyse the kinematics of the \H2~2.12\micron\ emission-line gas of the galaxies in our sample, derived from the \S\ data. We also include kinematic maps of the atomic gas for those galaxies which display \Brg\ emission in their spectra, and \HeI~2.06\micron\ in the case of \n4536, the only galaxy in which this line was observed. The emission-line gas kinematics is compared to that of the stellar component. 

\begin{table}
\caption{Properties of the sample galaxies.}
\label{t_prop}
\begin{tabular}{lcccccc}
\hline
Galaxy & D  & Nuclear & Scale & $i~^{\dagger} $ & PA$^{\dagger}$ & PA$_{\rm bar}$ \\
       & [Mpc] & activity$~^{\dagger\dagger}$ & [pc/\arcsec] & [deg]  &  [deg] &  [deg]\\
\hline
\n3351 & 10.0$^a$ & -- & 49 & 46 & 10  & 112$^b$ \\
\n3627 & 10.1$^a$ & L/S2 $^c$ & 49 & 65 & 175 & 161 \\
\n4501 & 16.5$^d$ & S2 $^c$ & 80 & 64 & 140 & --   \\
\n4536 & 14.9$^a$ & LLAGN $^e$ & 72 & 67 & 125 & --  \\
\n4569 & 16.5$^d$ & LLAGN $^f$ & 80 & 69 & 25  & 15$^g$   \\
\n4579 & 16.5$^d$ & L/S1.9 $^f$ & 80 & 40 & 95  & 58$^h$  \\
\hline
\end{tabular}

\medskip
$^{\dagger}$Derived from the analysis of the shape of outer galaxy isophotes (see details in Section~\ref{s_results}). 
$^{\dagger\dagger}$ L = LINER, S = Seyfert nucleus. \\
{\sc references:} $^{(a)}$\citet{Freedman2001}. $^{(b)}$\citet{Erwin2005}. $^{(c)}$Taken from NED. $^{(d)}$Mean Virgo cluster distance \citep{Mei2007}. $^{(e)}$\citet{McAlpine2011}. $^{(f)}$\citet{Ho1997}. $^{(g)}$\citet{Jogee2005}. $^{(h)}$\citet{Garc'ia-Burillo2005}.
\end{table}

\begin{table}
\centering
  \caption{PA of the LON derived from the HR and LR \S\ kinematics of the stars and gas in the sample galaxies.}
    \begin{tabular}{@{}ccccc@{}}
    \hline
\multirow{2}{*}{Galaxy}        & \multicolumn{2}{c}{Stars}       & \multicolumn{2}{c}{Gas} \\
                 &   HR   &   LR   & HR &  LR   \\
\hline
\n3351 &   --    &   $190\pm 3$     &  --  &  $192 \pm 3$  \\
\n3627 & $175 \pm 2$ &  $178 \pm 2$ & $130 \pm 5$ & -- \\
\n4501 & $140 \pm 3$ & -- & -- &  --  \\
\n4536 & $125.5 \pm 2.7$ & $123.5 \pm 1.2$ & $130 \pm 4$  & $121.5 \pm 1.5$  \\
\n4569 & $24 \pm 0.5$ & -- & -- &  -- \\
\hline
\end{tabular}
\label{t_PA}
\end{table}

\subsection{\n3351}\label{s_3351}

\begin{figure*}
\centering
\includegraphics[width=1.\textwidth]{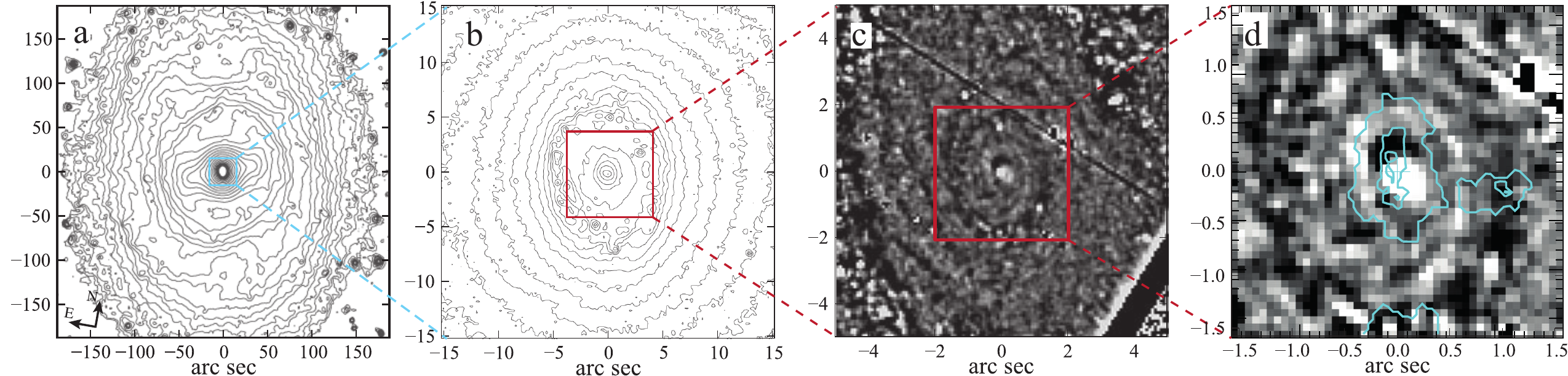}
\caption {IRAC 3.6\micron\ contours [panels (a) and (b)] and $HST$/NICMOS2 F160W unsharp mask [panels (c) and (d)] of \n3351. The contours on panel (d) show the \H2~2.12\micron\ flux distribution (90, 50 and 20 per cent of the flux maximum) measured in our HR \S\ data \protect\citepa{Mazzalay2013a}. The images are shown at the same spatial orientation as the \S\ data.}
\label{f_n3351_im}
\end{figure*}

\n3351 (M95) is an early-type barred galaxy [SB(r)b, \citet{deVaucouleurs1991}] in the Leo~I group, with little or no sign of interaction in its stellar component [panel (a) of Fig.~\ref{f_n3351_im}]. The circumnuclear region is dominated by a star-forming nuclear ring [panel (b) of Fig.~\ref{f_n3351_im}] observed at multiple wavelengths \citep[e.g.][]{Colina1997, Planesas1997, P'erez-Ram'irez2000, Jogee2005}. From the analysis of our SINFONI data, we showed in \citeta{Mazzalay2013a} that this ring, of radius $\sim 7$~arcsec, is also a source of \H2\ and intense \Brg\ line emission. Inside the nuclear ring, NICMOS images reveal tightly wound dust spirals [panels (c) and (d) of Fig.~\ref{f_n3351_im}]. Our SINFONI data show very weak (at noise level) \Brg\ emission inside the ring, while the strongest \H2\ emission comes from an elongated feature at the position of the nucleus, overplotted in contours in panel (d) of Fig.~\ref{f_n3351_im}.

To date, 2D kinematic observations covering the inner $\sim 20$~arcsec of \n3351 have been done in CO \citep{Jogee2005}, and also in \Hb\ and \OIII\ using the SAURON IFU \citep{Dumas2007}. Although CO observations show evidence of non-circular streaming motions in gas, the \Hb\ and \OIII\ velocity maps are regular, following the characteristics of the stellar velocity field, which is also regular within the inner $\sim 20$~arcsec \citep{Dumas2007}. In the SAURON data, the stellar velocity dispersion inside the nuclear ring is $\sim 90$\kms, compared to $\sim 110$\kms outside the ring. The \Hb\ velocity dispersion is low in the nuclear ring ($\sim 50$\kms), but it is slightly higher inside the ring. On the other hand, \citet{Jogee2005} report lower values of velocity dispersion for the cold molecular gas (i.e. CO): 20 and 25\kms\ in the ring and inside it, respectively.

\subsubsection{SINFONI kinematics}

\begin{figure}
\includegraphics[width=\columnwidth]{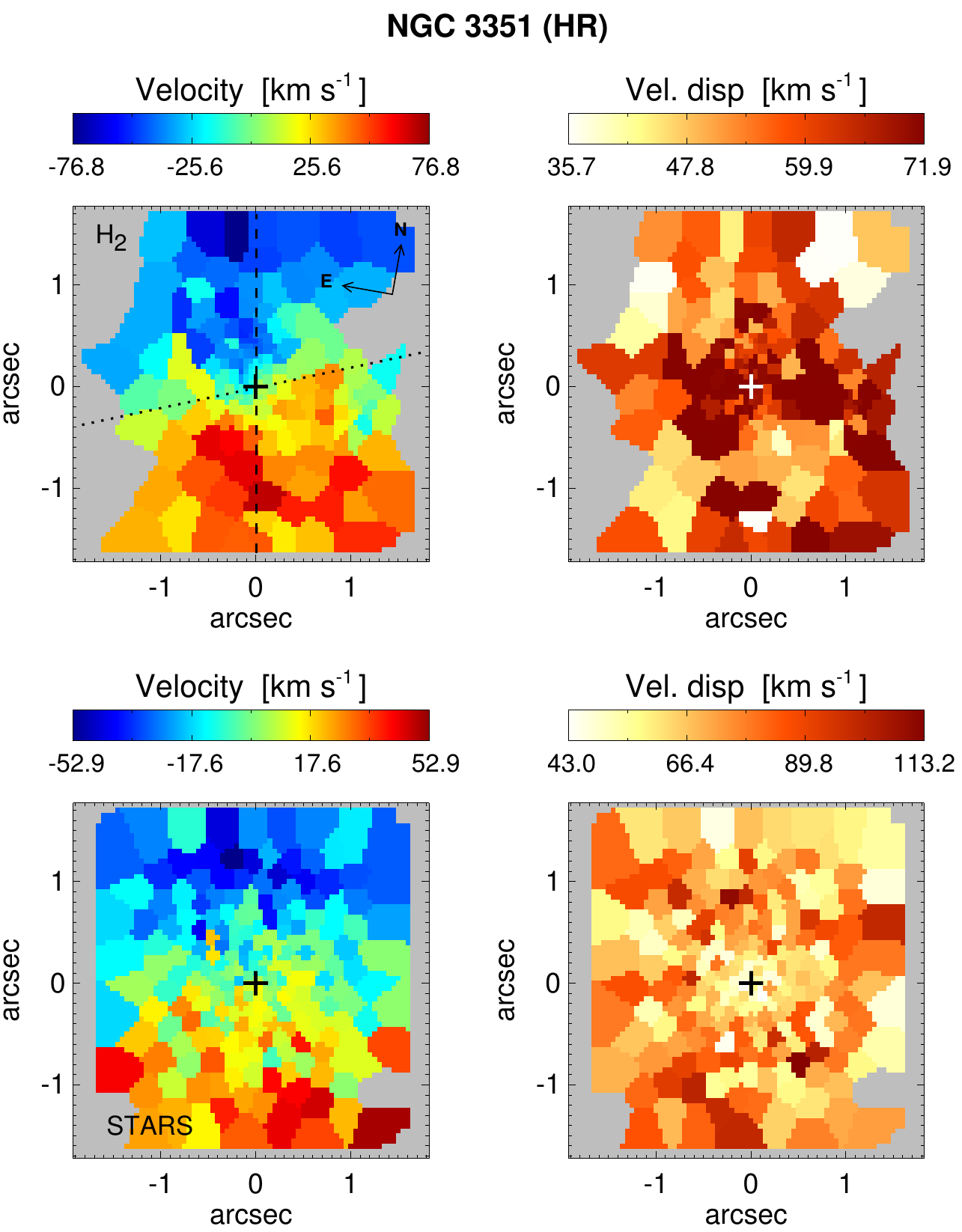}
\caption {2D maps of the \H2\ emission-line gas (upper panels) and stellar (lower panels) kinematics of \n3351 derived from the HR \S\ data. Regions where no reliable parameters were obtained are shown in grey. The spatial orientation, indicated in the upper-left panel, is the same for all the panels. The dashed line corresponds to the position of the major axis of the galaxy and the dotted line to the orientation of the stellar bar.}
\label{f_n3351_kin_100}
\end{figure}

\begin{figure}
\includegraphics[width=\columnwidth]{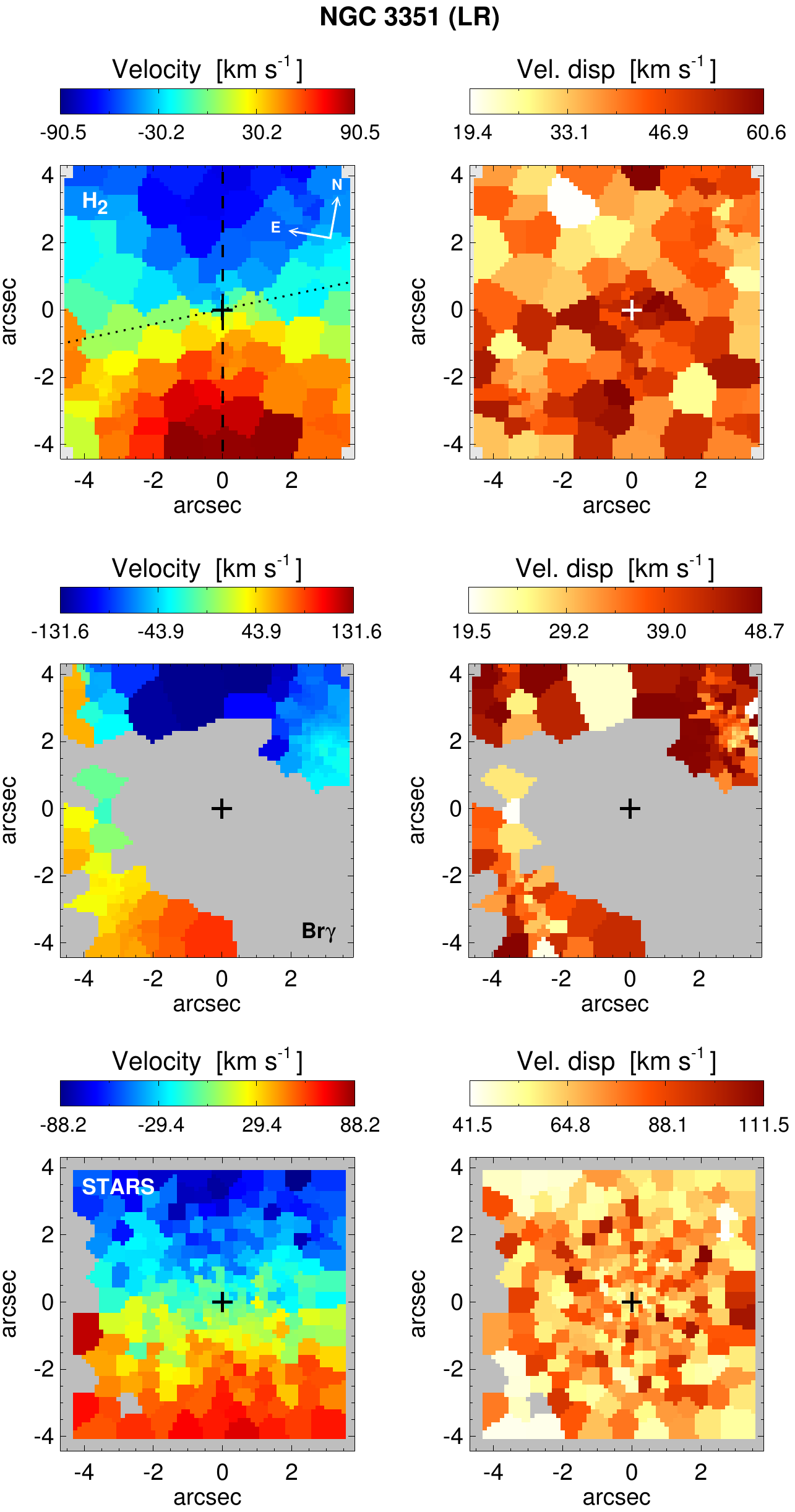}
\caption {2D maps of the \H2\ (upper panels) and \Brg\ (middle panels) emission-line gas kinematics and stellar kinematics (lower panels) of \n3351 derived from the LR \S\ data. Regions where no reliable parameters were obtained are shown in grey. The spatial orientation, indicated in the upper-left panel, is the same for all the panels. The dashed line corresponds to the position of the major axis of the galaxy and the dotted line to the orientation of the stellar bar.}
\label{f_n3351_kin_250}
\end{figure}

Our SINFONI \H2\ kinematic maps derived for the inner $3\times 3$~arcsec (HR map) and $8\times 8$~arcsec (LR map) of \n3351 are shown in the upper panels of Figs.~\ref{f_n3351_kin_100} and \ref{f_n3351_kin_250}, respectively. Although the low SNR displayed by the \H2~2.12\micron\ emission line required us to heavily bin the data, the velocity maps seem to show a global regular rotation pattern, with velocities of up to $\sim 90$\kms. Some deviations are present in the inner regions (especially noticeable in the HR map), where a hint of a (mild) twist is observed in the zero-velocity line. The velocity dispersion maps present no particular structure, with values between $\sim 25$--70\kms, which is consistent with the SAURON \Hb\ data, and higher than the CO dispersion. 

The stellar kinematic maps (lower panels of Figs.~\ref{f_n3351_kin_100} and \ref{f_n3351_kin_250}) show a rotational pattern aligned with that of the gas. In the inner $3\times 3$~arcsec, the maximum stellar velocity is only $\sim 50$\kms, significantly lower than the gas velocity. On the other hand, the measured velocity dispersion in stars, $\sigma \sim 50$--90\kms, is higher than that in the gas for the majority of the observed field, and consistent with the SAURON data.

Since a clear detection of \Brg\ emission only was possible in the region of the nuclear ring of the galaxy, which is located at the edges of the LR FOV, we have kinematic information for the ionized gas in only a small region of the FOV of the LR data (middle panels of Fig.~\ref{f_n3351_kin_250}). In general, the \Brg\ kinematic maps are consistent with the ones derived for the \H2\ emission-line gas. The \Brg\ emission-line gas located north of the nucleus seems to have a higher velocity than the molecular gas in this region. However, this high velocity should be viewed with caution, since they correspond to a region of weak \Brg\ emission and are probably affected by larger errors.

Applying the method described in Appendix~C of \citet{Krajnovic2006}, we measure from the LR data a position angle (PA) of the line of nodes (LON) of the molecular gas kinematics of $192 \pm 3$\deg, and $190 \pm 3$\deg\ for the stellar kinematics (Table~\ref{t_PA}).  
These values are similar to the corresponding values of $\sim 200$\deg\ and $190$\deg\ reported by \citet{Dumas2007} for the inner 4.3~arcsec of \n3351. Our measurements suggest that if there is a misalignment between gas and stars, it is very mild.
Given that the photometric PA derived from the outer disc is 10\deg\ (or $190$\deg\ for comparison with kinematics\footnote{The kinematic PA has periodicity of 360\deg\ while that of the photometric PA is 180\deg}, Table\ref{t_prop}), the observed kinematics is consistent with circular motion of gas and stars inside the nuclear ring. In particular, the stars do not show any signature of being closely trapped by elongated x1 orbits making up the bar, although this is expected given that the stellar velocity dispersion is much higher than the rotational velocity and the stellar isophotes are round inside the nuclear ring [see panel (b) of Fig.~\ref{f_n3351_im}]. 

\begin{figure}
\centering
\includegraphics[width=0.75\columnwidth]{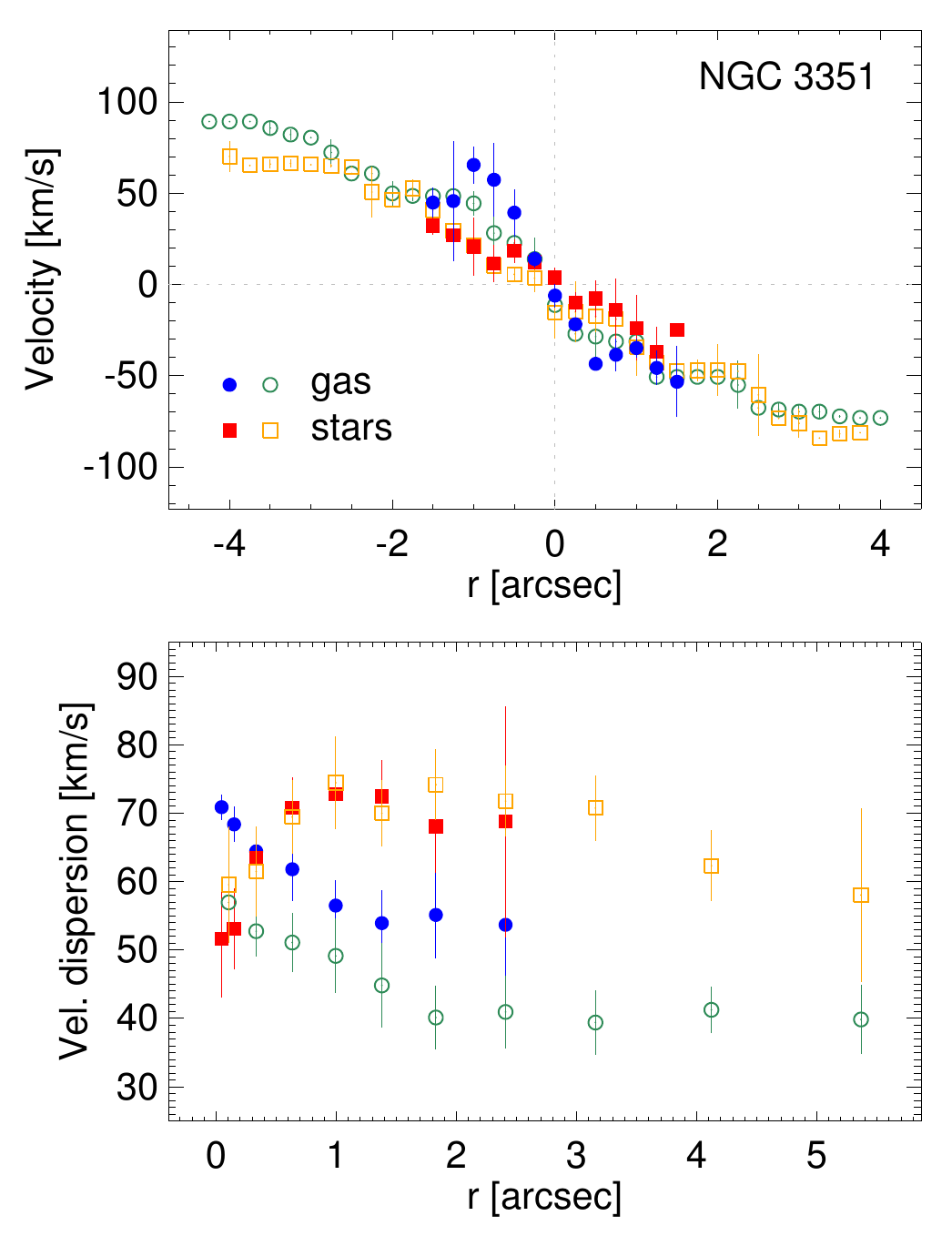}
\caption{\n3351 -- Upper panel: mean values of the velocity measured along the major axis of the galaxy (190\deg) in apertures of $0.25 \times 0.25$~arcsec; negative radii are south from the nucleus. Lower panel: mean velocity dispersion measured in annular apertures. In both panels blue filled and green open circles correspond to the values measured for the \H2\ gas in the HR and LR data, respectively. Red filled and orange open squares correspond to the stellar values measured in the HR and LR data, respectively. The error bars indicate the 3$\sigma$ error in the mean.}
\label{f_n3351_2D}
\end{figure}

The upper panel of Fig.~\ref{f_n3351_2D} shows the mean velocity of the \H2\ emission-line gas and stars calculated from the velocity maps for regions of $0.25 \times 0.25$~arcsec along the major axis of the galaxy, which is equivalent to a rotation curve obtained with a long slit. Note that our FOV captures only part of the overall rise of the rotation curve, which reaches almost 200\kms\ within the central kiloparsec \citep[e.g.][]{Mazzuca2011,Fabricius2012}.

The velocity dispersion, averaged over annuli, is shown for \H2\ emission-line gas and stars in the lower panel of Fig.~\ref{f_n3351_2D}. Within a radius of $\sim 0.3$~arcsec (15~pc), there is a clear drop in the velocity dispersion of stars (also seen in the HR map), reaching a minimum of $\sim 50$\kms. This is lower than the velocity dispersion of the gas in this region. Similar drops in stellar velocity dispersion have been observed in the centres of several galaxies \citep[e.g.][]{Emsellem2001a, M'arquez2003a, Falc'on-Barroso2006, Davies2007b, Peletier2007a, Comer'on2008, Nowak2010, Riffel2009a, Riffel2010b, Riffel2011b, Hicks2013}, and are usually associated with dynamically cold structures resulting from recent star formation, such as nuclear discs.

While the stellar velocity dispersion is (marginally) smaller than that displayed by the warm molecular gas, it is still much higher than that reported from CO observations \citep{Jogee2005}.
The relatively high values of velocity dispersion that we measure for the gas could be a consequence of the presence of SNe in the centre of the galaxy. As we found in \citeta{Mazzalay2013a}, the \H2\ line ratios in our SINFONI spectra of this galaxy are consistent with predominantly thermal excitation, probably produced by shocks associated to SNe. 

If the observed \H2\ velocity dispersion is representative of the majority of the molecular gas, then stars are not currently being formed out of gas in the innermost 15~pc, because stars cannot exhibit lower velocity dispersion than gas out of which they form. Stars probably \textit{will} form once this gas has cooled down and settled in a disc with lower velocity dispersion. 
In this scenario, star formation in the innermost 15~pc of \n3351 is probably episodic in nature. Evidence for multiple episodes of recent star formation in the central regions of \n3351 supports this picture. \citet{Sarzi2005} analysed \textit{HST}-STIS spectroscopy of the nucleus (using a $0.2 \times 0.25$~arcsec aperture, corresponding to $10 \times 12$~pc) and found evidence for multiple stellar populations, with a 10~Myr component contributing 25 per cent of the nuclear luminosity and a 100~Myr population another 20 per cent. The presence of a $\sim 10$~Myr  population in the nucleus is consistent with our analysis in \citeta{Mazzalay2013a} of the \Brg\ emission, where we estimated an upper age limit (ignoring any contributions from older populations) of 10.5~Myr in the inner $\sim 1$~arcsec (50~pc) for a stellar population capable of exciting \Brg\ emission. The presence of such a young population implies the occurrence of SNe, which could contribute to the gas kinematics as mentioned above. On the other hand, if the majority of the gas is characterized by the velocity dispersion measured from CO data ($\sim 25$ \kms), then such a restriction on current star formation is not applicable. 

\begin{figure}
\includegraphics[width=\columnwidth]{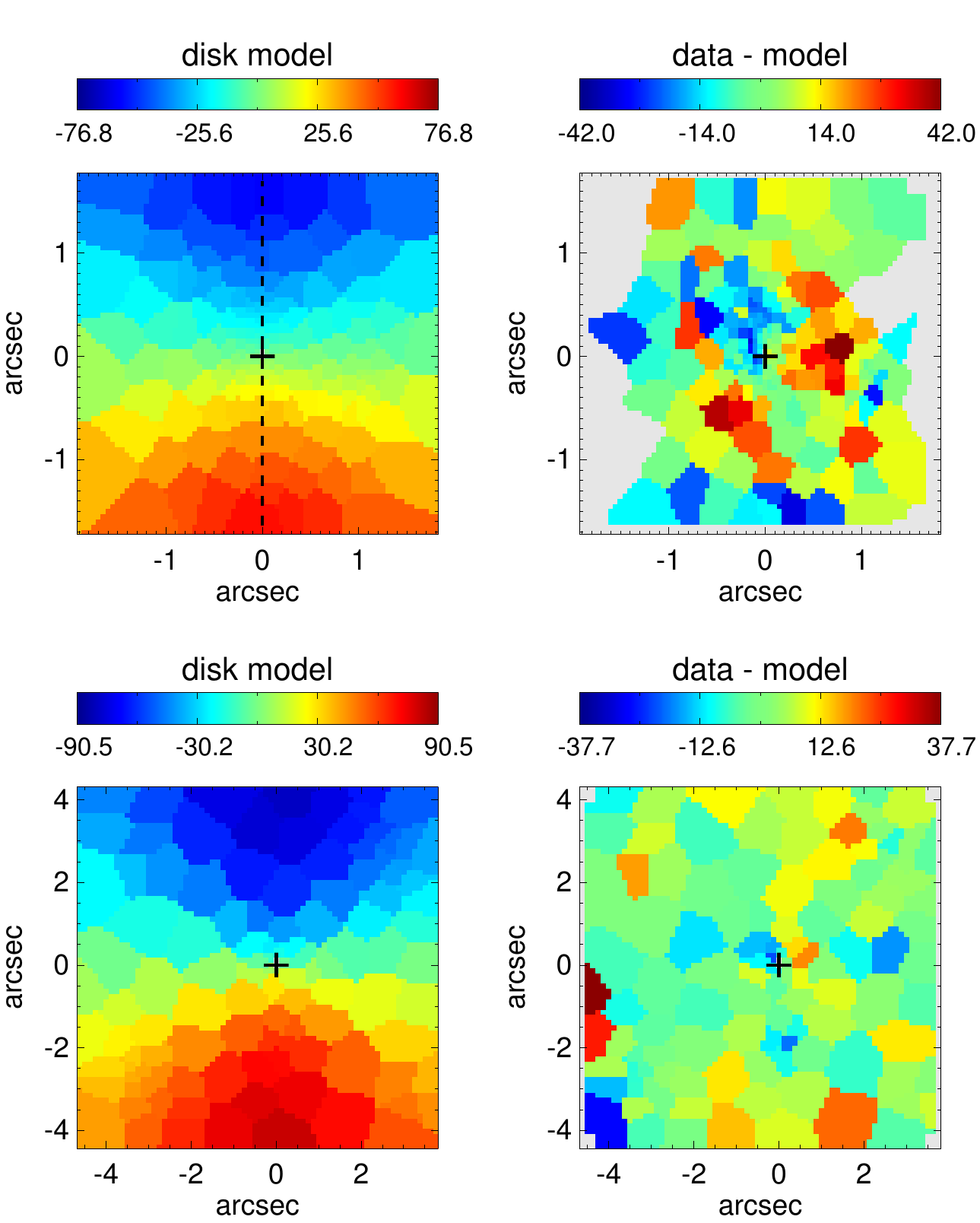}
\caption {Best-fitting disc model of the \H2\ velocity field and the velocity residuals of the inner $3 \times 3$~arcsec (HR data, upper panels) and $8 \times 8$~arcsec (LR data, lower panels) of \n3351. The dashed line in the upper-left panel indicates the position of the major axis of the galaxy.}
\label{f_n3351_disk}
\end{figure}

\subsubsection{2D kinematic analysis}

In order to estimate the amplitude and distribution of any deviations from the circular rotation in the \H2\ emission-line gas, we fitted a rotating exponential thin-disc model to the observed velocity maps (HR and LR) and subtracted it from the data. 
The circular velocity of an infinitesimally thin disc with an exponential mass distribution $\Sigma(r)=\Sigma_0~{\rm exp}(-r/r_d)$ can be written as
\[ v_{circ}^2(r)=4\pi G\, \Sigma_0\, r_d\, y^2 ~[I_0(y)\,K_0(y)-I_1\,K_1(y)]\,, \]
where $G$ is the gravitational constant, $y\equiv r/r_d$ and $I_n$ and $K_n$ are modified Bessel functions \citep[see eqs. 2.162--2.165 of][]{Binney2008}. The line-of-sight (LOS) velocity of the disc (of inclination $i$) can be expressed at each position $(x_s,y_s)$ on the sky, with the $x$-axis aligned with the LON, as  
\[ v_{\rm LOS}(x_s,y_s) = v_{circ}(r)~\frac{x_s}{r}~\sin i\,, \qquad r^2=\left(\frac{y_s}{\cos i}\right)^2+x_s^2. \]

Note that we are using a {\em thin}-disc model even though there is probably a substantial contribution to the gas kinetic energy from the velocity dispersion, which could produce a thick disc; see the discussion in Section~\ref{s_disc_thick}. A thorough description of the dynamics of the gas would also take into account this component, but this is beyond of the scope of the paper and the simple thin-disc model suffices for the purpose of finding deviations from circular rotation. We used the MCMC technique to explore the parameters space of the disc model (fixing its the inclination and PA to that of the galaxy, see~Table\ref{t_prop}) and obtained a unique solution that gives the best possible description of both the HR and LR data at the same time. Fig.~\ref{f_n3351_disk} shows the best-fitting model and the residual velocity maps. 
In general, the velocity residuals are of the order of $\pm 25$\kms\ or less, and do not form any coherent structure. These results support the hypothesis that the \H2\ emission-line gas is coming from a disc in circular rotation around the nucleus of \n3351.

\subsection{\n3627}\label{s_3627}

\begin{figure*}
\centering
\includegraphics[width=\textwidth]{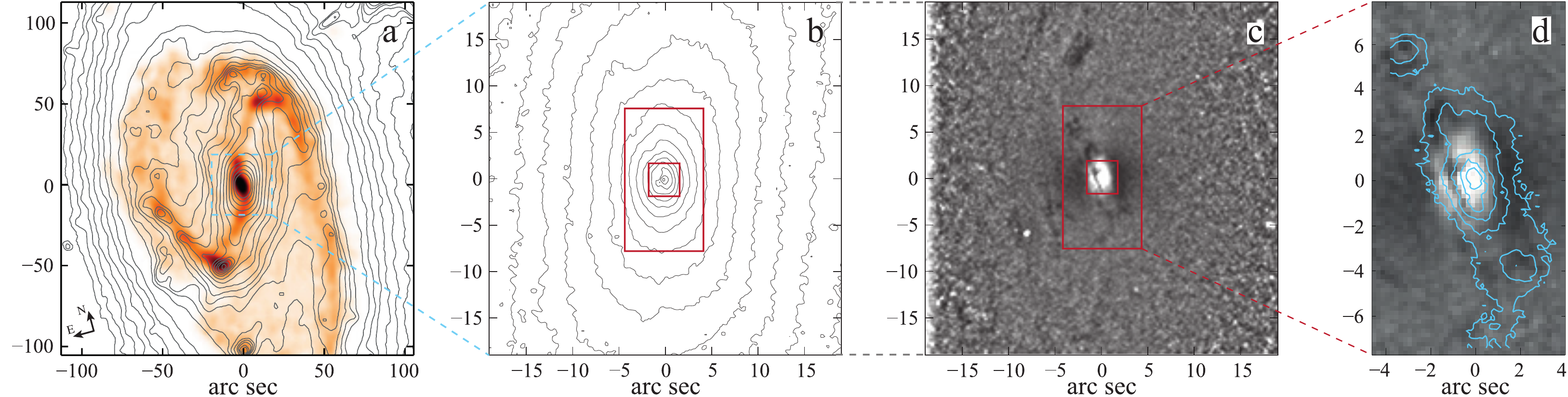}
\caption {\n3627 -- Panel (a): IRAC 3.6\micron\ contours overlaid on the CO BIMA image of \protect\citet{Helfer2003}. Panel (b): $HST$/NICMOS F160W contours. Panel (c) and (d): $HST$/NICMOS  F160W unsharp masks. The contours overploted in panel (d) correspond to the \H2~2.12\micron\ flux distribution derived from our LR \S\ data \protect\citepa{Mazzalay2013a}. The images are shown at the same spatial orientation as the \S\ data. The red boxes in panels (b) and (c) indicate the \S\ LR and HR FOVs.}
\label{f_n3627_im}
\end{figure*}

\begin{figure}
\centering
\includegraphics[width=\columnwidth]{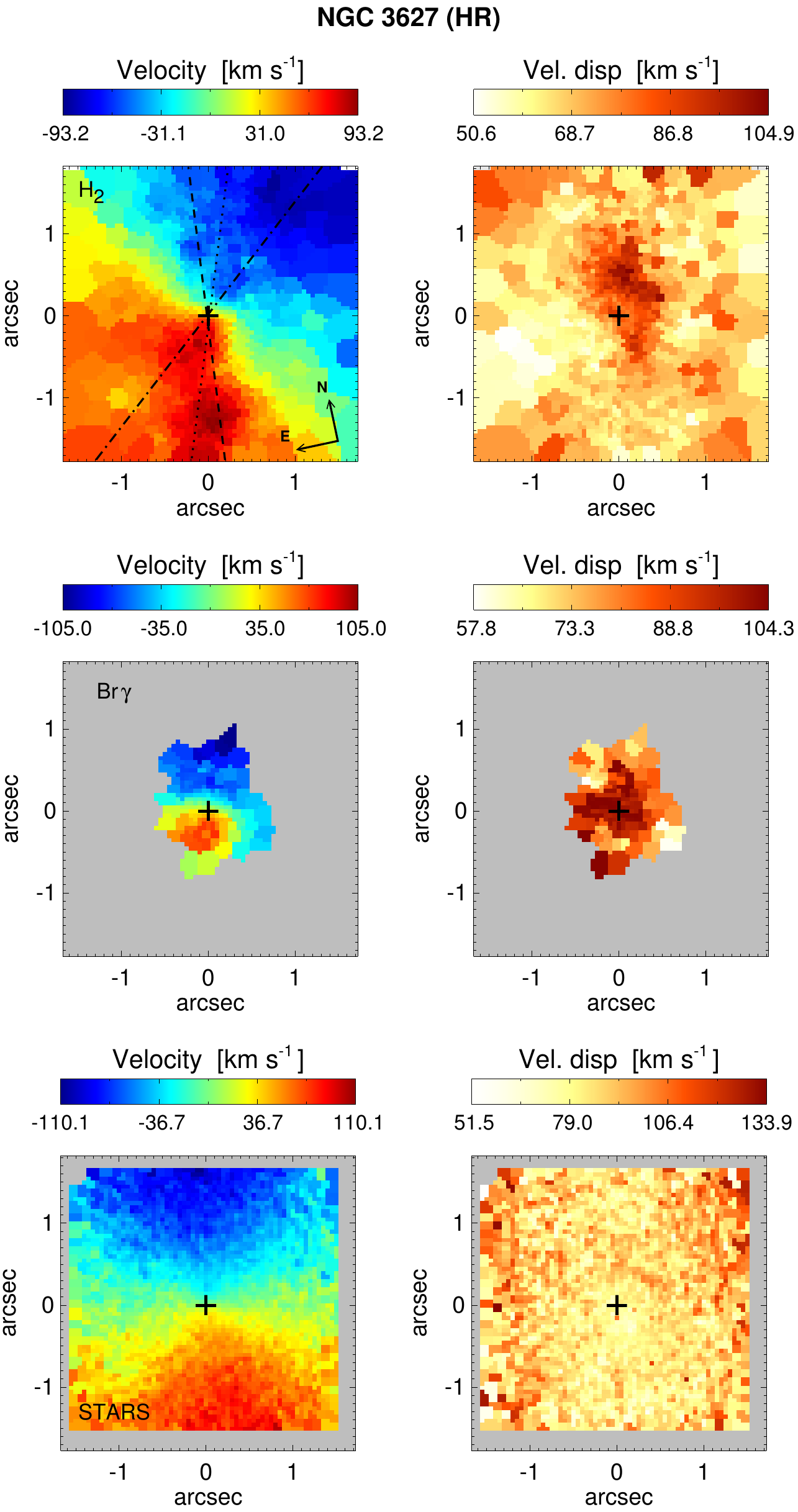}
\caption {2D maps of the \H2\ (upper panels) and \Brg\ (middle panels) emission-line gas kinematics and stellar kinematics (lower panels) of \n3627 derived from the HR \S\ data. Regions where no reliable parameters were obtained are shown in grey. The spatial orientation, indicated in the upper-left panel, is the same for all the panels. The dashed line in the \H2\ velocity map corresponds to the PA of the major axis of the galaxy (175\deg), the dotted line to the orientation of the stellar bar (161\deg) and the dot-dashed line to the PA of the \H2\ LON derived from the data (130\deg).}
\label{f_n3627_kin_100}
\end{figure}

\begin{figure}
\centering
\includegraphics[width=\columnwidth]{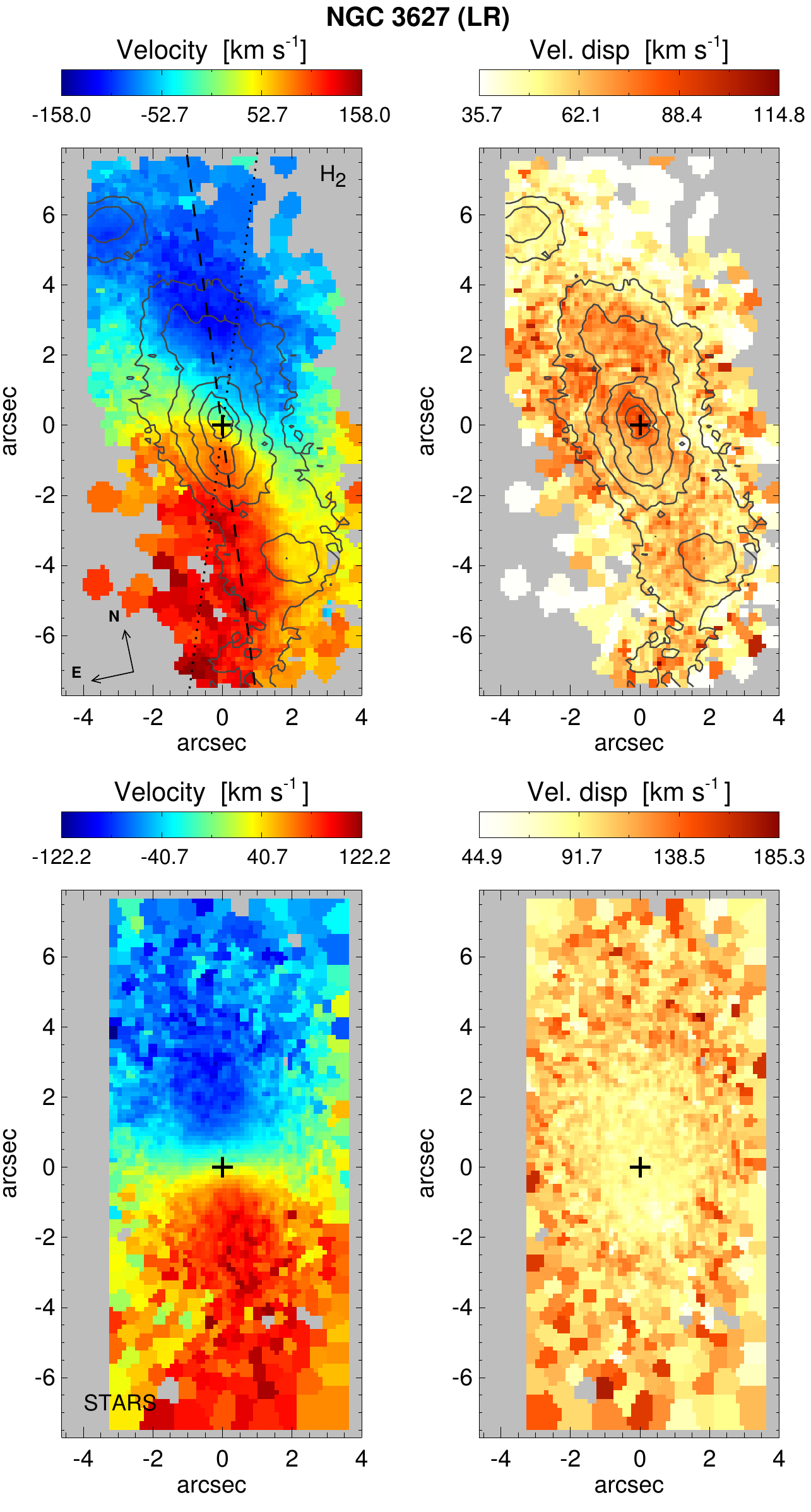}
\caption {2D maps of the \H2\ emission-line gas (upper panels) and stellar (lower panels) kinematics of \n3627 derived from the LR \S\ data. Regions where no reliable parameters were obtained are shown in grey. The spatial orientation, indicated in the upper-left panel, is the same for all the panels. The dashed line in the \H2\ velocity map corresponds to the PA of the major axis of the galaxy and the dotted line to the orientation of the stellar bar. The contours on the upper-panels show the \H2~2.12\micron\ flux distribution.}
\label{f_n3627_kin_250}
\end{figure}

\begin{figure}
\centering
\includegraphics[width=0.75\columnwidth]{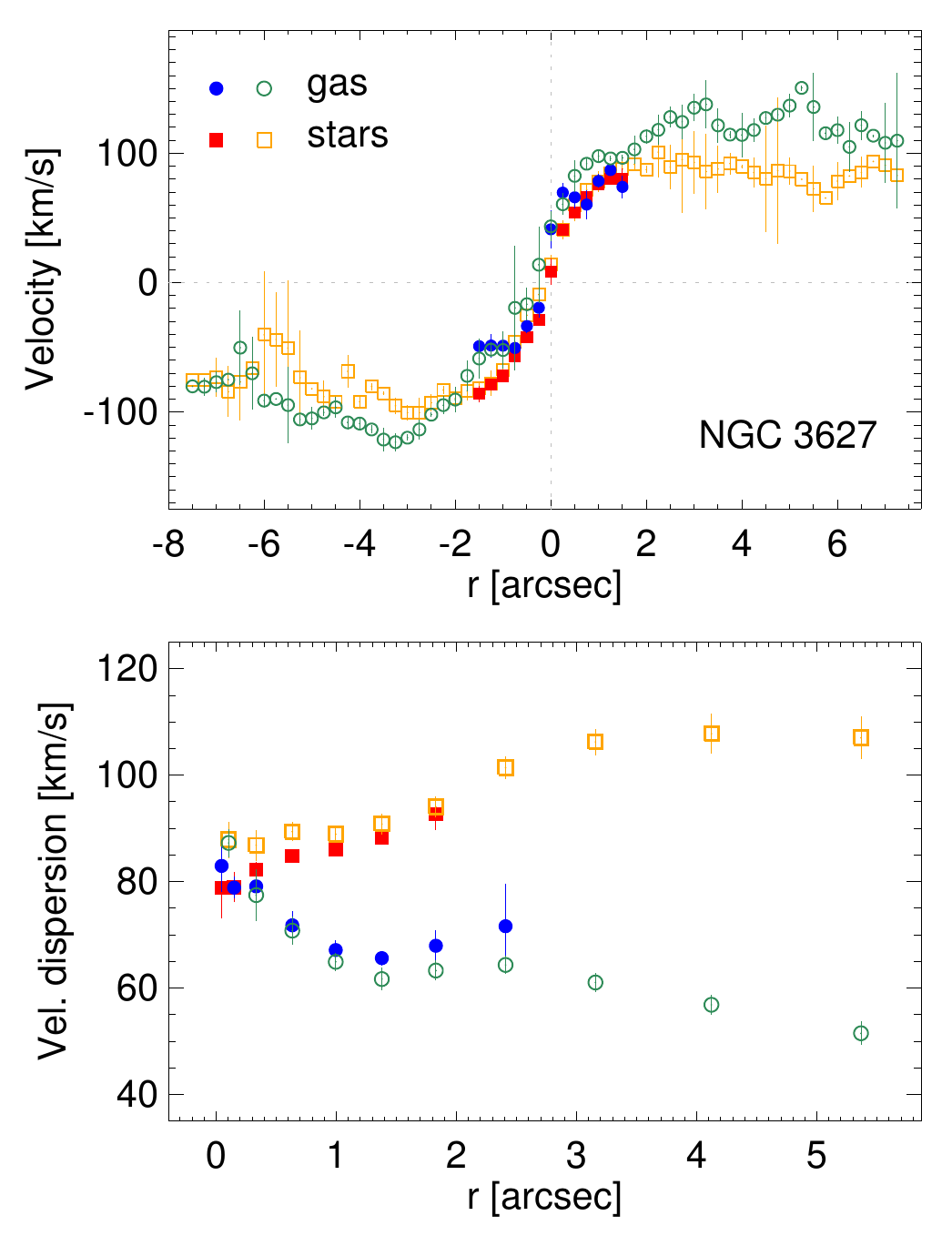}
\caption {\n3627 -- Upper panel: mean values of the velocity measured along the major axis of the galaxy (175\deg) in apertures of $0.25 \times 0.25$~arcsec; negative radii are north from the nucleus. Lower panel: mean velocity dispersion measured in annular apertures. In both panels blue filled and green open circles correspond to the values measured for the \H2\ gas in the HR and LR data, respectively. Red filled and orange open squares correspond to the stellar values measured in the HR and LR data, respectively. The  error bars indicate the 3$\sigma$ error in the mean.} 
\label{f_n3627_2D}
\end{figure}

\n3627 is a barred galaxy \nocite{deVaucouleurs1991}[SAB(s)b, de Vaucouleurs et al. 1991] which shows signatures of interaction with its neighbour \n3628 \citep[e.g.][]{Zhang1993, Afanasiev2005}. Panel (a) of Fig.~\ref{f_n3627_im} shows \textit{Spitzer} IRAC1 isophotes of the galaxy; the asymmetry between the western and eastern spiral arms is evidence of the distortions in the outer disc. Also visible is the strong bar, oriented almost vertically in the figure (we measure a PA = 161\deg\ for the outer part of the bar); the deprojected semi-major axis of the bar is $\sim 60$ arcsec \citep{Erwin2013}. The combination of boxy inner isophotes and narrow, slightly offset spurs in the outer part of the bar are indications that the bar has vertically buckled \cite[see the discussion in][]{Erwin2013}: the boxy isophotes in panel (b) of Fig.~\ref{f_n3627_im} are thus projections of the thick, `box/peanut structure', while the isophotes in the interior ($r \la 5$~arcsec) become rounder, which is evidence for a more axisymmetric central component (a bulge or pseudobulge; see \nocite{Afanasiev2005}Afanasiev \& Sil'chenko 2005 and Erwin et al. in preparation).

Because of the interaction and the presence of a bar, the orientation of \n3627 is somewhat uncertain, as the outer disc may not be intrinsically circular, and the strong bar may produce non-circular motions, even outside the bar itself. From a combination of ellipse fits to the outer isophotes of the IRAC1 image, the morphology of the bar \citep[see][]{Erwin2013} and fitting of the SINFONI HR stellar velocity field (see details below), we derive an overall PA for the galaxy's LON of $\approx 175\degr$, with a probable uncertainty of several degrees. This is roughly consistent with the estimates of \citet{Afanasiev2005}, \citet{Dicaire2008b} and \citet{Casasola2011}. From the shape of the outer isophotes (ellipticity $\sim 0.54$), we derive an inclination of 65\degr, assuming an intrinsically circular disc with thickness $c/a = 0.2$; this value is the same as that derived by \citet{Dicaire2008b} from the kinematic analysis of their large-scale H$\alpha$ data, but different from their photometric estimate of 57\degr\ from the shape of the outer isophotes. Therefore the inclination is uncertain to at least 5\degr.

This galaxy has been the target of numerous CO observations on different scales and resolutions. In panel (a) of Fig.~\ref{f_n3627_im}, we overplot the CO(1--0) emission from the BIMA observations of \citet{Helfer2003}, which show molecular gas associated with the spiral arms and along the leading edges of the bar \citep[see also][] {Regan2001,Sheth2002}. Higher-resolution observations by \citet{Casasola2011} show that the CO emission in the inner $20 \times 20$~arcsec of the galaxy is resolved into a nuclear peak and an elongated bar-like structure of $\sim 18$~arcsec length, oriented along ${\rm PA} \simeq 14$--15\degr. This is significantly tilted with respect to the orientation of the stellar bar ($161^{\circ}$; Table~\ref{t_prop}). The ionized gas morphology in the same region is tilted in the \textit{opposite} direction, along a PA of $\sim 152$\deg\ \citep{Dumas2007}.

The \H2\ morphology derived from our \S\ observations of the inner $8 \times 15$~arcsec of the galaxy [Paper\,I, see also the contours on panel (d) of Fig.~\ref{f_n3627_im}] is remarkably consistent with the CO morphology observed on larger scales by \cite{Casasola2011}. In the \S\ \H2\ flux distribution maps, we see further morphological details in the nucleus. The molecular bar-like structure seen by \citet{Casasola2011} is resolved into a nuclear structure with length $\sim 6$~arcsec, elongated approximately along the north-south direction, and two hot-spots located at about 6.5~arcsec NE and 4.5~arcsec SW from the nucleus, along PA $\simeq 17\degr$. The PA of the \H2\ emission varies from $\sim 10$--15\deg\ near the outer part of our LR FOV -- consistent with the orientation seen by \citet{Casasola2011} -- to $\sim 0$\deg\ for $r < 3$~arcsec. Panels (c) and (d) of Fig.~\ref{f_n3627_im} show unsharp masks of the NICMOS3 F160W image, emphasizing regions of strong dust extinction. In panel (d), we overplot the \H2\ morphology derived from the LR data on top of the unsharped masked image. From this, we can see that the NE blob of \H2\ emission is associated with patchy dust, and that the innermost, circumnuclear dust lanes have the same orientation as the elongated nuclear \H2\ emission, suggesting a close connection between the dust and the \H2\ emission-line gas.

\subsubsection{SINFONI kinematics}\label{s_3627_ss1}
The \H2\ emission-line gas kinematics derived from the HR and LR \S\ data is presented in the upper panels of Figs.~\ref{f_n3627_kin_100} and \ref{f_n3627_kin_250}, respectively. Additionally, the middle panels of Fig.~\ref{f_n3627_kin_100} show the kinematics of the ionized gas, traced by the \Brg\ emission-line. Since no \Brg\ emission was observed in regions outside those covered by the HR data, we only show the kinematics of the atomic gas derived from this data set. In the lower panels of Figs.~\ref{f_n3627_kin_100} and \ref{f_n3627_kin_250}, we show the stellar kinematics derived from the HR and LR \S\ data, respectively. The observed stellar kinematics is regular in both HR and LR data. Applying the method described in \citet{Krajnovic2006}, we determined a kinematic PA for the stars in the $3 \times 3$~arcsec HR field of $175^{\circ} \pm 2^{\circ}$, and in the $8 \times 15$~arcsec LR field of $178^{\circ} \pm 2^{\circ}$ (Table~\ref{t_PA}). These values are consistent with that derived by \citet{Dumas2007} for the stars in the innermost 6~arcsec of their SAURON FOV (175\degr); note, however, that these values are different from the global values reported by these authors.

Numerous studies of molecular and ionized gas kinematics on large scales \nocite{Dicaire2008b,Casasola2011}(e.g. H$\alpha$: Dicaire et al. 2008; CO: Casasola et al. 2011) and on scales of $r \la 15$~arcsec \nocite{Afanasiev2005,Dumas2007,Casasola2011}(e.g. [N\,{\sc ii}]: Afanasiev \& Sil'chenko 2005; H$\beta$ and [O\,{\sc iii}]: Dumas et al. 2007; CO: Casasola et al. 2011) show evidence for strong non-circular motions within the bar. Similarly, in our \S\ data the \H2\ emission-line gas kinematics significantly differs from stellar kinematics in both HR and LR. The HR \H2\ velocity map shows a global rotation pattern that is tilted with respect to the one observed for the stars. This kinematic misalignment indicates that the gas cannot be in circular rotation coplanar with the stars. Using the same method as we used for the stellar kinematics, we determined a kinematic PA for the molecular gas in the inner $3 \times 3$~arcsec of $130 \pm 5$\deg. This corresponds to a misalignment of the kinematic axis of molecular gas of $45$\deg\ with respect to the PA of the galaxy (see  Fig.~\ref{f_n3627_kin_100}). The stars--gas misalignment in the innermost arcseconds of \n3627 has also been observed in ionized gas. \citet{Dumas2007} reported a PA of about 137\deg\ for the innermost 6~arcsec of their \OIII\ SAURON data, in agreement with our estimation for the molecular gas. 

A closer look to the HR velocity map reveals that the general rotation pattern is not followed by the gas in the innermost $r \lesssim 0.3$~arcsec (or 15~pc), where a twist in the zero-velocity line is observed. The measured PA of the gas in this region is $\sim 145$\deg, which is between the PA of the galaxy and the kinematic PA of the gas measured in the entire HR FOV (Tables~\ref{t_prop} and \ref{t_PA}). This change in the zero-velocity line is supported by the observed \Brg\ emission-line gas kinematics (middle-left panel of Fig.~\ref{f_n3627_kin_100}). Even though the compactness and weakness of the \Brg\ emission allow us to obtain kinematic information of the ionized gas only on a small region of the HR FOV, in general, it seems to be consistent with that of the molecular gas.

A more complete view of the complex \H2\ molecular gas kinematics can be seen in the maps constructed from the LR \S\ data, shown in the upper panels of Fig.~\ref{f_n3627_kin_250}. The gas velocity field is not regular. Particularly noticeable are the outer parts of the velocity map, where on the left-hand side the negative near-zero isovelocity lines (around the green-to-blue transition) seem to deviate up as one moves away from the centre of the FOV and then down again right at the edge of the FOV; on the right-hand side, the positive near-zero isovelocity lines (around the green-to-yellow transition) deviate sharply downward, with a suggestion of a similar reversal upward right at the edge of the FOV. This `stretching' of the isovelocity lines, roughly parallel to the LON, occurs towards the two blobs of emission observed in the northern and southern regions of the \H2\ distribution. It is not possible to obtain a precise value for the gas LON from this dataset, since the weakness of the \H2\ emission line near the edges of the FOV means that we are missing key information. However, from the map in Fig.~\ref{f_n3627_kin_250} it is possible to place the LON of the gas in this region at an angle approximately in between the one measured for the gas in the inner $3 \times 3$~arcsec (PA=130\deg) and the major axis of the galaxy, PA=175\deg. Moreover, the zero-velocity line (PA $\sim 45$\deg\ within a radius of 1.5~arcsec) is not perpendicular to the LON, which indicates non-circular motions throughout the field.

The upper panel of Fig.~\ref{f_n3627_2D} shows the velocity curves along the major axis of the galaxy for the \H2\ emission-line gas and stars. A steep rise in both curves out to a distance of $\sim 1.5$~arcsec from the nucleus can be seen. Beyond this point, the curves flatten, with a velocity around $\pm 90$\kms\ for the stars and about $\pm 130$\kms\ in the case of the molecular gas. It is worth remembering that these velocity curves are measured along the galaxy major axis (or stellar LON), which does \textit{not} coincide with the LON of the gas in this region. Nevertheless, both curves seem to agree quite well.

The velocity dispersion of the \H2\ emission-line gas varies between $\sim 40$--100\kms\ over the FOVs of the HR and LR \S\ datasets (upper-right panels of Figs.~\ref{f_n3627_kin_100} and \ref{f_n3627_kin_250}). The highest values are observed in the innermost 1~arcsec. A peculiar asymmetric distribution can be seen in the HR map, where the higher velocity dispersion gas is located towards the north and north-west from the nucleus. As in the case of \n3351, there is a drop in the velocity dispersion of the stars as one goes toward the centre of \n3627. However, in \n3627 the drop is not as steep or abrupt as in \n3351 (this can be better seen in the lower panel of Fig.~\ref{f_n3627_2D}), and the stars and gas seem to share a common velocity dispersion $\sigma \sim  80$\kms\ at the centre. The gas shows the opposite behaviour to that of the stars, with its velocity dispersion increasing toward the centre.

\subsubsection{2D kinematic analysis}\label{s_3627_ss2}

\begin{figure*}
\centering
\includegraphics[width=0.7\textwidth]{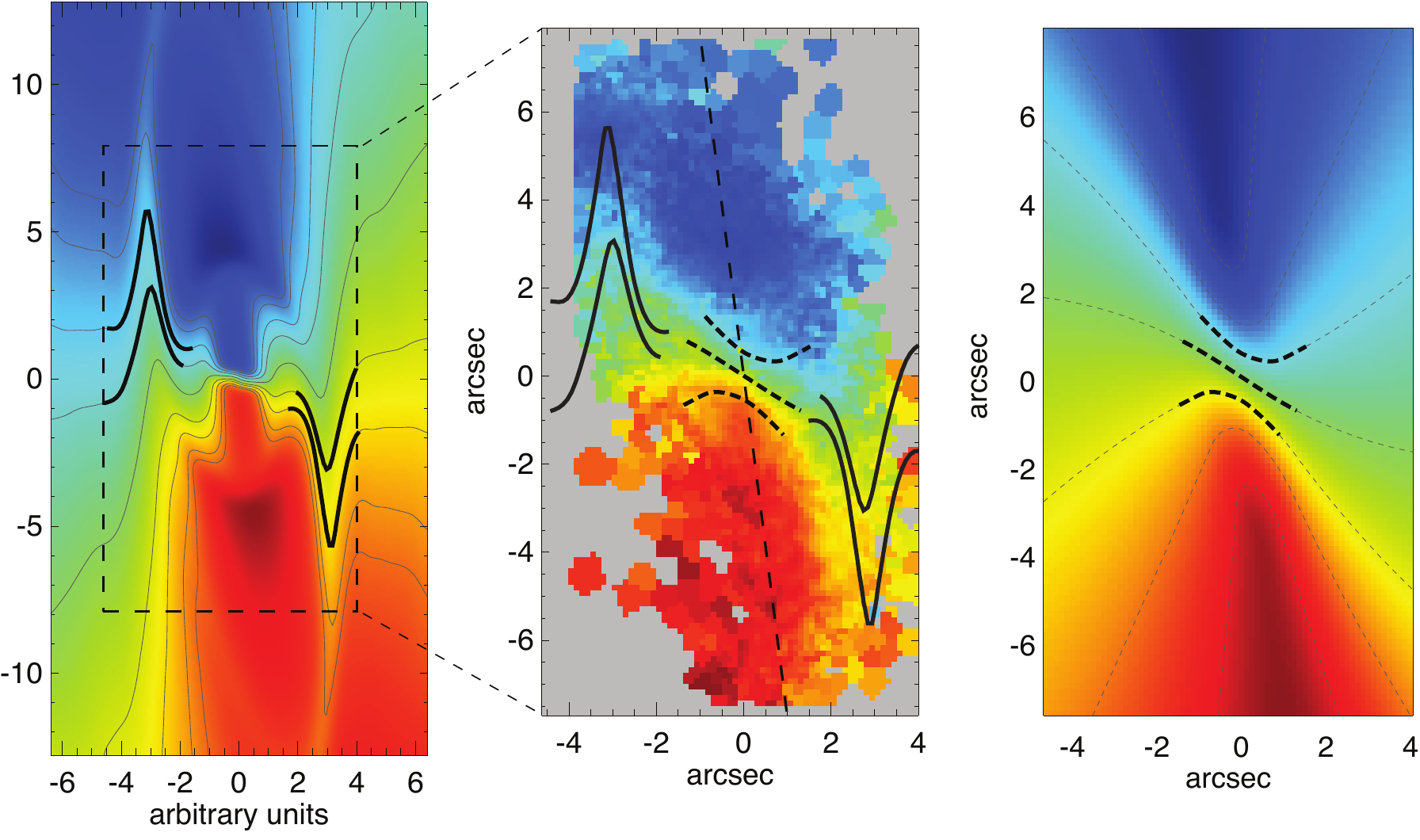}
\caption {Comparison of models of gas flow with the \H2\ kinematics in \n3627. The left panel shows the LOS velocity field of hydrodynamical model of gas flow in a barred galaxy (Model S20 of Maciejewski et al. 2002), viewed with the same orientation as  the galaxy and resized to approximately the same spatial scale, so that the dashed box corresponds to the SINFONI LR FOV. Isovelocity lines are indicated by solid lines. The right panel shows an oval-flow model (see Section 3.2.2 and the Appendix), oriented in the same way. Isovelocity lines are overploted in dashed lines, with the zero-velocity line going through the centre of the model. The thick segments of the isovelocity lines in each model indicate features which are similar to the observed data; they are superimposed on the \H2\ LR velocity map of \n3627 in the middle panel to indicate our suggested interpretation (see Section~\ref{s_3627_ss2} for details). The long-dashed line marks the LON.}
\label{f_n3627_modelS20}
\end{figure*}

As the gas motions in \n3627 are far from circular, fitting a model of a disc in circular motion, as we did for \n3351, will provide no insight. On the other hand, hydrodynamical models of gas flow in strong bars do show strong non-circular motions and significant twists in the isovelocity lines. Therefore we first compare our observation with hydrodynamical simulations in order to interpret the observed velocity field. In the left-hand panel of Fig.~\ref{f_n3627_modelS20}, we show the LOS velocity from the simple, isothermal, single-phase hydrodynamical Model S20 of \citet{Maciejewski2002a}. We plot this velocity field using the same orientation as \n3627, i.e. we take the inclination, major-axis PA and stellar-bar PA from Table~\ref{t_prop}, and North is in the same direction as in the middle panel. We stress that this is a generic hydrodynamical model of gas flow in a bar, not something tailored to our observations of \n3627; we only adjusted the arbitrary length unit so that it roughly corresponds to one arcsec in the \n3627 data. In this scaling, the size of the \S\ LR FOV is marked with the dashed box in the left panel of Fig.~\ref{f_n3627_modelS20}. In the outer part of this FOV, at the distance of $\sim 2$ units from the centre, the model's negative isovelocity lines turn up to the left as one moves out from the centre, and the positive isovelocity lines similarly turn down to the right. These deviations are followed, at slightly larger radii, by abrupt reversals, with the isovelocity lines turning sharply down on the left and up on the right. These portions of the isovelocity lines are marked with the thick solid lines in the left-hand panel of Fig.~\ref{f_n3627_modelS20}, and are overplotted on the LR \S\ data in the middle panel. The behaviour of these isovelocity lines in the model is qualitatively similar to that reported in Section~\ref{s_3627_ss1} for the data: both the sharp curvature of isovelocity lines around the blue-to-green transition on the left, and around the green-to-yellow transition on the right, along with the reversals further out (more clearly visible in the left-hand side of the observed velocity field). We note that the isovelocity lines in the data seem to be smoother than the overplotted ones from the model. There are several possible reasons for this (in addition to the fact that our data has lower SNR as one moves away from the centre of the galaxy). First, the model is not specifically tailored to match \n3627. Second, the model is highly idealized, and the inclusion of a multi-phase medium, turbulence and star formation will likely tone down the sharp twists. Finally, the model is two-dimensional, with the twists happening on scales $\sim$50~pc. This is on the order of the disc scale height, and so projection effects are likely to smear out the sharp appearance of the twists in the model. 

In the model, the shape of the isovelocity lines, marked by the thick solid lines in the left-hand and middle panels of Fig.~\ref{f_n3627_modelS20}, is due to shocks in the gas flow. This argues for the presence of shocks in the \H2\ emission-line gas in the regions marked by the thick solid isovelocity lines. Other kinematic evidence for shocks is possibly present but weak: the \H2\ velocity-dispersion panel in Fig.~\ref{f_n3627_kin_250} shows slightly higher dispersions in the upper left and lower right than in the opposite corners of the field of view, consistent with the approximate shock regions. We note that the two blobs of \H2\ emission observed north and south of the central elongated structure (top-left panel of Fig.~\ref{f_n3627_kin_250}) appear spatially coincident with the postulated shocks; this is possibly consistent with a scenario of denser gas immediately downstream from the shocks. We also showed in Paper\,I that the \H2\ line ratio in these regions is consistent with shock excitation.  

In the analysis above, we used a hydrodynamical model of gas flow in a bar, in which shocks in the gas along the bar, offset from the bar’s major axis, are present; the signatures of these shocks appear to be present in the LR \H2\ data. However, in the inner 2~arcsec, the flow in the hydrodynamical model settles onto a nuclear ring or spiral and becomes effectively circular, and so cannot reproduce the clear misalignment that we observe in \n3627 between the LON of the central gas kinematics (130\deg; dot-dashed line in the top-left panel of Fig.~\ref{f_n3627_kin_100}) and the PA of the galaxy (175\deg; dashed line in the same panel).

We can, however, attempt to reproduce such a misalignment with an \textit{oval} flow. This kind of flow can arise naturally in non-axisymmetric gravitational potentials, such as those resulting from a bar, and it is known to induce changes in the observed PA of the LON \citep[see][and references therein]{Teuben1991}. In the Appendix~\ref{s_app1}, we show that the zero-velocity line of an oval flow shifts towards the major axis of the flow as the ellipticity of the flow increases (see Fig.~\ref{f_ap2}). If such a flow were oriented along the major axis of the stellar bar (PA = 161\deg), then the zero-velocity line would be shifted towards this PA, to values larger than the PA of 85\deg\ expected for a circular flow (given the LON of \n3627 at 175\deg). However, we observe a shift of the zero-velocity line to a \textit{smaller} PA: the PA of the zero-velocity line in the HR FOV is 45\deg. Therefore, an oval flow along the stellar bar is ruled out, as already pointed out by \citet{Afanasiev2005} based on lower-resolution data. However, we note that the \H2\ emission of \n3627 is extended not along the bar, but along PA of 0--17\deg\ (see the upper panels of Fig.~\ref{f_n3627_kin_250} and the references at the beginning of this section). If a central oval flow were elongated along the major axis of the \H2\ emission, then its zero-velocity line would be dragged from the PA of 85\deg, expected for a circular flow, toward the PA of the \H2\ emission's major axis, i.e. towards smaller values. This kinematics is in agreement with the observed PA of the zero-velocity line in the HR FOV being 45\deg.

In the right-hand panel of Fig.~\ref{f_n3627_modelS20}, we present the LOS velocity in a simple analytical model of an oval flow that reproduces the angle between the LON and the zero-velocity line in the innermost 1.5~arcsec of \n3627, plotted for the orientation of \n3627, i.e., using the inclination and the PA of the LON from Table~\ref{t_prop}, with North in the same direction as in the middle panel. The formulae for this model are given in the Appendix~\ref{s_app2}, and in this realization we assume that the ellipticity of the flow is constant with radius, because we observe that the zero-velocity line in the \H2\ data is almost straight within the radius of 1.5~arcsec. We place the corotation radius of the flow at the end of the stellar bar, at 60~arcsec. Figure rotation of the flow curves the zero-velocity line at larger radii, but it has little effect deep inside the corotation, within the radius of 1.5~arcsec. Therefore, for a given angle between the major axis of the flow and the LON of the galaxy, a unique value of the ellipticity of the flow will yield the desired PA of the zero-velocity line (see Eq.~A1). If we adopt a PA of 0\deg\ for the major axis of the oval flow, coinciding with the major axis of the \H2\ distribution at $r < 2$~arcsec, then a flow with an axial ratio 0.58 produces a zero-velocity-line PA of 45\deg\ -- which reproduces the observed value (e.g. upper-left panel of Fig.~\ref{f_n3627_kin_100}). Thus the model in the right-hand panel of Fig.~\ref{f_n3627_modelS20} is not fitted to the observed velocity field, but only constrained by the observed PA of the observed zero-velocity line. The velocity profile in this model is based on the rotation curve in Fig.~\ref{f_n3627_2D}: it increases up to the radius $r_{\rm flat}$=4~arcsec, beyond which it assumes a constant value $v_{\rm flat}$=300\kms (corresponding to 174\kms\ on the major axis of the oval flow).  
Our adopted constraints imply that the angle between the major axis of the flow and the LON in the galaxy plane is $\alpha=-12$\deg, while the angle between the major axis of the flow and the zero-velocity line in the galaxy plane is $\phi=58$\deg. 

The zero-velocity line and two neighbouring isovelocity lines within the 1.5~arcsec radius are marked with thick dotted lines in the right-hand panel of Fig.~\ref{f_n3627_modelS20}, and overplotted on the data in the middle panel. The match with the data suggests that an oval flow is indeed present in the innermost parts of \n3627. Although we argued for the presence of shocks at larger galactocentric radii (from comparing features present in the data with the hydrodynamical model), these shocks most likely do not extend into the innermost, oval-flow-dominated region. In this region, the angle between the stellar bar and the major axis of the oval flow is 42\deg\ in the disc plane, which implies strong torque from the stellar bar on the \H2\ flow, and possibly gas inflow as a result (see Casasola et al. 2011 for inflow estimates).

\subsection{\n4501}\label{s_4501}

\begin{figure*}
\centering
\includegraphics[width=0.8\textwidth]{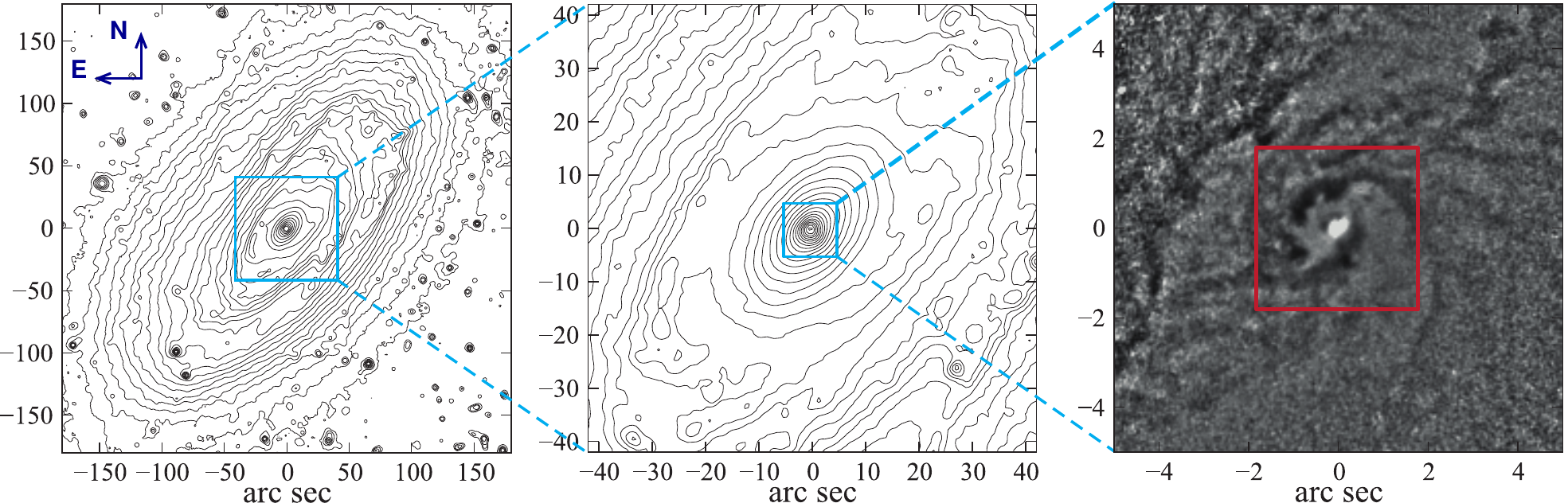}
\caption{IRAC 3.6\micron\ contours (left and middle panels) and $HST$/WFPC2 F547M unsharp mask (right panel) of \n4501. The red box in the right panel indicates the \S\ FOV. The images are shown at the same spatial orientation as the \S\ data.}
\label{f_n4501_im}
\end{figure*}

\n4501 is one of the largest spiral [SA(rs)b, \citet{deVaucouleurs1991}] galaxies in the Virgo Cluster, with good evidence that it is currently undergoing ram pressure stripping which has led to the truncation of its outer gas disc near the optical ($R_{25}$) radius \citep[e.g.][]{Vollmer2008b,Vollmer2009a}. Although the galaxy is usually considered unbarred (and is dominated in both the optical and NIR by strong, irregular spiral arms), there is perhaps some tenuous evidence for a bar aligned near the minor axis of the galaxy in the \textit{Spitzer} IRAC1 isophotes (left and middle panels of Fig.~\ref{f_n4501_im}), at a PA of $\sim 80$\deg, with an observed radius of $\sim 25$~arcsec. The elliptical inner region visible in the middle panel (radius $\sim 15$~arcsec) has the same orientation and almost the same ellipticity as the outer disc (0.4 versus 0.52), and is most likely not a bar but rather a highly flattened bulge or some kind of disky pseudobulge, as suggested by \citet{Fisher2008,Fisher2010}. Our analysis of the outer-disc isophotes suggests a PA of 140\deg\ and an ellipticity of 0.52, which corresponds to an inclination of 64\deg, assuming a circular disc with thickness $c/a = 0.2$. Both of these measurements are consistent with previous morphological and kinematic analyses, which yield PAs of 140--141\deg\ and inclinations of 60--65\deg\   \citep[e.g.][]{Barbera2004,Wong2004a,Chemin2006a}. 

In the right-hand panel of Fig.~\ref{f_n4501_im}, we show a close-up of the inner $10 \times 10$~arcsec, using an unsharp mask of an \textit{HST}-WFPC2 F547M image to indicate nuclear dust lanes (the region covered by our SINFONI FOV is marked by the red box). Comparison with the full WFPC2 image suggests that these dust lanes may be a continuation of the global spiral pattern in the galaxy \citep[see][]{Onodera2004}. Based on the dust distribution, NE corresponds to the near side of the galaxy. 
High-resolution interferometric observations of the $^{12}$CO(1-0) emission in the inner 5~kpc of the galaxy show spiral arms extending out from the nucleus associated with the spiral dust lanes, along with a concentration of gas in the centre \citep{Onodera2004}. The central molecular gas concentration is high, especially for a galaxy that does not contain a strong bar \nocite{Sakamoto1999b}(e.g. Sakamoto et al. 1999; Paper\,I). Based on the analysis of the non-circular motions along the molecular arms, \citet{Onodera2004} suggested that this concentration arises from spiral-driven gas transfer. The link between the molecular gas and the dust lanes is also visible on smaller scales: in the inner $3\times3$~arcsec, marked by the red box in the right-hand panel of Fig.~\ref{f_n4501_im}, there is a close spatial coincidence between the dust lanes and \H2\ emission from the circumnuclear spiral structure in our \S\ data \citepa[see figs.~4 and 5 of][]{Mazzalay2013a}. 
Two spiral arms can be outlined in the dust morphology, with the majority of the northern arm and some of the southern arm spatially coincident with \H2\ emission in our data.

\subsubsection{SINFONI kinematics}

\begin{figure}
\includegraphics[width=\columnwidth]{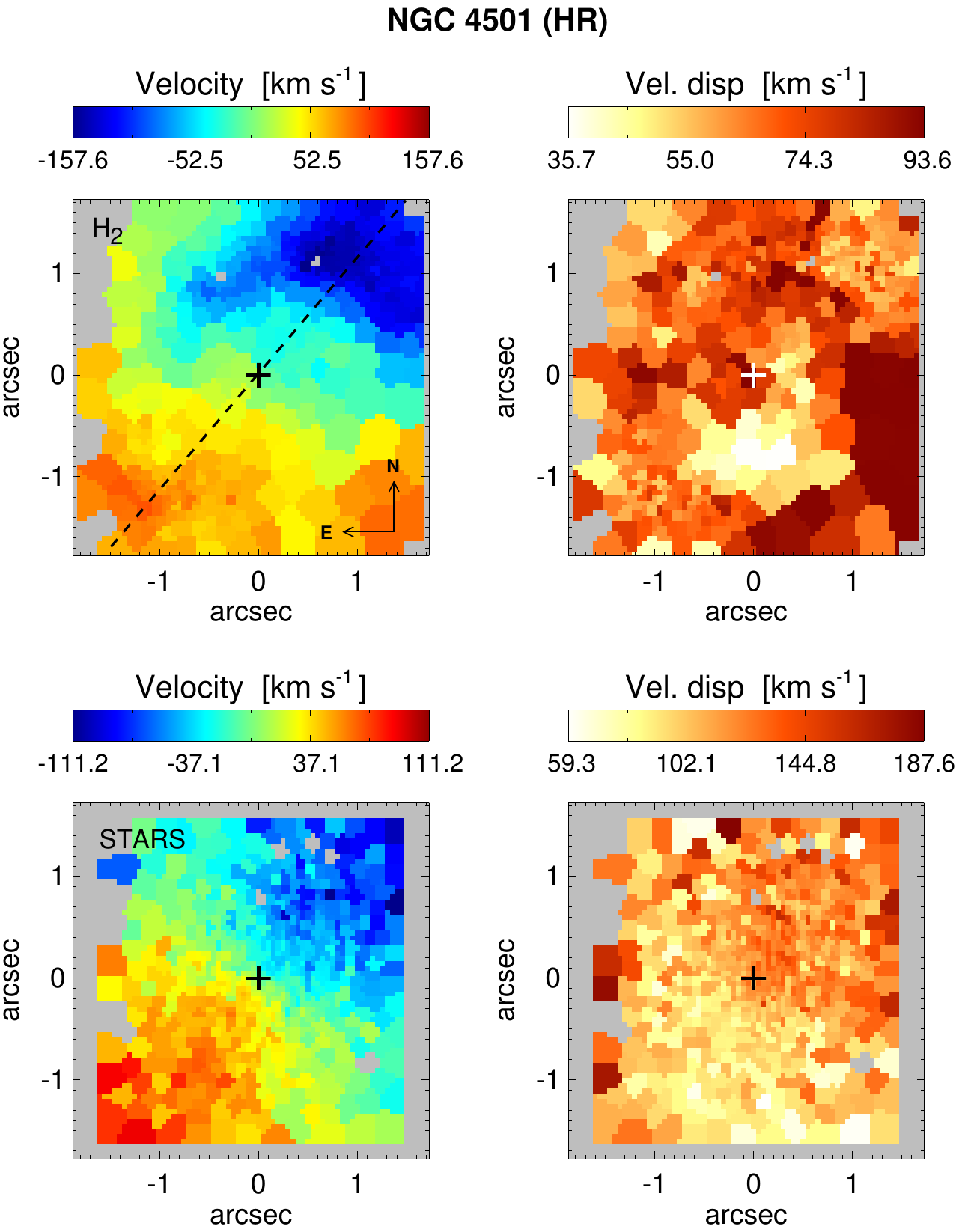}
\caption {2D maps of the \H2\ emission-line gas (upper panels) and stellar (lower panels) kinematics of \n4501 derived from the HR \S\ data. Regions where no reliable parameters were obtained are shown in grey. The spatial orientation, indicated in the upper-left panel, is the same for all the panels. The dashed line in the upper-right panel indicates the orientation of the major axis of the galaxy.}
\label{f_n4501_kin}
\end{figure}

\begin{figure}
\centering
\includegraphics[width=0.75\columnwidth]{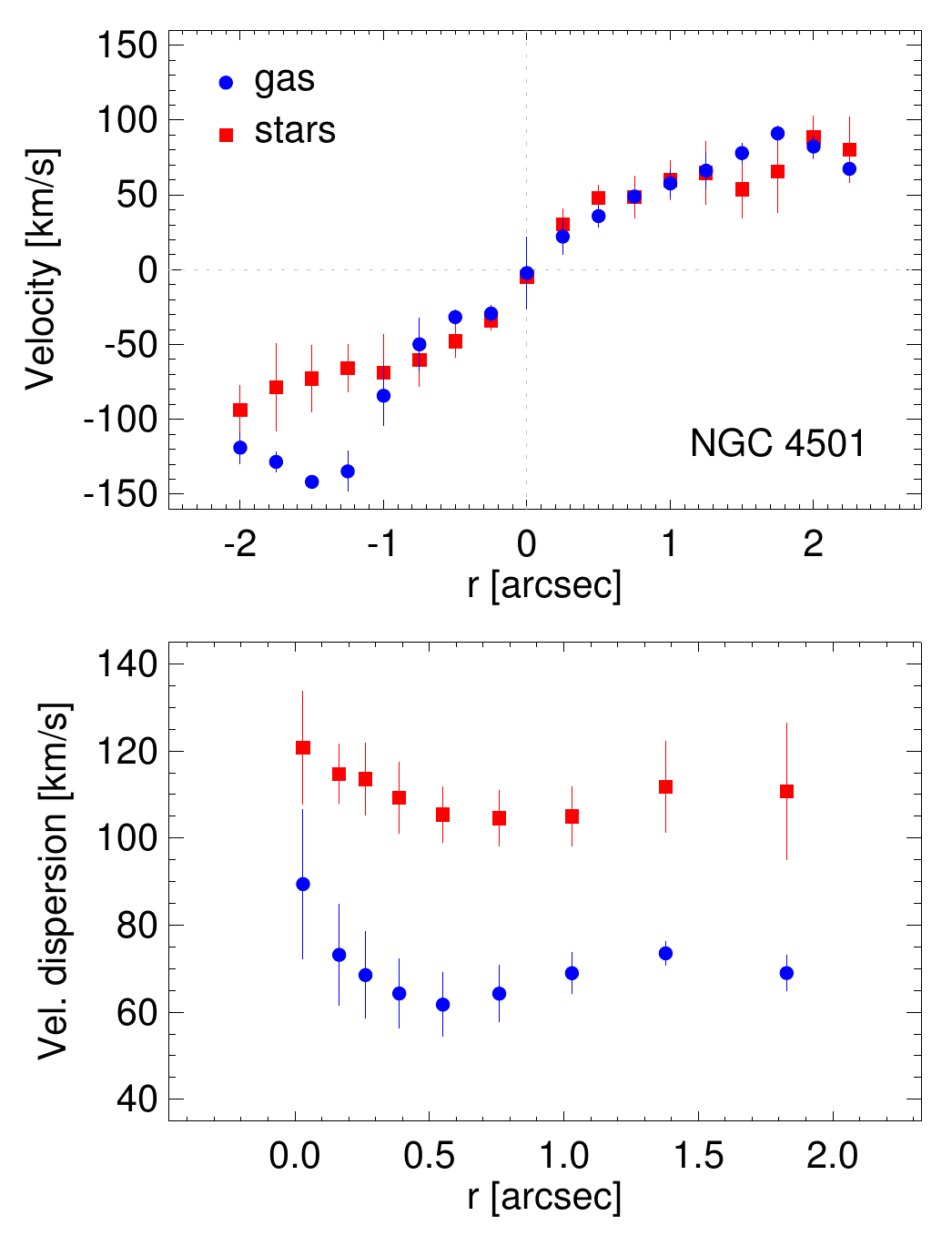}
\caption {\n4501 -- Upper panel: mean values of the velocity measured along the major axis of the galaxy (140\deg) in apertures of $0.25 \times 0.25$~arcsec; negative radii are NW from the nucleus. Lower panel: mean velocity dispersion measured in annular apertures as a function of the semimajor axis. In both panels blue circles correspond to the values measured for the \H2\ gas and red squares to those for the stars. The error bars indicate the 3$\sigma$ error in the mean.}
\label{f_n4501_2D}
\end{figure}

The \S\ spectra of \n4501 display emission lines from molecular hydrogen only, with no sign of emission from ionized gas. Fig.~\ref{f_n4501_kin} shows the \H2~2.12\micron\ emission-line gas and stellar kinematics maps derived for this galaxy. Additionally, Fig.~\ref{f_n4501_2D} shows the mean velocity measured along the major axis of the galaxy in regions of $0.25 \times 0.25$~arcsec and the velocity dispersion averaged over annuli for the molecular gas and stars. The stellar kinematic maps show a regular rotation pattern, with the LON at ${\rm PA}= 140 \pm 3$\deg, consistent with the PA of the major axis of the galaxy. A symmetric velocity curve is observed along this PA, reaching maximum velocities of $\sim \pm 100$\kms\ at the edges of our FOV, at $\sim 2$~arcsec (160~pc) from the centre. High stellar velocity dispersion is measured in the inner $3 \times 3$~arcsec of the galaxy, with a mean value of 108\kms. This value is higher than the maximum rotational velocity observed.  
The velocity dispersion averaged over annuli marginally increases inwards (lower panel of Fig~\ref{f_n4501_2D}), the opposite trend to that seen in \n3351 and \n3627, reaching $\sim 120$\kms\ at the centre. Note that this trend is weaker than the peculiar gradient in the 2D velocity-dispersion field (lower-right panel of Fig.~\ref{f_n4501_kin}), where the dispersion goes from $\sim 80$\kms\ in the SE area corner to $\sim 130$\kms\ in the NW corner. 

The \H2\ emission-line gas velocity field is dominated by rotation, which is in the same direction as the stars. However, strong signatures of non-circular motions are present in the gas velocity field; especially noticeable is the blueshifted  kinematic arc or spiral observed towards the NW (hereafter blue kinematic spiral). This feature  coincides spatially with the northern spiral arm in the \H2\ morphology seen in the \S\ FOV \citepa[see fig.~5 of][]{Mazzalay2013a} and with the northern spiral dust lane (right-hand panel of Fig.~\ref{f_n4501_im}). The blue kinematic spiral also affects the gas velocity curve along the major axis of the galaxy (upper panel of Fig.~\ref{f_n4501_2D}), which at $r \simeq -1.5$~arcsec shows an excess of about $50$\kms\ when compared to the stellar velocity curve. The gas velocity dispersion map, which has a mean value of 69\kms\ measured over the entire \S\ FOV, does not show any particular features associated with this kinematic perturbation. Deviations from the mean dispersion value are observed south from the nucleus ($\sigma \sim 35$\kms) and on the SW edge of the FOV ($\sigma \sim 90$\kms). However, these regions display relatively weak \H2\ emission, therefore these low and high values could be an artefact of the low SNR of the region. The gas velocity dispersion curve shown in the lower panel of Fig.~\ref{f_n4501_2D} shows a radial trend similar to that of the stars but with lower values ($\sigma \sim 60$--90\kms). As in the case of \n3351 and \n3627, the gas velocity dispersion increases towards the centre. 

\begin{figure}
\includegraphics[width=\columnwidth]{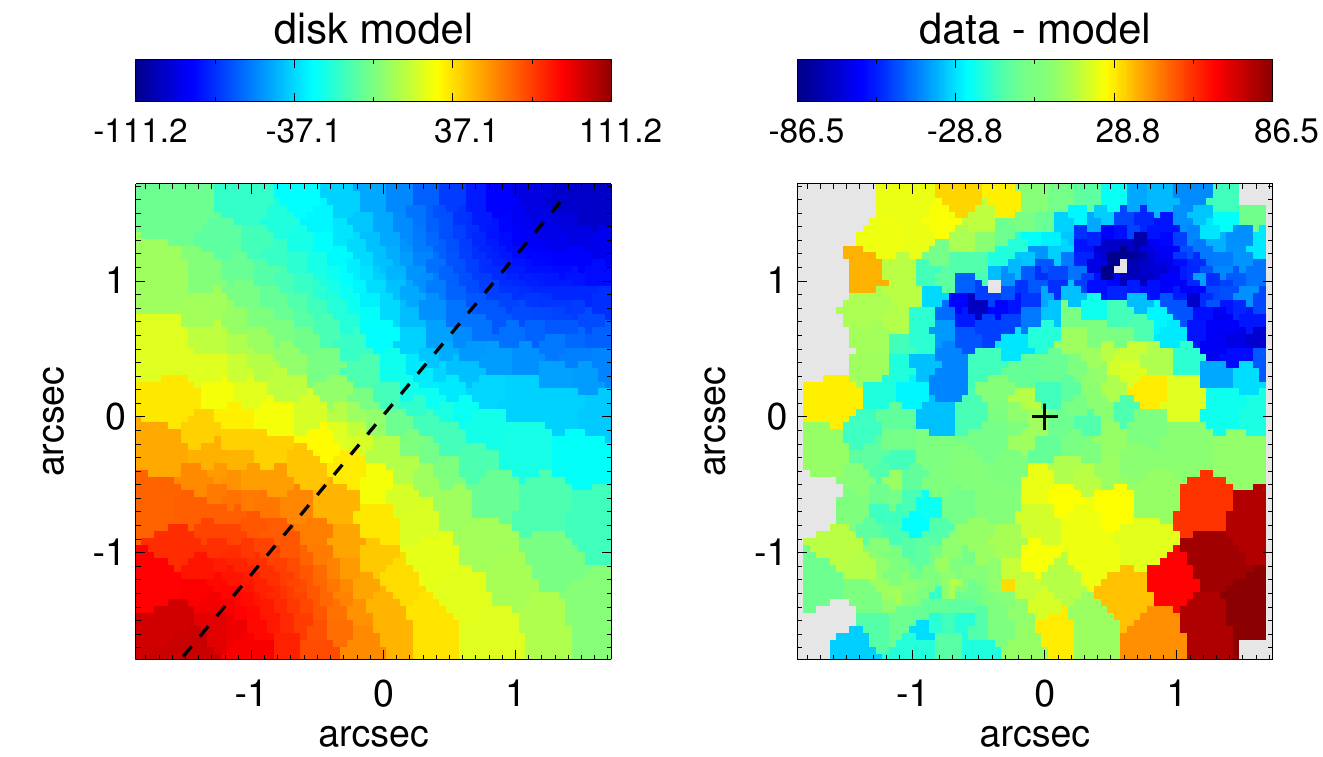}
\caption {Left panel: best-fitting disc model of the stellar velocity field of \n4501. Right panel: velocity residuals of the \H2\ emission-line gas velocity with respect to the stellar disc model. The dashed line indicates the position of the major axis of the galaxy (PA = 140\deg). }
\label{f_n4501_disk140}
\end{figure}

\subsubsection{2D kinematic analysis}

In order to isolate the gas non-circular motions in the nucleus of \n4501, we subtracted the underlying rotational pattern. Note that this pattern would be biased if we were to fit a disc model to the gas velocity field (as we did in the case of \n3351) because it would be affected by the velocity excess associated with the blue kinematic spiral arm. This single feature would increase the values of the rotation curve, which when subtracted from the gas velocity field would then yield artificial residuals in the SE part of the field. On the other hand, if we subtract the fit to the \textit{stellar} kinematics, which is close to symmetrical along the LON, we can avoid this problem. The assumption that the underlying rotational pattern in the gas velocity field is well described by the one given by the stars is supported by the similarity between the gas and stellar velocity curves along the major axis of the galaxy in the SE corner, where gas flow is less affected by non-circular motions (upper panel of Fig.~\ref{f_n4501_2D}). Consequently, we fitted a rotating exponential thin-disc model to the observed stellar velocity field (assuming an inclination $i= 64^{\circ}$ and a kinematic PA of 140\deg) and subtracted it from the observed gas velocity field.

The disc model and the velocity residuals in the gas are shown in Fig.~\ref{f_n4501_disk140}. The blue kinematic spiral is even more pronounced in this residual velocity field. Gas velocities in the spiral are up to 85\kms\ more negative than in the surrounding area. The blue kinematic spiral in the \H2\ velocity map of this galaxy, together with the morphology of the molecular gas and the presence of spiral-dust lanes, appear to make this galaxy a good candidate for hosting a nuclear spiral, something which has been proposed as a way for transporting gas from kiloparsec scales down to the active nucleus \citep[e.g.][]{Englmaier2000, Maciejewski2004b, Maciejewski2004a}, and which has been observed in IFS data of some galaxies (e.g. \n1097 -- Fathi et al. 2006, Davies et al. 2009; \n6951 -- Storchi-Bergmann et al. 2007)\nocite{Fathi2006a,Davies2009a,Storchi-Bergmann2007}. 

However, several observational facts advise against a nuclear-spiral density wave interpretation in the case of \n4501. First, although the blue kinematic spiral in velocity residuals spatially coincides with the northern spiral dust-lane (right-hand panel of Fig.~\ref{f_n4501_im}), the southern dust lane is not associated with any kinematic signature in our \S\ data.  
If the observed kinematic spiral is a wave phenomenon, the blueshifted arm should be accompanied by a redshifted arm on the other side of the galaxy centre (see e.g. fig.~7 of Davies et al. 2009). 
A possible explanation for the absence of the redshifted arm could be reddening effects. However, there are no signs of  extinction in our reconstructed $K$-band \S\ image.  
Second, although in the spiral wave the photometric and kinematic arms usually do \textit{not} coincide, in \n4501 the blue kinematic spiral arm perfectly overlaps with the strongest \H2\ emission. This close overlapping indicates that we are seeing a material arm with some intrinsic peculiar velocity, rather than a wave propagating in the gas. Third, for density waves, kinematic and photometric arms should intersect on the minor axis of the galaxy, where maximum density occurs close to maximum inflow velocity. 
However, taking into account that the NE is the near side of \n4501, what we are seeing where the photometric arm crosses the minor axis on this side is a maximum {\it outflow}. 
This outflow cannot be reconciled with a spiral wave, but \textit{is} consistent with a material arm, possibly one extending away from the galaxy plane. Finally, the appearance of this kinematic feature as being similar to a spiral arm may be affected by projection in this highly inclined galaxy. On the sky, the radial coordinate of the dust filaments and of the kinematic feature appear to change monotonically with radius, which is characteristic for a spiral. But after deprojection, there is no monotonic change any more, and the dust features, as well as the kinematic spiral, take on more of an oval appearance. We conclude that the kinematic spiral observed in the inner 2~arcsec of \n4501 is not a spiral density wave, but rather a material feature: a gas filament moving with respect to the rotating galaxy disc. As gas clouds are sheared due to differential rotation, careful analysis is needed in order to distinguish spiral density waves from material arms.

\subsection{\n4536}\label{s_4536}

\begin{figure*}
\centering
\includegraphics[width=0.8\textwidth]{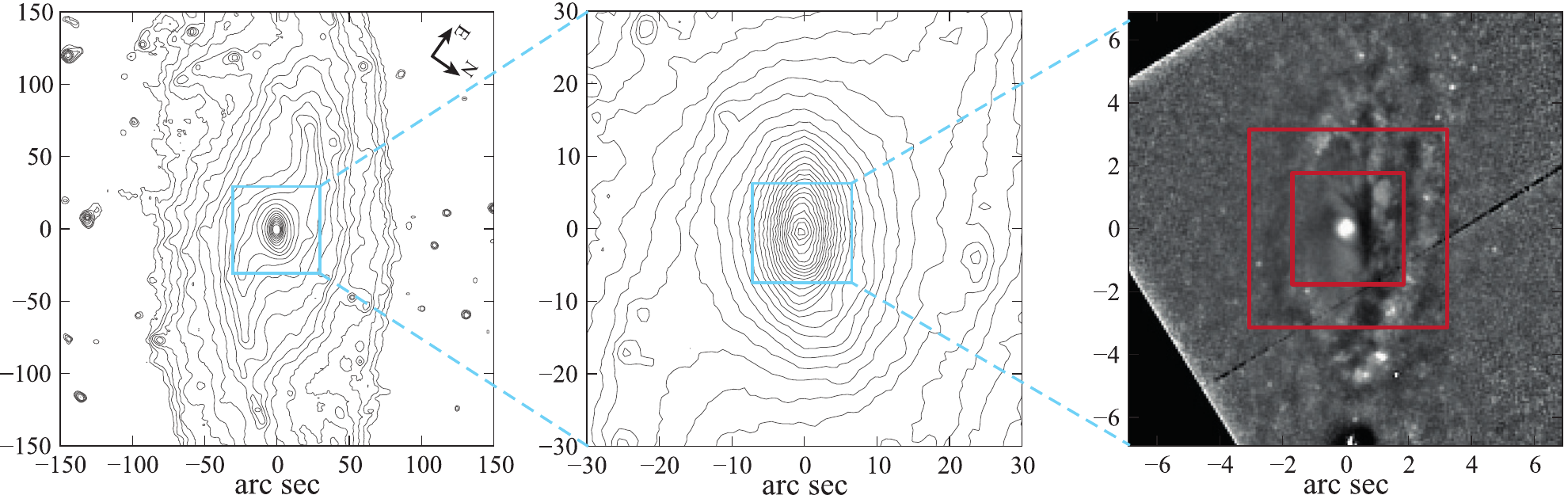}
\caption{IRAC 3.6\micron\ contours (left and middle panels) and $HST$/NICMOS2 F160W unsharp mask (right panel) of \n4536. The images are shown at the same spatial orientation as the \S\ data. The red boxes in the right panel indicate the \S\ LR and HR FOVs. }
\label{f_n4536_im}
\end{figure*}

\n4536 is another large spiral galaxy in the Virgo Cluster, classified as SAB(rs)bc \citep{deVaucouleurs1991} and notable for two very prominent spiral arms.  In contrast to \n4501, it does not show any clear signatures of ram pressure stripping of its H\,{\sc i} gas, though a weak interaction with its neighbour \n4533 has been suggested \citep{Chung2009}. Fig.~\ref{f_n4536_im} shows \textit{Spitzer} IRAC1 isophotes of the whole galaxy (left-hand panel) and a close-up of the inner $60 \times 60$~arcsec region (middle panel); the right-hand panel shows an unsharp mask of the innermost $10 \times 10$~arcsec region (using an \textit{HST} NICMOS2 F160W image) with our SINFONI FOVs marked.

Our analysis of the outer-disc isophotes (using SDSS and IRAC1 images) suggests a PA for the LON of $125^{\circ} \pm 5^{\circ}$ and an ellipticity of $0.6 \pm 0.04$; for an assumed disc thickness of $c/a = 0.2$, this indicates an inclination of $69 \pm 3$\deg. These values are very consistent with previous estimates from large-scale kinematics: e.g. \citet{Chemin2006a} found a kinematic PA of $301 \pm 2\degr$ (or $121 \pm 2\degr$ for comparison with the photometric) and $i = 68 \pm 3\degr$ from their H$\alpha$ velocity field, while \citet{Kuno2007} report PA $= 124$\deg\ and $i = 70$\deg\ from their CO observations.

Although \n4536 is nominally classified as weakly barred, the evidence for a bar is rather ambiguous. Two strong stellar spiral arms, visible in the IRAC1 image, span approximately $\sim$90\deg\ in azimuth; they can be traced in to a distance of $\sim 20$--25~arcsec from the centre, which sets an upper limit to the size of any possible bar. Further in, there is a clear elliptical inner region dominating the light at $r \lesssim 15$~arcsec, as in the case of \n4501. This structure has an orientation of $\sim 122$\deg\ and is thus aligned to within $\sim 3$\deg\ of the outer disc; its ellipticity ($\sim 0.45$) is somewhat lower than that of the outer disc, which suggests that it may be either a flattened bulge or a thick inner disc (or a combination of the two), rather than a bar. We note that high-resolution CO observations are consistent with molecular gas located in a circumnuclear disc of $\approx 10$~arcsec radius \citep{Sofue2003b, Jogee2005}, again suggesting that the central regions of the galaxy are largely axisymmetric.

\begin{figure}
\centering
\includegraphics[width=\columnwidth]{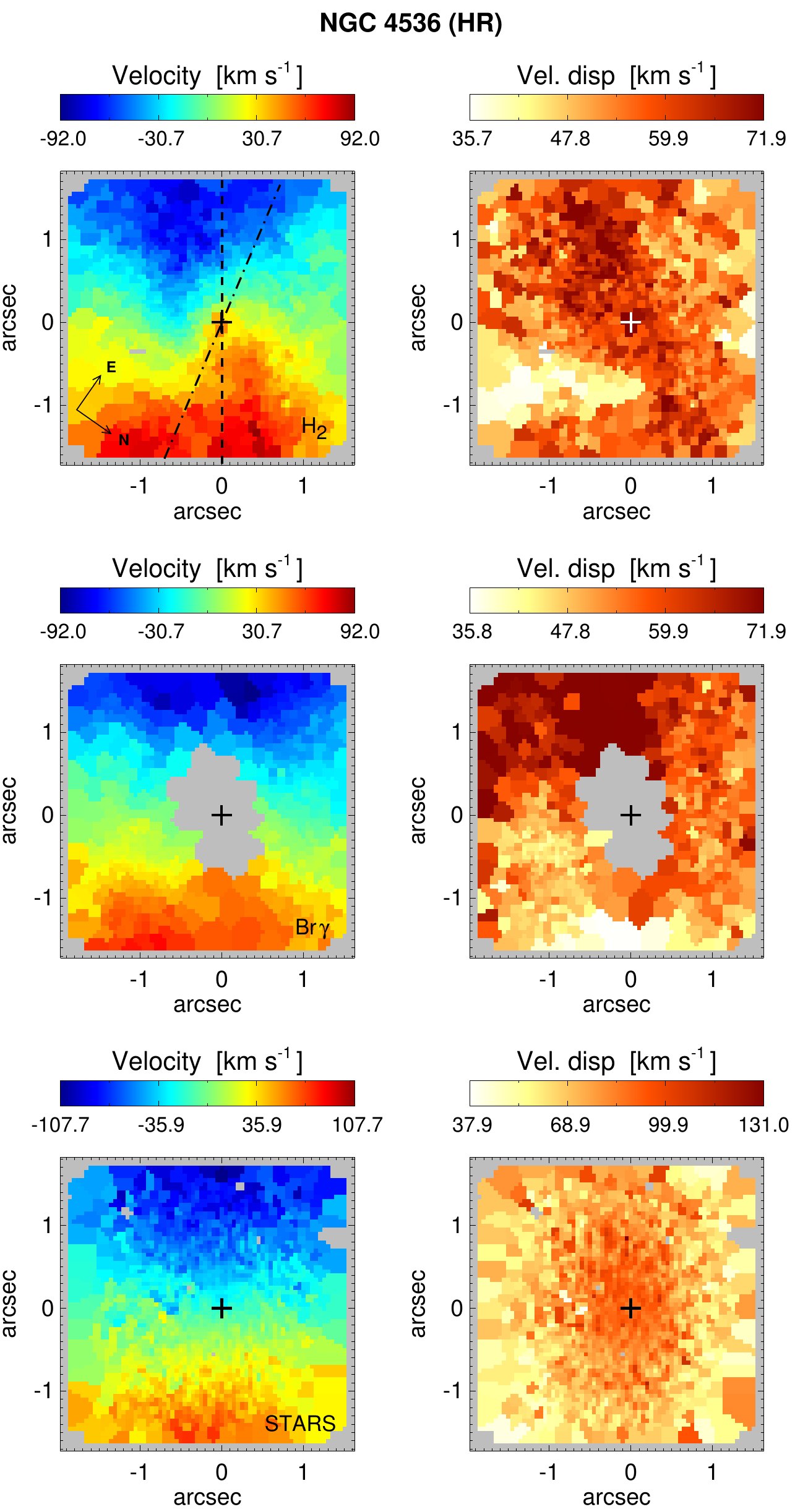}
\caption {2D maps of the \H2\ (upper panels) and \Brg\ (middle panels) emission-line gas kinematics and stellar kinematics (lower panels) of \n4536 derived from the HR \S\ data. Regions where no reliable parameters were obtained are shown in grey. The spatial orientation, indicated in the upper-left panel, is the same for all the panels. The dashed line corresponds to the PA of the major axis of the galaxy and the dot-dashed line indicates the PA = 158\deg\ of the oval flow (see Section~\ref{s_4536_ss2} for details).}
\label{f_n4536_kin_100}
\end{figure}

\begin{figure}
\centering
\includegraphics[width=\columnwidth]{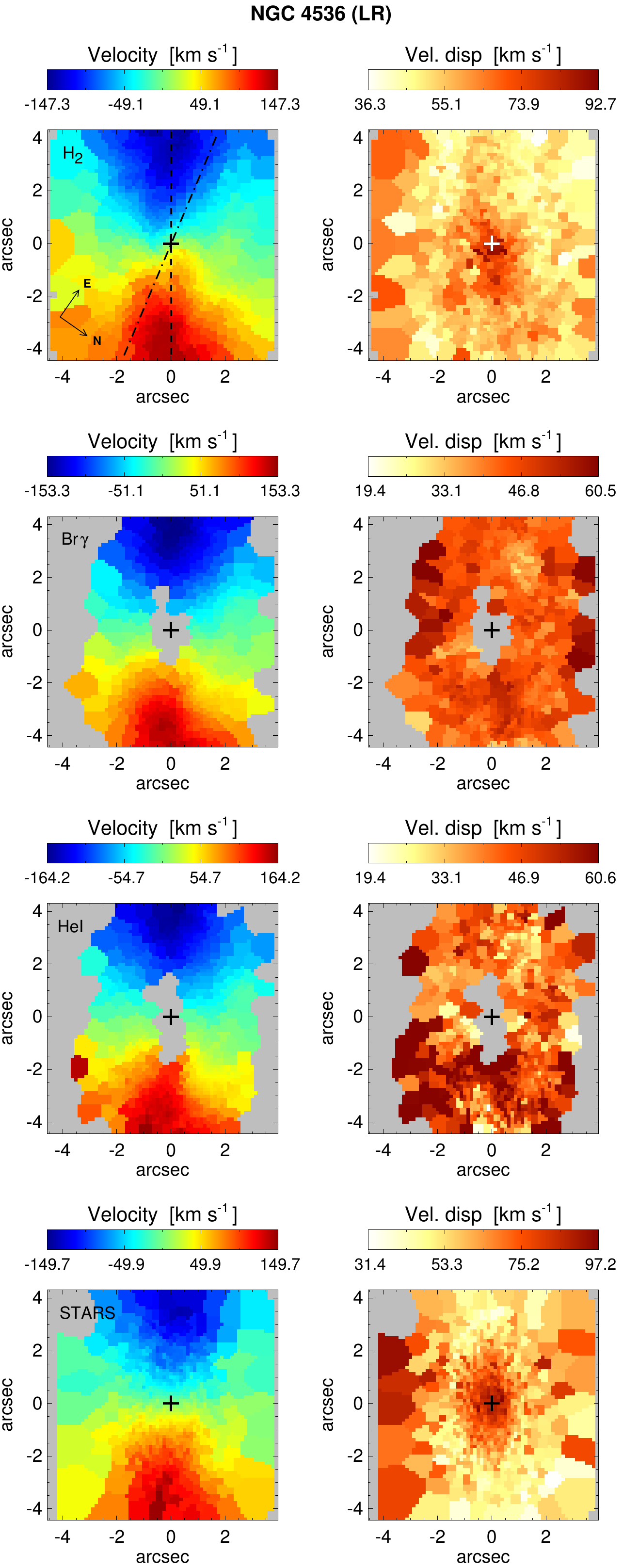}
\caption {Kinematic maps of the emission-line gas and stars of \n4536 derived from the LR \S\ data. These correspond, from top to bottom, to \H2, \Brg, \HeI\ and stars. Regions where no reliable parameters were obtained are shown in grey. The spatial orientation, indicated in the upper-left panel, is the same for all the panels. The dashed line corresponds to the PA of the major axis of the galaxy and the dot-dashed line indicates the PA = 158\deg\ of the oval flow (see Section~\ref{s_4536_ss2} for details).}
\label{f_n4536_kin_250}
\end{figure}

\begin{figure}
\centering
\includegraphics[width=0.746\columnwidth]{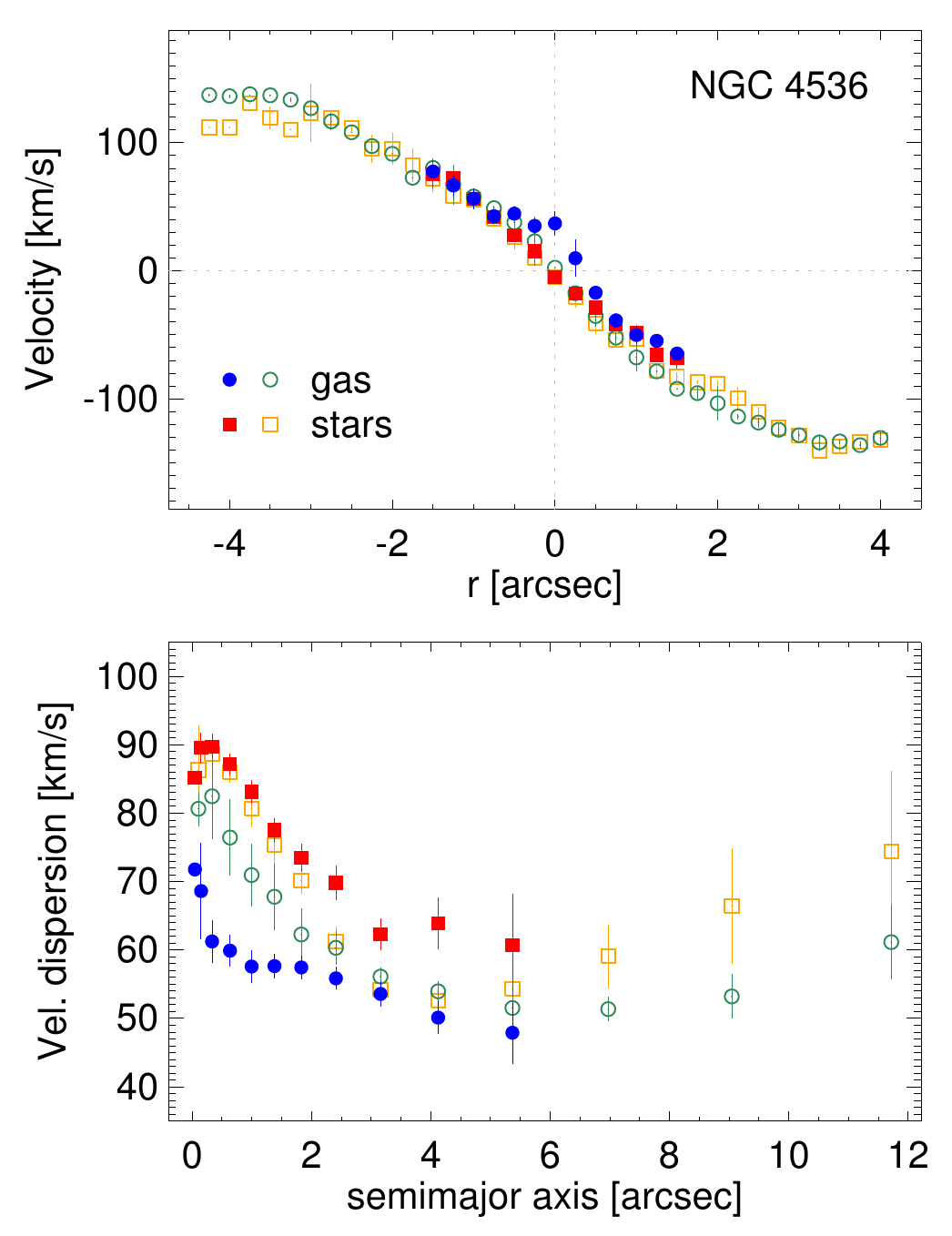}
\caption {\n4536 -- Upper panel: mean values of the velocity measured along the major axis of the galaxy (125\deg) in apertures of $0.25 \times 0.25$~arcsec; negative radii are NW from the nucleus. Lower panel: mean velocity dispersion measured in elliptical apertures as a function of the semimajor axis. In both panels blue filled and green open circles correspond to the values measured for the \H2\ gas in the HR and LR data, respectively. Red filled and orange open squares correspond to the stellar values measured in the HR and LR data, respectively. The error bars indicate the 3$\sigma$ error in the mean.}
\label{f_n4536_2D}
\end{figure}

\begin{figure}
\includegraphics[width=\columnwidth]{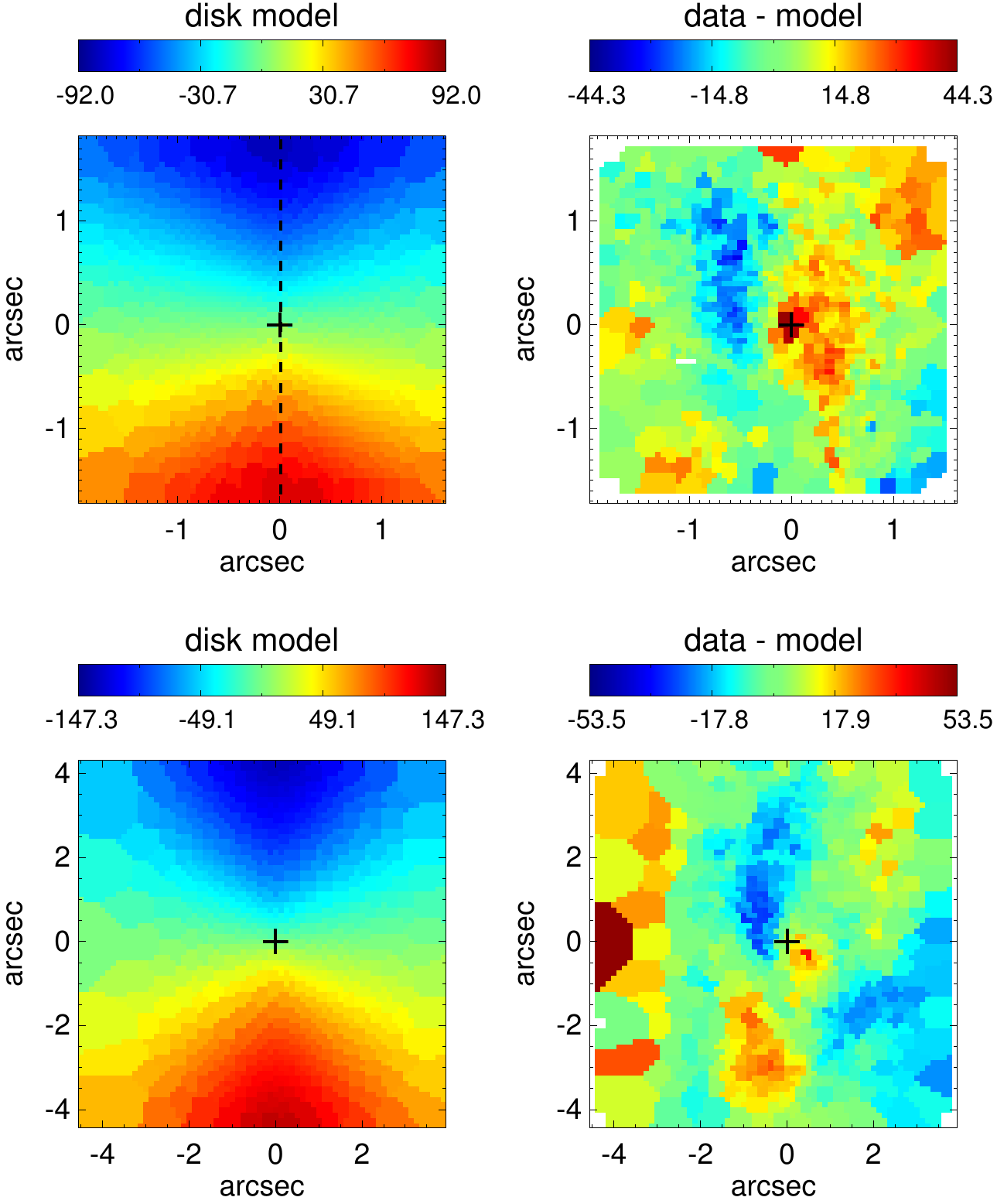}
\caption {Best-fitting disc model of the \H2\ velocity field and velocity residuals of the inner $3 \times 3$~arcsec (HR data, upper panels) and $8 \times 8$~arcsec (LR data, lower panels) of \n4536. The dashed line in the upper-left panel indicates the position of the major axis of the galaxy.}
\label{f_n4536_disk}
\end{figure}
\begin{figure}
\includegraphics[width=\columnwidth]{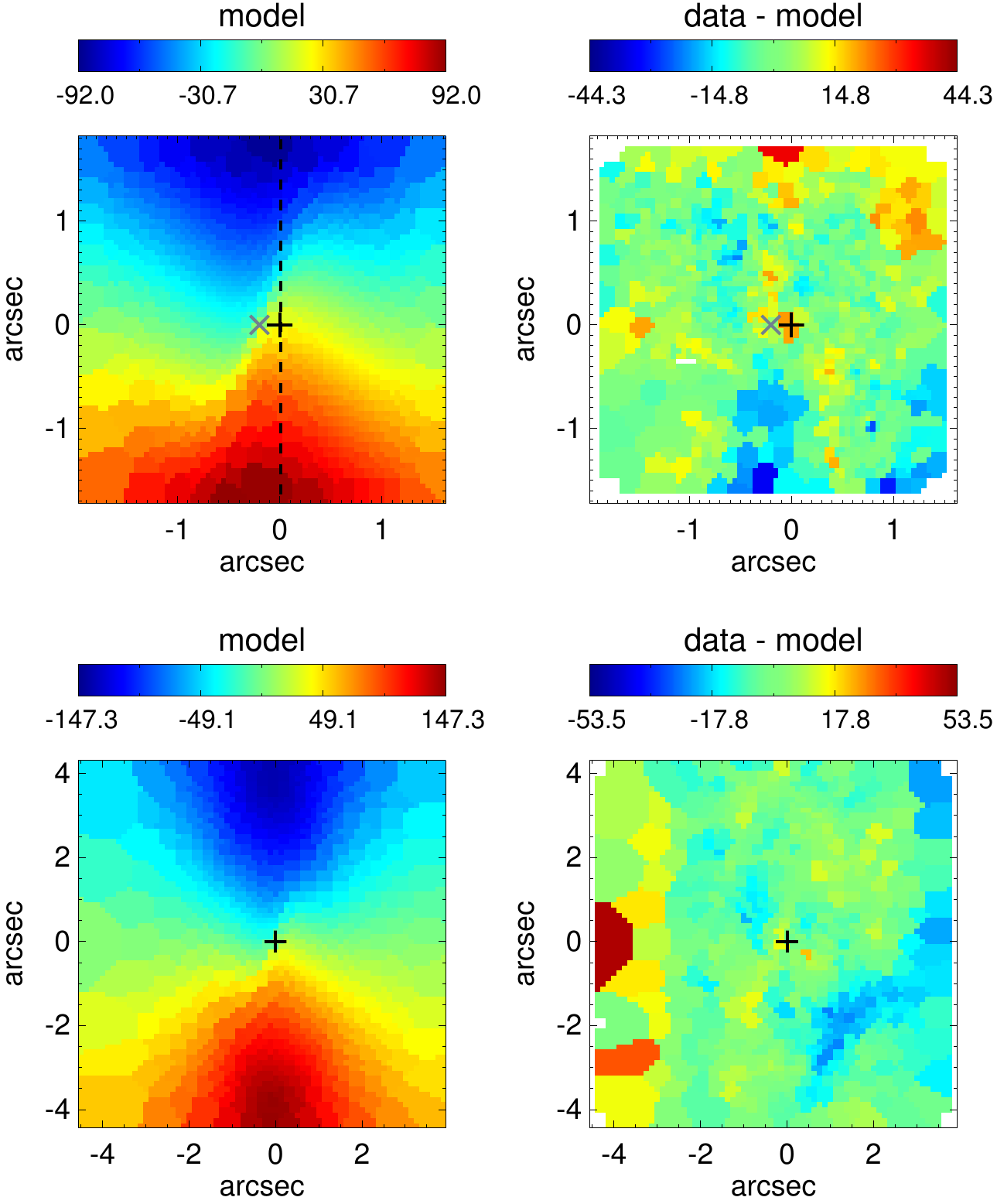}
\caption {Best-fitting oval flow model of the \H2\ velocity field and velocity residuals of the inner $3 \times 3$~arcsec (HR data, upper panels) and $8 \times 8$~arcsec (LR data, lower panels) of \n4536. The dashed line in the upper-left panel indicates the position of the major axis of the galaxy. The grey cross in the upper panels indicates the position of the kinematic centre and the black plus sign the position of the maximum of the $K$-band continuum. }
\label{f_n4536_bar}
\end{figure}

There is some weak evidence for non-circular gas motions in the inner 10~arcsec of \n4536: e.g. the possibly twisted inner contours of the \Ha\ velocity field \citep{Chemin2006a} and the slight ($\sim 15$\deg) offset between the zero-velocity line of the CO observations of \citet{Jogee2005} and the minor axis of the galaxy (and of the inner CO distribution). In addition, a nuclear ring of radius $\sim 3.7$~arcsec is present in the NICMOS data, as can be seen in the right-hand panel of Fig.~\ref{f_n4536_im}. This ring also shows up in ionized emission in our \S\ data \citepa{Mazzalay2013a}, indicating active star formation present in the ring. Since such rings are usually associated with the inner Lindblad resonances of bars (but see Comer\'on et al. 2010 for some examples of nuclear rings in unbarred galaxies)\nocite{Comer'on2010}, this is possible (indirect) evidence for a bar in \n4536. If there really \textit{is} a bar in this galaxy, then it is probably oriented close to the minor axis, with projection effects compressing it to form the slightly boxy isophotes which lie between the inner parts of the main spiral arms and the highly elliptical isophotes noted above. In the case, the hypothetical bar would have an observed radius -- near or along the galaxy \textit{minor} axis -- of $\sim 10$--14~arcsec ($\sim 30$--40~arcsec deprojected).

\subsubsection{SINFONI kinematics}

\n4536 was observed with \S\ in two different configurations, covering the inner $\sim 3 \times 3$ and $\sim 8 \times 8$~arcsec of the galaxy with high- and low-resolution spatial sampling, respectively. In \citeta{Mazzalay2013a} we show that while the ionized gas (traced by \Brg\ and \HeI~2.06\micron) is mainly located in a ring-like feature of radius $\sim 3.7$~arcsec ($\sim 270$~pc), the bulk of the warm molecular gas is concentrated in the inner $\sim 2$~arcsec ($\sim 150$~pc), with weaker emission along the galaxy major axis, possibly associated with the nuclear ring \citepa[see fig.~6 of][]{Mazzalay2013a}.
Figs.~\ref{f_n4536_kin_100} and \ref{f_n4536_kin_250} show the HR and LR kinematic maps of the molecular gas, ionized gas and stars in the inner regions of \n4536 derived from our observations. 
Both the gas and stellar velocity fields are dominated by rotation, although a strong kink in the zero-velocity line near the centre can be seen in both the LR and HR \H2\ velocity maps. The low SNR displayed by \Brg\ and \HeI\ emission lines inside the star-forming ring did not allow us to derive the kinematics of the ionized gas in the centre, where the deviations from circular rotation are observed in the \H2\ data. Aside from the \H2\ zero-velocity twist observed in the inner $\sim 0.8$~arcsec radius, the LON of both gas and stars seem to be aligned. To quantify this we used the same method as for the previous galaxies, which for the LR $8\times 8$~arcsec data yields the following kinematic PAs: $122.0 \pm 1.5$\deg\ for the \Brg\ emission-line gas, $121.5 \pm 1.5$\deg\ for the \HeI\ emission-line gas, $121.5 \pm 1.5$\deg\ for the \H2\ emission-line gas and $123.5 \pm 1.2$\deg\ for the stars. To within measurement accuracy these values are equal and consistent with the photometric PA of the galaxy $125 \pm 1.2$\deg, as well as with the PA of the LON derived from \Ha\ and CO observations \citep[e.g.][]{Chemin2006a, Kuno2007}. For the inner HR $3\times 3$~arcsec data, the derived kinematic PA for the stars, $125.5 \pm 2.7$\deg, is consistent with the LR data, while the PA for the \H2\ gas, $130.5 \pm 3.8$\deg, is slightly departing from the rest of the estimates because the velocity field of \H2\ gas is strongly affected by the non-circular motions in HR data. The derived PAs are summarized in Table~\ref{t_PA}.  

The upper panel of Fig.~\ref{f_n4536_2D} shows the velocity curve of the \H2\ emission-line gas and the stars along the major axis of the galaxy (PA = 125\deg). Both increase smoothly out to a distance of 3--4~arcsec from the nucleus, reaching a maximum velocity of about $\pm 140$\kms.
Some differences between the warm molecular gas and stellar curves are observed close to the nucleus ($r \lessapprox 0.5$~arcsec) in the HR data, where the isovelocity lines of the gas are clearly twisted in Fig.~\ref{f_n4536_kin_100}.
Our velocity curves are consistent with those derived previously from the kinematics of the ionized (\Ha) and cold molecular (CO) gas of \n4536 on larger scales (and lower spatial resolutions), where they are characterized by a steep gradient (reaching $\sim 150$--200\kms\ at a radius of $\sim 2$--4~arcsec) after which they become flat \citep[e.g.][]{Rubin1997,Sofue2003a,Sofue2003b,Chemin2006a,Daigle2006}.
Interestingly, this transition occurs at the radius of the circumnuclear star-forming ring observed with \S.

The right-hand panels in Figs.~\ref{f_n4536_kin_100} and \ref{f_n4536_kin_250} show the velocity dispersion maps for the gas and stars derived from the \S\ HR and LR data. The velocity dispersion of the ionized gas, measured from the LR maps, is $45 \pm 6$\kms\ for \Brg\ and $45 \pm 10$\kms\ for \HeI; these values come from the nuclear ring region, since that is where we observe the emission from ionized gas. The low SNR of the \HeI{} line emission leads to a stronger spatial variation than in the case of the \Brg{} emission, reflected in larger standard deviation above. In the HR data, the signal in \HeI\ is too weak, and so we do not show this map. The HR \Brg\ velocity dispersion is apparently much higher in the SE part of the ring than in the NW part, but the accuracy of our measurements is compromised by the weakness of the emission from these regions. 
In the rest of this paragraph, we focus on the velocity dispersions of the \H2\ emission-line gas and the stars. As in the case of \n4501, both increase towards the centre of the galaxy, with the stellar velocity dispersion reaching 90\kms{} and the molecular gas dispersion reaching 80\kms. It can be seen in the \H2\ gas and stellar LR maps of Fig.~\ref{f_n4536_kin_250} that the high velocity dispersion regions are elongated along the major axis of the galaxy, extending to a radius of $\sim 2$~arcsec ($\sim$150~pc); note that this is \textit{inside} the nuclear star-forming ring. Further out, in the region roughly coincident with the nuclear ring, the velocity dispersion drops to $\sim$50\kms\ in both stars and \H2\ gas. Additionally, there is a hint of an increase in the \H2\ gas and stellar velocity dispersion on the right and left edges of the LR FOV, outside the nuclear ring. This suggests a correlation between the star-forming ring and the velocity dispersion of both the \H2\ emission-line gas and the stars in \n4536, in the sense that the regions of low velocity dispersion are spatially coincident with the ring. This correlation is particularly evident in the LR map of stellar velocity dispersion, indicating that we are seeing a cold stellar component associated with the nuclear ring.

The lower panel of Fig.~\ref{f_n4536_2D} shows the velocity dispersion of stars and the \H2\ gas averaged over elliptical annuli as a function of the semimajor axis; using elliptical annuli is more appropriate for cold components with disc-like geometry. The axis ratio and PA of the elliptical apertures correspond to projected circles in the galactic disc, with PA and inclination listed in Table~1. 
Note that for semimajor axis larger than $\sim 4$~arcsec only part of the elliptical annuli falls within the \S\ FOV. 
In the resulting radial profile of the stellar velocity dispersion, there is a clear minimum at a radius of $\sim$4 arcsec, with a value of $\sim$52\kms\ in the LR data. This minimum is not well reproduced in the HR data, which do not extend to large enough radii. The minimum in the \H2\ gas velocity dispersion is not as pronounced, and may be located further out, at $\sim$6 arcsec radius; its variation within elliptical annuli is larger than for stellar velocity dispersion (Fig.~\ref{f_n4536_kin_250}). At $r \lesssim 1$~arcsec the \H2\ velocity dispersion derived from the LR data is higher by $\sim 10$\kms\ than that obtained from the HR data. This is probably the result of the non-circular bulk motions observed in these regions combined with the lower spatial resolution provided by the LR data. Our observation that the regions of low velocity dispersion in \H2\ gas, and particularly in stars in \n4536, are spatially coincident with the nuclear star-forming ring agrees with the general picture of nuclear rings, where stars form from cold molecular gas, and consequently the young stellar population has low velocity dispersion and remains confined to the ring.

\subsubsection{2D kinematic analysis}\label{s_4536_ss2}

Fitting an exponential rotating disc model to the \H2\ velocity field using the inclination and PA of the large-scale disc (left-hand panels of Fig.~\ref{f_n4536_disk}) suggests that the gas flow is dominated by circular motions. 
However, the residuals after subtracting the model LOS velocity from the data (right-hand panels of Fig.~\ref{f_n4536_disk}) are significant (amplitude of $\sim 50$\kms) and coherent over large areas, indicating deviations from this simple picture and suggesting that other motions can significantly contribute to the observed velocity field. 
The residuals do not display any signs of spiral structure (even a partial or pseudo-spiral structure like that seen in \n4501), hence a mode different than a nuclear spiral is taking place in \n4536. Closer inspection of the \H2\ velocity field in Figs.~\ref{f_n4536_kin_100} and \ref{f_n4536_kin_250}, together with the velocity residuals in Fig.~\ref{f_n4536_disk}, indicates that the residuals arise because of the twist in the zero-velocity line and isovelocities around it at $r \lesssim 0.8$~arcsec, which can be best seen in the HR data. As shown in the Appendix, such a twist in the isovelocity lines can be caused by an oval flow. Since the zero-velocity line is consistent with circular motion at larger radii, we deduce that the ellipticity of the flow should be highest at the centre, decreasing outwards until it reaches zero (i.e. circular flow) at some finite radii. Having this in mind, we fitted the observed \H2\ velocity fields with the variable-axis-ratio model described in the Appendix, where the axial ratio of the flow at semimajor axis $a$ is $a/r_{\rm CR}$ for $a<r_{\rm CR}$, and equal to 1 for $a>r_{\rm CR}$, with $r_{\rm CR}$ being the corotation radius of the oval figure of the flow.

\begin{figure*}
\centering
\includegraphics[width=0.55\textwidth]{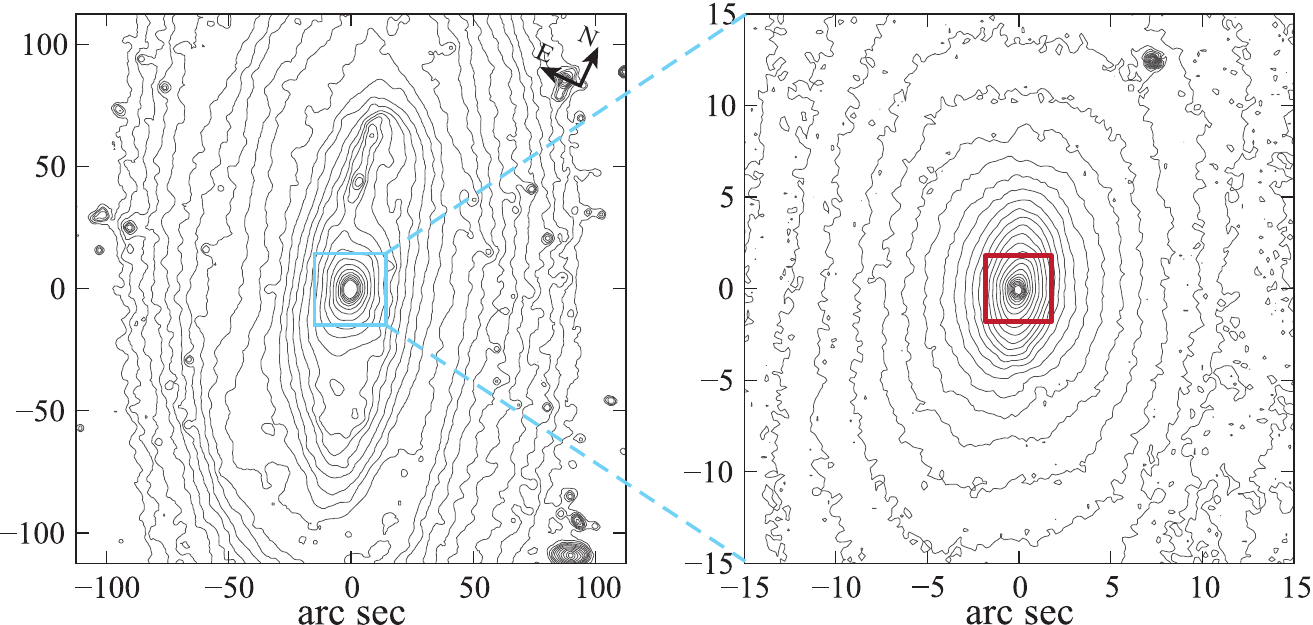}
\caption{IRAC 3.6$\mu$m contours (left panel) and \textit{HST}-NICMOS3 F160W contours (right panel) for \n4569. The red box in the right panel indicates the \S\ FOV.}
\label{f_n4569_im}
\end{figure*}

We performed this oval flow fit with the MCMC technique, using the same LON PA and inclination as the galaxy (Table~\ref{t_prop}). The model that gives the best description of both the HR and LR velocity fields was obtained for the following parameters: the angle between the major axis of the oval flow and the LON of the galaxy $\alpha = 133$\deg\ in the galaxy plane (158\deg\ in the plane of the sky); corotation radius $r_{\rm CR} = 3.7$~arcsec; radius beyond which the rotation curve becomes flat $r_{\rm flat}=3.7$~arcsec; and rotation velocity of the flat part of the rotation curve $v_{\rm flat}=153$\kms. The left-hand panels of Fig.~\ref{f_n4536_bar} show the LOS velocity field of the best-fitting model and the right-hand panels show the residuals after subtracting this model from the data. The oval flow model does a good job in reproducing the non-circular motions observed in the \H2\ emission-line gas in the inner $\sim 1.5$~arcsec of \n4536, including accurately reproducing the zero-velocity twist in the data. The amplitude of the LOS velocity residuals are notably lower than in the case of the circular rotation model, with values $\lesssim 15$\kms. Higher values are observed at the edges of the FOVs, where the measurements errors are higher.

Attempts to increase $r_{\rm CR}$ beyond its best-fitting value produced significantly poorer fits, which indicates that if the model of the oval flow is correct, the figure of the flow rotates quite rapidly, with corotation radius at or just inside the star-forming nuclear ring. We also tried to modify the model by adding a component of radial inflow, but this had no effect on the LOS velocity field for inflow velocities as large as 15\kms. Therefore, we conclude that the data are consistent with only a moderate gas inflow. Thus our model suggests that the gas flow inside the nuclear ring is oval-shaped and rapidly rotating, consistent with an oval gas flow in a nuclear bar located inside the nuclear ring. As the nuclear ring could in turn be caused by gas flow in a large-scale bar, \n4536 might possibly be a double-barred galaxy. We note that there is no indication of a nuclear bar in the NICMOS2 image, though the strong dust lanes could obscure a weak bar.

Note that in order to achieve a good fit to the observed velocity field, the kinematic centre of the model had to be shifted from the maximum of the continuum flux by $\sim 0.2$~arcsec towards the SW, as indicated by the grey cross in the upper panels of Fig.~\ref{f_n4536_bar}. The need for a shift persisted when fitting variations of the model described above. As \n4536 is highly inclined, and the shift of the centre is in the direction perpendicular to the LON,  this shift might be caused by a projection effect, particularly if contributions to the stellar-continuum flux and to the \H2\ emission line have different vertical distributions. 

Once the twist in the central isovelocity lines is modelled by the oval flow, a blueshifted stripe of $\sim -20$\kms\ remains in the LR residual map in the lower-right panel of Fig.~\ref{f_n4536_bar}. This feature is also visible in the disc model residuals. We do not see any clear relation between this feature and any morphological structure observed in the gas and/or dust, although it might be related to the brightest \Brg\ and \HeI\ knot observed in the \S\ LR FOV, which is spatially coincident with the strongest residuals in the blueshifted stripe at $\sim (1.5, -2.5)$~arcsec.

\subsection{\n4569}\label{s_4569}

\n4569 is a large barred spiral in the Virgo Cluster, notable for exhibiting very strong effects of ram-pressure stripping, with the H\,{\sc i} in the disc confined to within $\sim 1/3$ of the optical radius \citep[e.g.][]{Vollmer2004a,Chung2009}. There is also evidence for an extended bipolar outflow from the nuclear regions, possibly due to a recent starburst \citep[e.g.][]{Chyzy2006}. \citet{Vollmer2009a} suggested that the galaxy underwent a major stripping event $\sim 300$~Myr ago, possibly when it passed closest to M87 in its orbit and experienced a strong degree of ram-pressure stripping. In Fig.~\ref{f_n4569_im}, we show log-spaced isophotes from a Spitzer IRAC1 image from SINGS \citep{Kennicutt2003a}, along with a close-up of the central regions based on an \textit{HST} NICMOS2 F160W image. The latter shows the inner boxy region of the bar and a more elliptical structure dominating the inner $r < 5$ arcsec.

\begin{figure}
\centering
\includegraphics[width=\columnwidth]{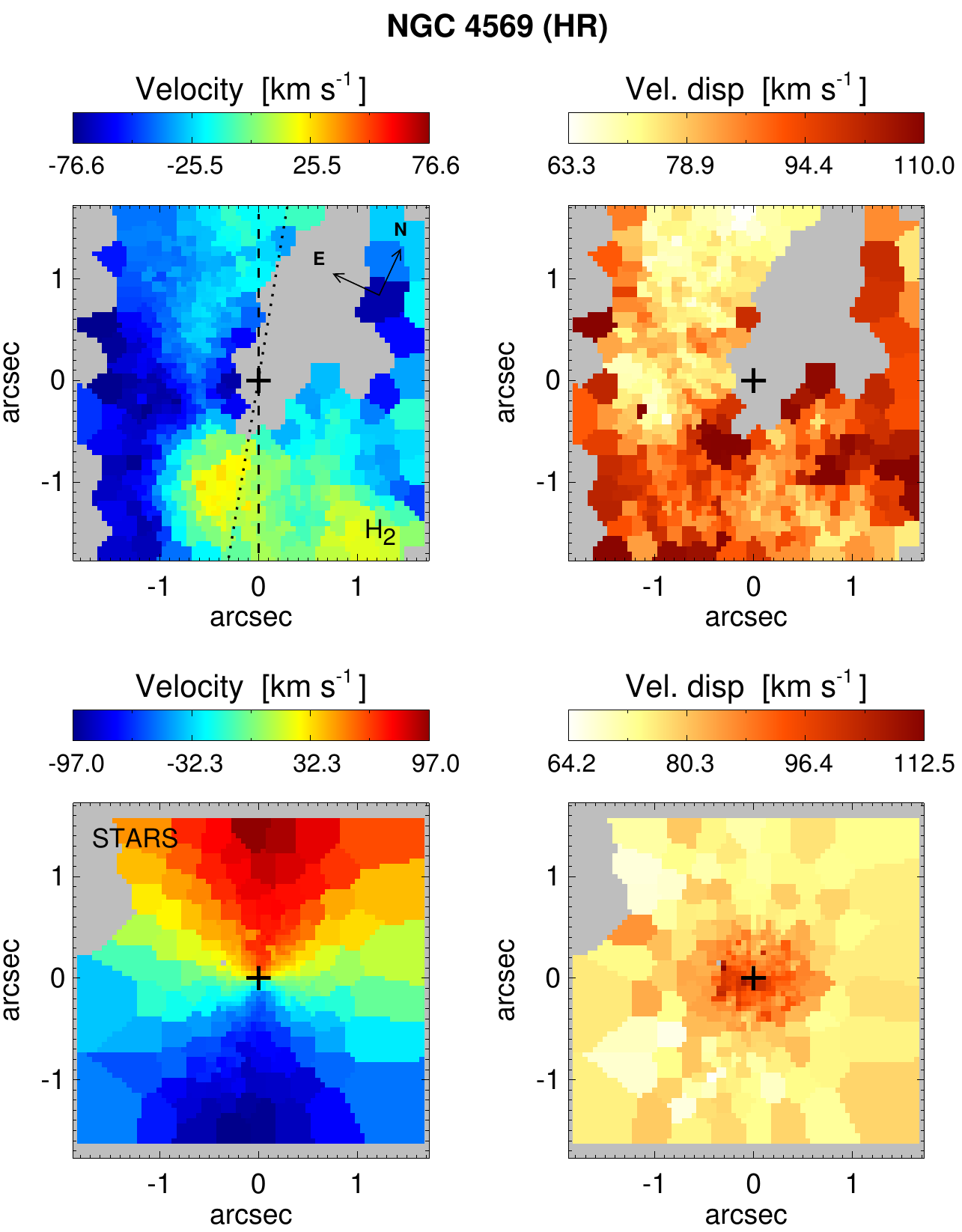}
\caption {2D maps of the \H2\ emission-line gas (upper panels) and stellar (lower panels) kinematics of \n4569 derived from the HR \S\ data. Regions where no reliable parameters were obtained are shown in grey. The spatial orientation, indicated in the upper-left panel, is the same for all the panels. The dashed line corresponds to the position of the major axis of the galaxy and the dotted line to the orientation of the stellar bar.}
\label{f_n4569_kin}
\end{figure}

\begin{figure}
\centering
\includegraphics[width=0.75\columnwidth]{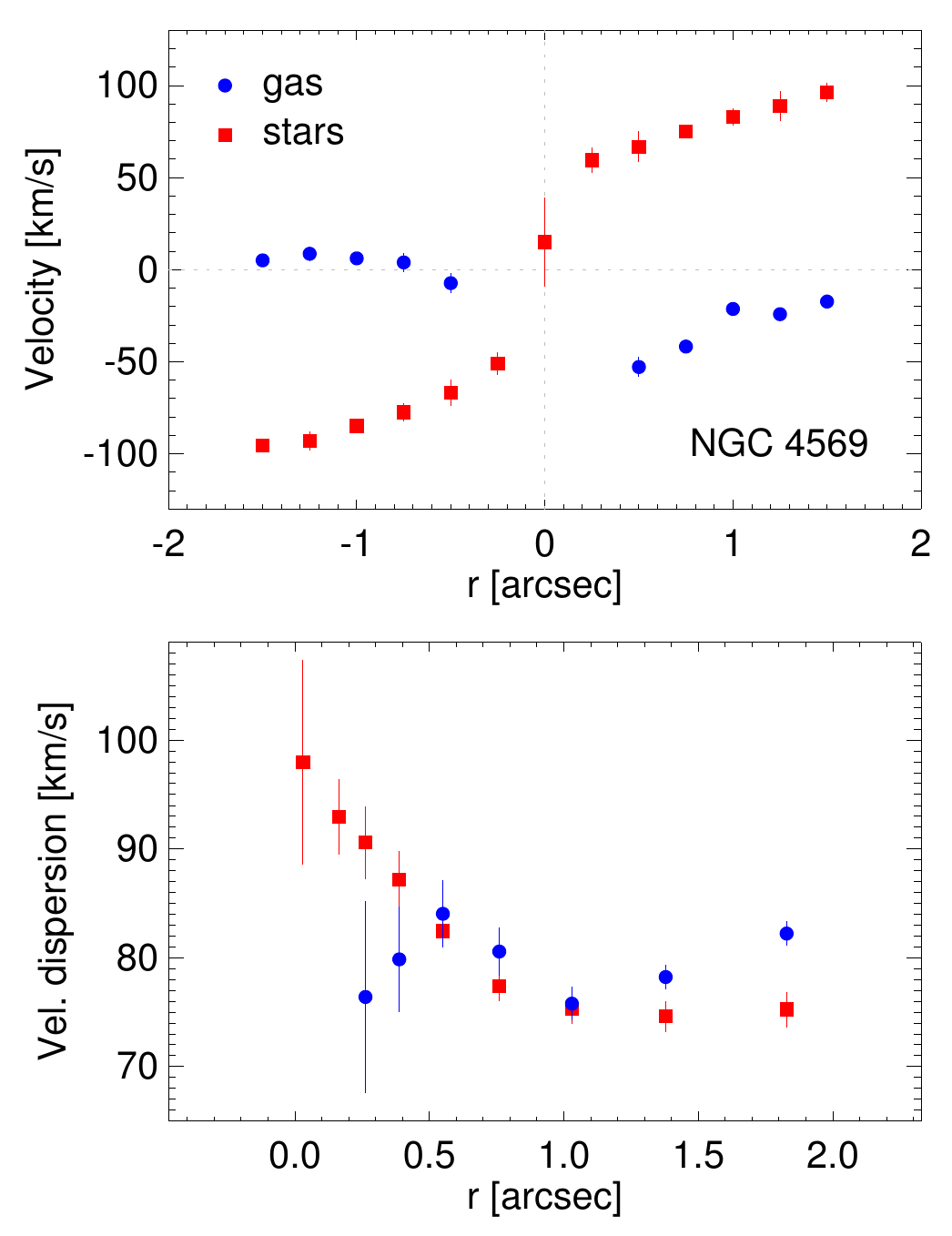}
\caption {\n4569 -- Upper panel: mean values of the velocity measured along the major axis of the galaxy (25\deg) in apertures of $0.25 \times 0.25$~arcsec; negative radii are south from the nucleus. Lower panel: mean velocity dispersion measured in annular apertures. In both panels blue circles correspond to the values measured for the \H2\ gas and red squares to those for the stars. The error bars indicate the 3$\sigma$ error in the mean.}
\label{f_n4569_2D}
\end{figure} 

Since the kinematics of the truncated H\,{\sc i} data may be affected by the bar -- and since CO observations \citep[e.g.][]{Jogee2005,Boone2007} show that the galaxy's molecular gas is only found within the bar \textit{and} shows multiple velocity components and non-circular motions -- we rely on the shape of the outer disc for determining the overall galaxy orientation. From the IRAC1 image, we measure an ellipticity of 0.60, suggesting an inclination of $\approx 69\degr$; our best estimate for the PA of the LON is $\sim 25\degr$ (this agrees very well with the kinematic PA of the inner stellar kinematics from our \S\ data; see below).

The \S\ data allowed us to map the warm \H2\ gas corresponding to a small portion of the molecular bar observed in CO \citepa[the inner $3\times 3$~arcsec,][]{Mazzalay2013a}. 
Figs.~\ref{f_n4569_kin} and \ref{f_n4569_2D} show the kinematic maps and velocity curves along the major axis of the galaxy derived for the \H2\ emission-line gas and the stars. 
The stellar kinematics maps show a very well defined velocity field, characterized by a regular rotation pattern, with a velocity amplitude of about 100\kms. There is a rapid increase in velocity along the major axis of the galaxy, reaching $\sim 60$\kms\ at a distance of only 0.25~arcsec from the nucleus. Such a steep rise is not observed in any other galaxy in our sample. The stellar velocity dispersion is about 75\kms\ for $r > 0.9$~arcsec, and smoothly increases towards the centre to a value of $\sim 100$\kms. The PA of the LON, determined in the same way as the previous galaxies, is $24.0 \pm 0.5$\deg; consistent with the PA of the major axis of the galaxy.

In contrast to what is observed for the stars, the velocity and velocity dispersion maps of the \H2\ gas exhibit a complex structure, with no clear signs of rotation. Due to the very low SNR of the \H2\ line at the nucleus and on the north-west quadrant of the \S\ FOV, it was not possible to derive \H2\ kinematics in these regions. A similar drop in emission is also observed in CO, and is possibly related to the compact starburst located at the centre, which could have consumed or expelled the gas from these regions \citep[][]{Boone2007}.
The bulk of the \H2\ emission-line gas, mainly located on the east side of the nucleus close to the galaxy minor axis, show the highest blueshifted velocities, up to $\sim -75$\kms. Given that the east side is apparently the far side of the galaxy \citep[i.e. the side that appears as the less extinguished,][]{Boone2007}, this could indicate that the molecular gas with brightest \H2\ emission is inflowing in the galaxy plane. However, if vertical motions dominate in that region, then the observed velocity indicates a vertical \textit{outflow}.  
The gas near the major axis of the galaxy has velocities close to the systemic velocity of the galaxy or slightly redshifted ($v < 20$\kms) on the south and blueshifted ($v > -50$\kms) on the north (Figs.~\ref{f_n4569_kin} and \ref{f_n4569_2D}). These values are much lower than the stellar rotation curve at the same location, and if they arise because of gas rotation in the galaxy plane, then that rotation is in the opposite direction from the stellar rotation. 
The \H2\ velocity dispersion spans a range of values similar to that observed for the stellar component, $\sim 70-110$\kms, but its spatial distribution is very different, with most of the high dispersion gas located south of the nucleus and the low dispersion gas towards the east, where most of the \H2\ emission occurs.
To summarize, gas flow in the centre of \n4569 is highly perturbed, and possibly counter rotating. Recent starburst activity \citep[e.g.][]{Keel1996,Ho1997,Ho2001a,Maoz1998,Barth2000,Alonso-Herrero2000,Tschoke2001,Gabel2002} may be responsible for such dynamics. We do not attempt any further dynamical interpretation.

\subsection{\n4579}\label{s_4579}

\begin{figure*}
\centering
\includegraphics[width=0.55\textwidth]{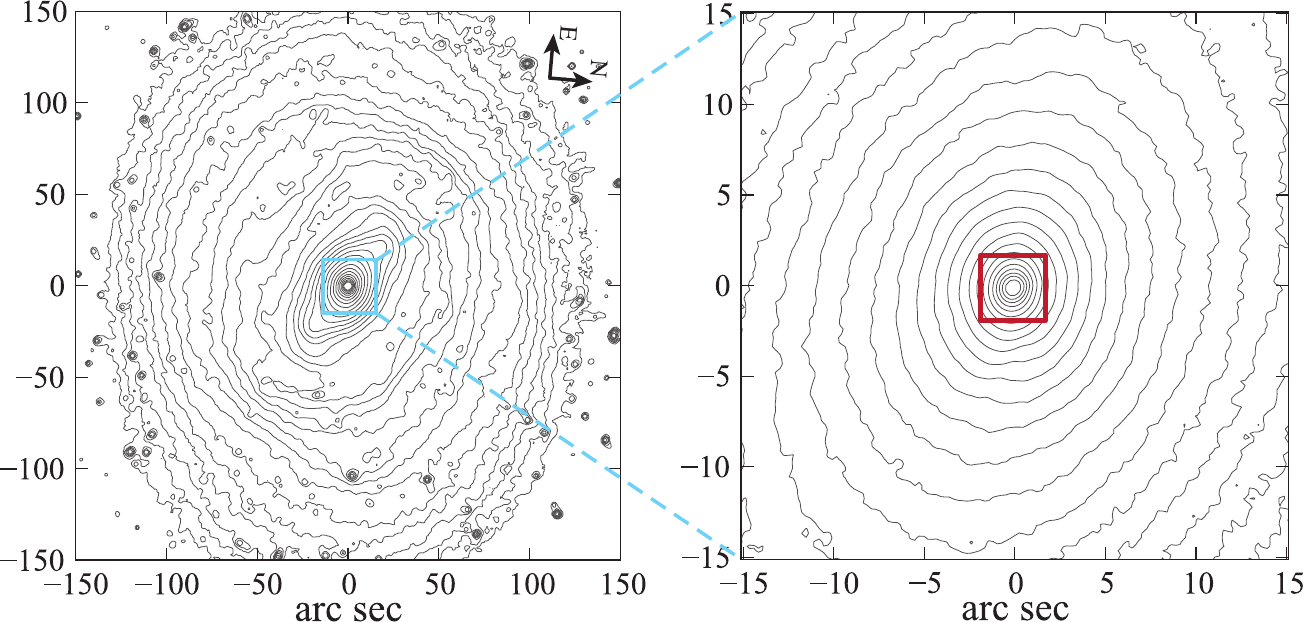}
\caption{IRAC 3.6$\mu$m contours (left panel) and WHT-INGRID $K$-band contours (right panel; image from Knapen et al. 2003) for \n4579. The red box in the right panel indicates the \S\ FOV.}
\label{f_n4579_im}
\end{figure*}

\begin{figure*}
\centering
\includegraphics[width=13.5cm]{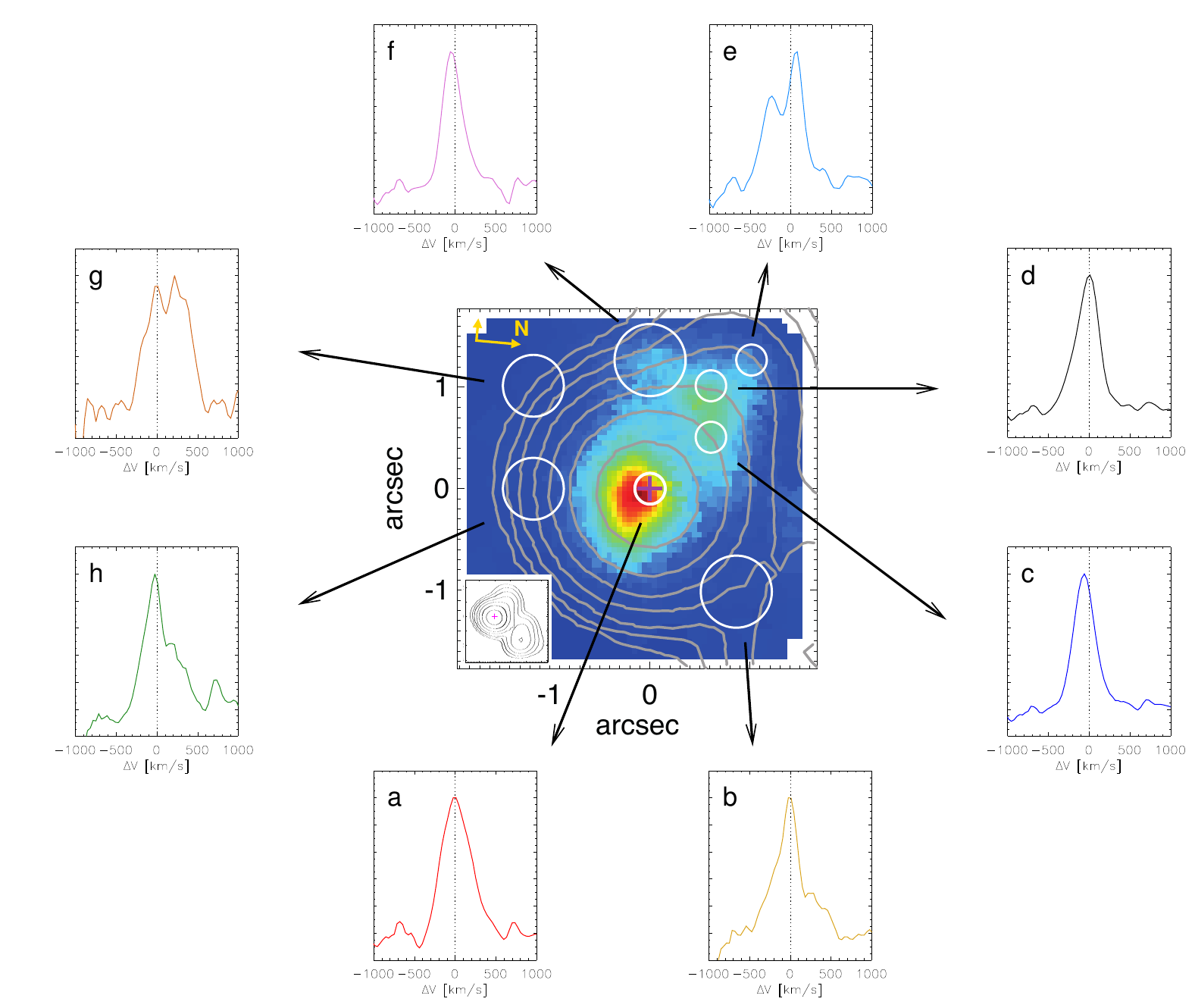}
\caption {\H2\ emission-line profiles of \n4579 in the galaxy's rest frame velocity space. The spectra were extracted from different spatial regions  indicated by white circles on the \H2\ flux distribution map. The grey contours correspond to the VLA 20~cm emission taken from \protect\citet{Ho2001b}. The jet-like radio structure can be better seen in the inset on the lower-left corner, which shows the contours of the same 20~cm emission but over the entire FOV detected, $\sim 5\times5$~arcsec. The $y-$axis of the map corresponds to a PA of 95\deg, approximately the PA of the galaxy's major axis.}
\label{f_perfiles}
\end{figure*}

\begin{figure*}
\centering
\includegraphics[trim = 2cm 2.5cm 2.2cm 10.5cm, clip=true, width=\textwidth]{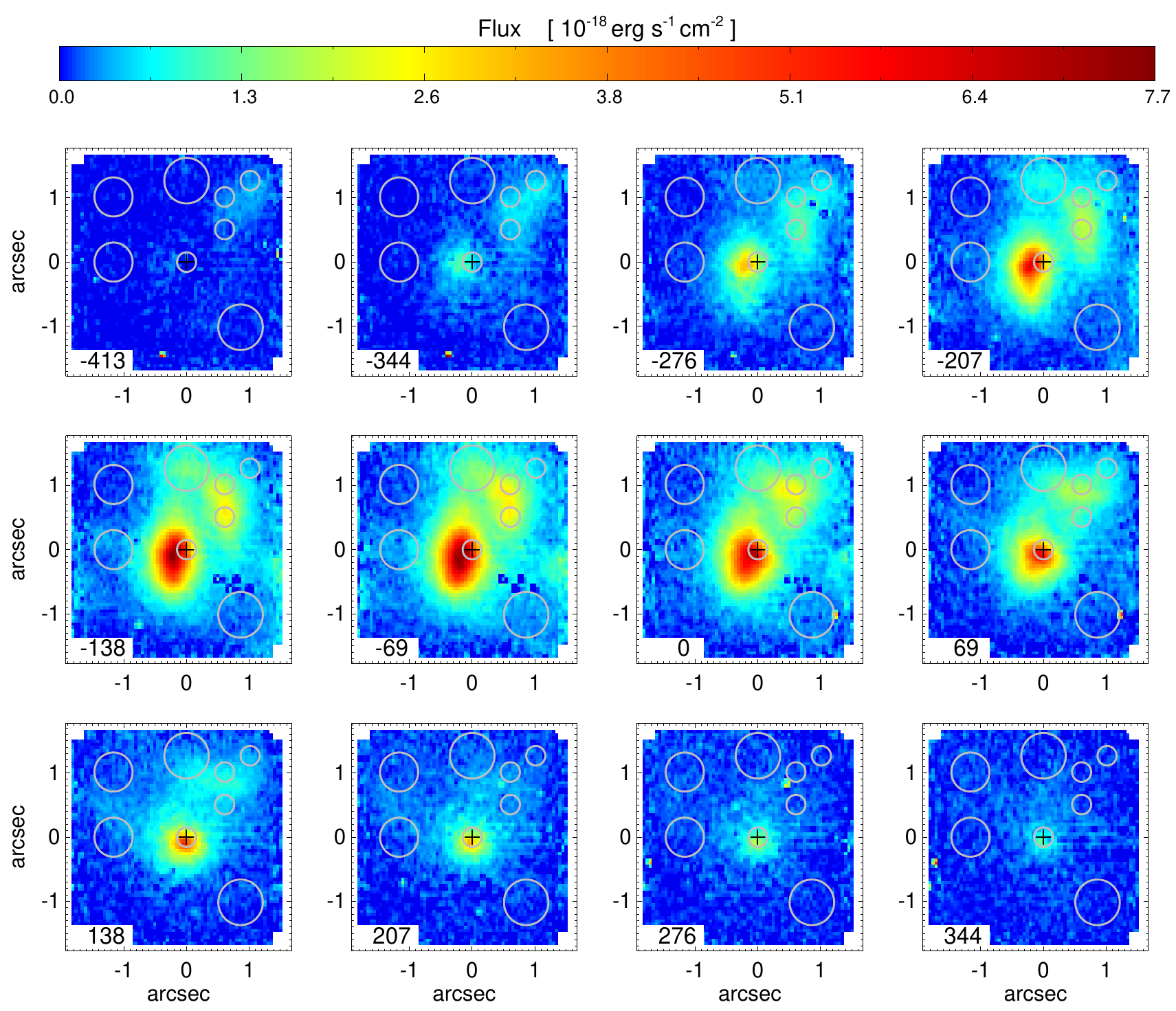}
\caption {\H2~2.12\micron\ channel maps of \n4579. The centre of each velocity bin of 69\kms\ is indicated in the lower-left corner of the corresponding panel. Negative/positive values correspond to blueshifted/redshifted velocities relative to the systemic velocity of the galaxy. The spatial orientation of each panel is the same as in Fig.~\ref{f_perfiles}, with the $y-$axis approximately aligned with the major axis of the galaxy. Gray circles indicate the spatial regions from where the individual spectra showed in Fig.~\ref{f_perfiles} were extracted.}
\label{f_channels}
\end{figure*}

\n4579 is a strongly barred spiral [SAB(rs)b, \citet{deVaucouleurs1991}] in the Virgo Cluster; there is some evidence for possible ram-pressure stripping of the outer gas disc, though the H\,{\sc i} kinematics is regular \citep[e.g.][]{Chung2009} and the effects are nowhere near as extensive as in the case of \n4569. It harbours an active nucleus classified as LINER or Seyfert~1.9 by \citet{Ho1997};  \textit{HST} spectroscopy shows clear broad-line H$\alpha$ emission \citep[e.g.][]{Barth2001c} and there is evidence for compact radio jets emanating from the nucleus \citep[see, e.g.][]{Ulvestad2001a}. 

In Fig.~\ref{f_n4579_im}, we show log-spaced isophotes from a Spitzer IRAC1 image from SINGS \citep{Kennicutt2003a}, along with a close-up of the central regions using the WHT-INGRID $K$-band image of \citet{Knapen2003}. From the outer isophotes in the IRAC1 image, we measure a PA for the LON of 95\degr{} and an ellipticity of 0.22; the latter implies an inclination of $\approx 40\degr$.  These estimates agree well with the kinematic analysis of the H$\alpha$ velocity field by \citet{Chemin2006a}, who found PA $= 91 \pm 3\degr$ and $i = 44 \pm 8\degr$.

\n4579 was observed with \S\ in its HR mode, covering the inner $\sim 3\times 3$~arcsec of the galaxy. The spectra exhibit intense \H2\ emission lines; no evidence of ionized gas was observed. As we mention in Section~\ref{s_gaskin}, the \H2\ emission-line profiles display strong asymmetries and double peaks, hence no kinematic maps were constructed for \n4579. Instead, in this section we analyse the \H2\ emission line profile in different regions of the galaxy, as well as the \H2\ channel maps.  

The complexity and spatial variation of the emission-line profiles can be seen in Fig.~\ref{f_perfiles}, which shows the \H2~2.12\micron\ emission-line profiles corresponding to different spatial regions, as indicated on the \H2\ flux distribution map \citepa{Mazzalay2013a}. 
At least two main velocity components are superimposed along the LOS in most regions. None of these components seems to match the circular rotation in the plane of the galaxy observed for the stars \nocite{Dumas2007}(e.g. Dumas et al. 2007; Erwin et al. in preparation), which, assuming that the northern side is the one closer to the observer (as suggested by the dust distribution, \nocite{Pogge2000} Pogge et al. 2000), is rotating anticlockwise. Note that if the gas followed the stellar kinematics, we would see redshifted (blueshifted) velocities towards the east (west). 
The \H2\ emission-line profile at the nucleus [panel (a) of Fig.~\ref{f_perfiles}] peaks at the systemic velocity of the galaxy and has the largest FWHM (about 420\kms). The shape of the profile is not well reproduced by a Gaussian function, and is probably the result of blending of two or more (narrower) components with bulk velocities close to the systemic velocity of the galaxy, but slightly redshifted and blueshifted. 
Towards the NE, where most of the intense extra-nuclear \H2\ emission is observed, the line profiles are single-peaked (with $\Delta v \lesssim -50$\kms\ and FWHM $\sim 300-350$\kms) and slightly asymmetric [weak blue or red wings can be observed in the spectra of panels (d) and (f)].
Further away from the nucleus, a strong double-peaked emission-line is observed [panel (e)]. The difference of the maxima of these components is of 300\kms, with one component strongly blueshifted ($\Delta v \sim -225$\kms\ and FWHM $\sim 260$\kms) and the other redshifted ($\Delta v \sim 25$\kms\ and FWHM $\sim 230$\kms). 
Additional multicomponent line profiles are observed towards the SE, S and NW of the nucleus [panels (b), (g) and (h)]. These profiles are characterized by a relatively narrow component at about the systemic velocity of the galaxy ($\Delta v \lesssim -50$\kms\ and FWHM $\sim 250$\kms) and a broader redshifted component ($\Delta v \sim 260-300$\kms\ and FWHM $\sim 350$\kms).

Additional information about the molecular gas kinematics in \n4579 can potentially be obtained from analysis of the \H2\ emission-line distribution in velocity slices or channel maps. Fig.~\ref{f_channels} shows the channel maps constructed by integrating the \H2~2.12\micron\  flux in velocity bins of 69\kms\ along the line profile. Each panel represents the flux distribution of a velocity bin centred at the velocity indicated in the lower-left corner. Positive/negative values indicate redshifted/blueshifted velocities relative to the systemic velocity of the galaxy. 
Gas with velocities between approximately $-400$ and $350$\kms\ are observed.   
The bulk of the \H2\ emission, located close to the nucleus in the SW quadrant, is mostly blueshifted or close to the systemic velocity of the galaxy ($\Delta v$ = -207 to 69\kms). At velocities outside this range, the \H2\ emission is more concentrated towards the nucleus. Approaching/receding gas with velocities of up to -344/344\kms\ is observed at the centre, consistent with the nuclear spectrum showed in panel (a) of Fig.~\ref{f_perfiles}. 
The \H2\ emission located further from the nucleus towards the NE does not reach positive velocities as high as in the central regions (it is only clearly observed up to $\Delta v$=207\kms) and displays mostly blueshifted velocities. Weak emission is observed at the velocity bin centred at $\Delta v=-413$\kms\ around 1~arcsec NE from the nucleus; no clear detection of emission at these velocities is seen in the nucleus. As suggested by the \H2\ emission-line profiles, the bulk of the \H2\ emission is not following the rotation pattern seen for the stars.  
The multi-velocity nature of the \H2\ seen with the line profiles [see, for example, panels (b), (g) and (h) of Fig.~\ref{f_perfiles}] is not clearly observed in the channel maps, since correspond to regions where the \H2\ emission is weak. Some hints of the double peak structure observed at $\sim 1.5$~arcsec NE from the nucleus [panel (e) of Fig.~\ref{f_perfiles}] can be seen in the channel maps: going from lower to higher velocities, the intensity in this region has a first peak at $\Delta v=-276$\kms, then drops slightly, and has a second peak $\Delta v=0$\kms. 

The observed line profiles and channel maps are not easy to reconcile with a simple velocity pattern in the plane of the galaxy (e.g. circular rotation or in-plane radial outflow/inflow). Instead, they can be explained as extra-planar gas outflows, as previously proposed by \citet{Garc'ia-Burillo2005, Garc'ia-Burillo2009} to explain the CO kinematics in the inner $\sim 2$~arcsec of this galaxy.  
The multiple components of \H2\ emission observed with \S\ can be interpreted as motion towards and away from the galaxy centre projected onto the LOS. This picture is supported by the presence of a jet-like radio structure and its spatial correlation with the \H2\ emission observed in the galaxy. We included in Fig.~\ref{f_perfiles} the VLA 20~cm contours taken from \citet{Ho2001b}, which shows a radio-jet--like structure resolved into two components, one associated with the nucleus and a second weaker source located towards the NW (see the inset on the lower-left corner of the \S\ FOV in Fig.~\ref{f_perfiles}). It can be seen that the nuclear radio source is asymmetric and extended towards the NE, where intense off-centre \H2\ emission is observed, suggesting a relation between the two media. Additionally, the \H2\ emission observed with \S\ resembles that seen in high-resolution UV, \Ha\ and \OIII\ \textit{HST} images \citepa{Mazzalay2013a}. The UV and \Ha\ emission was analysed in detail by \citet{Comer'on2008}, who concluded that the emission in the inner 100~pc radius of the galaxy is most likely the result of the interaction of the radio jet with the ambient gas.

\section{Discussion}\label{s_discussion}

\subsection{Velocity fields: types of gas flow in galactic nuclei}\label{s_flows}

In the previous section we presented and analysed the kinematics of the \H2\ emission-line gas of a sample of six nearby spiral galaxies. Only one of these galaxies (\n3351) shows kinematics on nuclear scales consistent with circular motion in a disc. The remaining five galaxies display more complex motion on scales down to the resolution limit (10--15~pc).  Departures from planar circular motion have conventionally been analysed in terms of warped discs (tilted-ring fitting, e.g. Neumayer et al. 2007)\nocite{Neumayer2007}, radial flows \citep[e.g.][]{Wong2004a} or vertical outflows. Far less attention has been given to oval-like flows, which arise naturally in potentials without axial symmetry, like in bars. However, \citet{Spekkens2007} have already shown that oval flows can be misinterpreted as significant radial outflows or inflows, which generate serious continuity problems. 

An oval flow is in fact the simplest kind of flow that can account for the kinematics observed in the nuclei of \n3627 and \n4536. In \n3627, there is additional circumstantial evidence for an oval flow: the gas morphology is elongated for both cold and warm molecular gas (CO and \H2\ emission), and an oval flow which is aligned with the observed morphology also reproduces the observed kinematics; this oval flow is not elongated along the major axis of the stellar bar. Although the inner gas morphology in \n3627 is misaligned with the major axis of the bar, it is probably still driven by the bar. Note that \n3351 gives an example of an opposite situation, in which the gas flow in a bar settles on a nuclear ring of radius $\sim$350~pc, and inside the nuclear ring the gas flow is consistent with circular motion.  

Despite the example of \n3351, it is evident in our sample that circular gas flows inside nuclear rings are not a rule, either. In \n4536, which has a nuclear star-forming ring, we observe a clear twist in zero-velocity line in the \H2\ kinematics, which is inconsistent with circular motion in the galaxy plane. We interpret, and are able to successfully model, this twist as a signature of an oval flow. As the twist is present only inside the nuclear ring, only there the flow is non-circular. The data are better fitted by a rotating figure of flow with corotation radius approximately equal to the radius of the nuclear ring, which suggests that the oval flow inside the nuclear ring might be driven by an inner bar. 

In \n4501, a strong velocity perturbation, with an amplitude similar to the rotation velocity itself, is imposed on the circular motion over a large spatial extent. Although the perturbation occurs over an area with a shape suggestive of a spiral arm, we showed that it is inconsistent with motion in a nuclear spiral. Therefore in \n4501, we are possibly seeing a filament moving with high velocity against the background. The gas kinematics in \n4569 does not show any underlying circular motion, though we still can single out areas of coherent motion. Finally, in \n4579 the \H2\ emission line cannot be fitted by a single Gaussian, which indicates the existence of multiple gas components with different kinematics that are not spatially resolved by our data.  

In short, the observed gas kinematics in galactic nuclei of our sample is consistent with the following types of gas flow, in order of increasing complexity: (1) simple circular motion in a disc (\n3351); (2) non-circular (oval) motion in a disc (\n3627 and \n4536); (3)  streaming motion superimposed on circular rotation (\n4501); and (4) complex streaming motions in absence of circular rotation (\n4569 and \n4579). 
We note that although the fraction of galaxies with simple circular motion (type 1) can be small, a good fraction of galaxies may display other forms of ordered motion (like type 2), or underlying circular rotation (type 3), which are good tracers of the potential well in which they reside.

\subsection{Velocity dispersion: its origin and role in gas dynamics}\label{s_discdyn}

In all the galaxies of our sample, the observed LOS \H2\ velocity dispersion in the inner 1--2~arcsec ($\sim$50--160~pc) in radius, after correction for instrumental broadening, is high -- usually higher than 50\kms. High gas velocity dispersion in the innermost parsecs of galaxies has been noticed in earlier observations intended to measure central SMBH masses \citep[e.g.][]{Macchetto1997,vanderMarel1998,Verdoes2000,Barth2001b,Marconi2006,Shapiro2006,deFrancesco2006,Neumayer2006,Walsh2010}, and is now intensely studied in the context of the nuclear activity in galaxies \citep[][]{Hicks2009, Sani2012, Muller-S'anchez2013}. 
In the sample of nearby Seyfert galaxies of \citet{Hicks2009}, \H2\ velocity dispersions higher than 60\kms\ are also seen, though most of these observations are limited to the
inner 0.5--1.0 arcsec in radius. In a small sample of LLAGN galaxies studied by \citet{Muller-S'anchez2013}, the \H2\ dispersion rises to values $> 100$ \kms, but only in the inner 0.2--0.4 arcsec, and in fact the dispersion drops to values of $\sim 30$ \kms\ for radii $> 0.6$--0.9 arcsec. 
Similarly, rapid increases in the velocity dispersion of ionized gas in the innermost few tens of arcsecond of galaxies with central SMBHs have been observed, which can be partially ascribed to the smearing by the point spread function (PSF) of the rotation velocity gradient, but models indicate that in some galaxies a significant intrinsic velocity dispersion has to be present in order to account for the observed dispersion \citep[e.g.][]{vanderMarel1998,Verdoes2000,Barth2001b,Marconi2006}. 

In our data, the \H2\ emission-line gas display high velocity dispersions out to radii even larger than in the previous observations listed above: in \n3627 and \n4501 it remains above 60\kms\ out to almost 3~arcsec in radius, and for both \n3627 and \n4501 it remains $\gtrsim 50$ \kms\ out to at least 5~arcsec. Since this is far outside the regions where the effects of PSF are expected to be important, smearing of the rotation velocity gradient contributes negligibly to the observed velocity dispersion in our data, and what we observe is mostly the intrinsic velocity dispersion in the \H2\ emission-line gas. 

Currently, the physical nature of the intrinsic velocity dispersion of the gas in the centre of galaxies is poorly understood. 
Part of it can be attributed to thermal broadening which, for example, in the case of ionized hydrogen at 10\,000~K corresponds to $\sim 10$\kms. However, this can clearly only explain a small fraction of the intrinsic velocity dispersion usually derived for the gas in the inner tens of parsecs of galaxies. \nocite{vanderMarel1998}Van der Marel \& van den Bosch (1998) suggested that this dispersion can be caused by local turbulence, generated by processes such as stellar winds and supernova explosions. In this case, the dispersion can be considered dynamically unimportant. On the other hand, disc dynamics \textit{can} be affected by high velocity dispersion if the latter is due to random motion of individual gas clouds generated by their gravitational interaction \citep[e.g.][]{Verdoes2000,Neumayer2006}. In this case we call the velocity dispersion dynamical. In particular, if the gas rotates in a disc, then its average azimuthal velocity decreases when its velocity dispersion increases. The dynamical importance of the gas velocity dispersion can be accounted for via asymmetric drift corrections \citep[e.g.][]{Barth2001b} or use of the Jeans equations with a pressure-like term \citep[e.g.][]{Neumayer2006}. 

In our sample, the intrinsic velocity dispersion of the \H2\ emission-line gas reaches values of 80\kms\ or more. This is close to the dissociation limit of the \H2\ molecule: \citet{LeBourlot2002} calculated that the maximum shock speeds which can be attained prior to the collisional dissociation of \H2\ increase from 20-30\kms\ at high pre-shock densities (n$_{\rm H} \geq 10^6$~cm$^{-3}$) to 70--80\kms\ at low densities (n$_{\rm H} \leq  10^4$~cm$^{-3}$). Observationally, \citet{Giannini2008} confirmed these values, obtaining  a range for the \H2\ breakdown velocity of 70--90\kms\ in the Herbig--Haro object HH99B. This puts an upper limit on plausible values of velocity dispersion due solely to local turbulence; dynamical dispersions can be higher if cloud-cloud collisions are rare. 
So while the velocity dispersions seen in our sample galaxies \textit{could} be mostly (or even entirely) due to internal turbulence, we cannot rule out the possibility that the dispersion is partly dynamical.

When interpreting the observed values of the \H2\ velocity dispersion, one should keep in mind that the \H2\ emission comes from a small fraction of the gas; most of the mass is expected to reside in colder, dense clumps. 
\citet{Sani2012} showed that for the four Seyfert galaxies they observed, the velocity dispersion in HCN and HCO$^{+}$ emission, which traces dense gas, is systematically smaller than the dispersion in both \H2\ and the stars; the HCN and HCO$^{+}$ dispersion (ranging from 20 to 40 \kms) is about half of the stellar and \H2\ values within the innermost 20--100~pc radius. We observe a similar trend in \n3351, for which the measured velocity dispersion of the \H2\ emission-line gas in the inner $\sim 25$~pc radius is about twice that previously reported for CO observations (Section~\ref{s_3351}). 
This suggests that if the \H2\ velocity dispersion is entirely dynamical (i.e. not due to local turbulence), then the observed \H2\ rotation velocity should be lower than that of dense tracers (e.g. CO, HCN, HCO$^{+}$, etc.). 
However, in their analysis of \n3227, \citet[][]{Davies2006} found no difference between the rotational velocities of CO and \H2\ for this galaxy, even though the velocity dispersion of \H2\ was almost twice that of CO in the overlapping region. Future ALMA observations will hopefully provide the necessary data to resolve this issue, but at this stage we suggest caution when interpreting velocity dispersions of warm or ionized tracers as being entirely dynamical. While most of the observed dispersion in cold, dense tracers is possibly dynamical, as dense cores of clouds behave more like bullets and respond more to gravity than to hydrodynamical forces, the higher dispersion in warm or ionized tracers can be due to an additional contribution from local turbulence.

\subsection{The vertical structure of circumnuclear gas discs}\label{s_disc_thick}

In addition to affecting the azimuthal velocity of gas rotating in a disc, dynamical velocity dispersions can also imply a vertical thickening of the disc. There are different ways to relate the gas velocity dispersion and disc height $h_z$, though in practice, these can only produce rough estimations of the disc height based on simplifying assumptions about the physical system in study. For example, in the potential of a razor-thin stellar disc of surface density $\Sigma$, an isothermal gas disc with vertical velocity dispersion $\sigma_z$ has a scale height 
\begin{equation}
h_z = \frac{\sigma_z^2}{2 \pi G \Sigma}. 
\end{equation}
This is also the scale height of a mixture of stars and gas of equal vertical velocity dispersions $\sigma_z$ and total surface density $\Sigma$. An alternate approach is to use the ratio of observed rotation velocity -- assumed to be given by the deprojected LOS velocity along the LON ($v_{\rm dp} = v_{\rm LOS}/\sin i$) -- to the velocity dispersion, under the assumption that at radius $R$,
\begin{equation}
\frac{v_{\rm dp}}{\sigma_z} \sim \frac{h_z(R)}{R},
\end{equation}
in which case values of this ratio $< 1$ indicate a thick disc.

For their sample of nine Seyfert galaxies, \citet{Hicks2009} used both of the two approaches outlined above to obtain the disc approximate scale height of the galaxies.
Employing the isothermal-sheet approximation, and assuming isotropic intrinsic velocity dispersions in the \H2\ gas, they used the observed LOS velocity dispersion to estimate the scale heights of the \H2\ discs at a 30~pc radius from the galaxy centre, which coincides with the average radial half width at half maximum (HWHM) of the \H2\ emission in their sample. The average scale height they derived is 40~pc, effectively equal to the \H2\ emission scale length in the disc plane. Using the second approach listed above, they found results similar to those obtained with the first approach: $v_{\rm dp}/\sigma$ at the radius of 30~pc is $0.9 \pm 0.3$, consistent with the nuclear \H2\ discs being thick at HWHM of the \H2\ radial distribution. 

However, it is important to note that both approaches are strictly correct only in the limit of thin discs ($h_z \ll R$), where the radial variation of the potential is negligible compared to the vertical variation; in addition, in the second approach the left- and right-hand sides are equal only for a massless, thin disc residing in a Keplerian potential. When the estimated vertical scale height and the radial extent are approximately equal, as the results of \citet{Hicks2009} would seem to suggest, then the approximations are not longer valid.  A second issue is the question of how much of the observed velocity dispersion is dynamical, versus how much is due to, e.g. turbulence within clouds. Taking into account the smaller velocity dispersion in colder and denser gas traced by the HCN molecular emission, \citet{Sani2012} suggested that the gaseous nuclear discs of radii 20--80~pc are much less thick, with a vertical-to-radial size ratio of about 1/4.

\begin{figure}
\includegraphics[trim = 2.8cm 2cm 5cm 16.5cm, clip=true,width=\columnwidth]{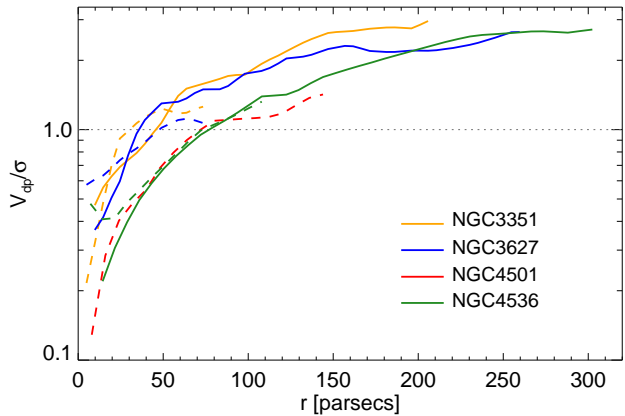}
\caption{Ratio of the \H2\ deprojected velocity to the velocity dispersion as a function of the distance to the nucleus of each galaxy indicated in the lower-right corner. Dashed and solid lines correspond to the values estimated from the HR and LR data, respectively.}
\label{f_v_sigma}
\end{figure}

\begin{table}
\centering
  \caption{Deprojected velocity ($v_{\rm dp}$) and velocity dispersion ($\sigma$) of the \H2\ emission-line gas and stars at a distance of 50~pc from the nucleus of the galaxies.}
    \begin{tabular}{@{}cccccccc@{}}
    \hline
\multirow{2}{*}{Galaxy}    &    & \multicolumn{3}{c}{Stars}       & \multicolumn{3}{c}{Gas} \\
  &&   $v_{\rm dp}$   &   $\sigma$   & $v_{\rm dp}/\sigma$ & $v_{\rm dp}$   &   $\sigma$   & $v_{\rm dp}/\sigma$     \\
\hline
\multirow{2}{*}{\n3351} & HR & 31.0 & 72.8 & 0.4 & 69.6 & 56.5 & 1.2 \\
\smallskip
                        & LR & 38.5 & 74.4 & 0.5 & 52.5 & 49.1 & 1.1 \\

\multirow{2}{*}{\n3627} & HR & 80.6 & 86.0 & 0.9 & 68.4 & 67.1 & 1.0 \\
\smallskip
                        & LR & 77.3 & 89.0 & 0.9 & 84.3 & 64.9 & 1.3 \\
\smallskip
\n4501                  & HR & 52.5 & 105.2 & 0.5 & 42.4 & 62.3 & 0.7 \\

\multirow{2}{*}{\n4536} & HR & 42.1 & 86.4  & 0.5 & 41.4 & 59.5 & 0.7 \\
\smallskip
                        & LR & 48.0 & 85.0  & 0.6 & 51.1 & 75.4 & 0.7 \\

\n4569                  & HR & 120.8 & 75.0 & 1.6 &  --  &  --  &  -- \\
\hline
\end{tabular}
\label{t_kin}
\end{table}

\citet{Muller-S'anchez2013} make further use of $v_{\rm dp}/\sigma$ as an indicator of the disc thickness, suggesting that this ratio increases with radius more rapidly in LLAGN than in Seyfert galaxies (on average $v_{\rm dp}/\sigma=1$ at $r\sim 30$ and $\sim 75$~pc, respectively), but slower than in starbursts and inactive galaxies ($v_{\rm dp}/\sigma=1$ at $r\sim 5$~pc), which they interpret as different thickness of the nuclear disc at various stages of nuclear activity and a possible evolutionary sequence of AGN activity and SMBH growth.
The galaxies in our sample cover a range of nuclear activity, including starburst (\n3351), LLAGN (e.g. \n4536) and Seyfert (e.g. \n4501); see column~3 of Table~\ref{t_prop}. This gives us the opportunity to compare our data with the scenario proposed by \citet{Muller-S'anchez2013}. Fig.~\ref{f_v_sigma} shows the ratio of the deprojected velocity to the velocity dispersion of the \H2\ emission-line gas as a function of the distance from the nucleus for the four galaxies in our sample that display ordered velocity fields. We also include in Table~\ref{t_kin} the deprojected velocity, velocity dispersion and $v_{\rm dp}/\sigma$ for both the stars and \H2\ gas, measured at a distance of 50~pc from the nucleus of these galaxies.
Two of our galaxies display $v_{\rm dp}/\sigma=1$ at radii similar to those proposed by M\"uller-S\'anchez et al.: the LLAGN \n3627 shows $v_{\rm dp}/\sigma=1$ at $r\sim 35$~pc and the Seyfert~2 galaxy \n4501 at $r\sim 75$~pc. However, the other two galaxies for which $v_{\rm dp}/\sigma$ was possible to estimate, \n3351 and \n4536, disagree with the range of values derived by these authors. The first galaxy, probably best classified as a starburst, has $v_{\rm dp}/\sigma=1$ at a relative large radius ($r\sim 30$~pc) compared to the non-AGN galaxies in M\"uller-S\'anchez et al.'s sample; it is much more similar to  their LLAGNs. The second galaxy, a LLAGN, does not reach $v_{\rm dp}/\sigma=1$ until $r\sim 75$~pc, which is consistent with the Seyfert galaxies in M\"uller-S\'anchez et al.'s sample instead of the LLAGNs.
Thus, only two of four galaxies in our sample follow the scenario proposed by M\"uller-S\'anchez et al.. So far, this scenario has been tested in only a handful of objects (sixteen including our galaxies), and additional observations are needed to derive more robust conclusions.

\begin{figure*}
\includegraphics[width=0.75\textwidth]{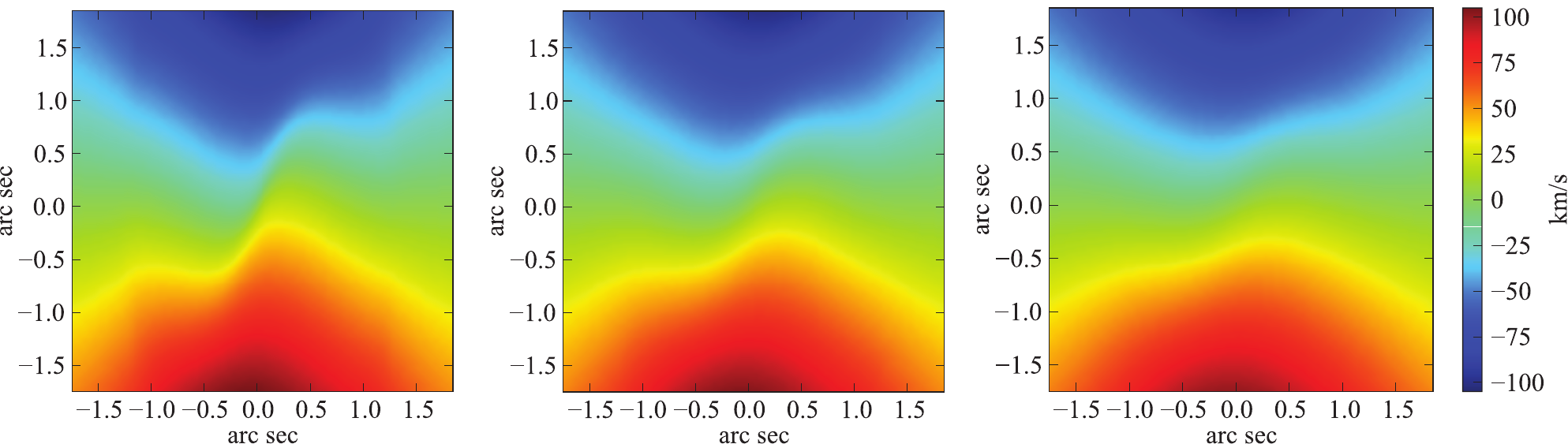}
\caption{Projections of a geometrically thick and optically thin model disc. We use the best-fitting model velocity field from our fit to the \H2\ HR velocity field of \n4536 (see Section~\ref{s_4536} and Figs.~\ref{f_n4536_kin_100} and \ref{f_n4536_bar}) for the planar velocity field, assume cylindrical rotation, and view the model at an inclination of 67\deg; the emissivity-density profile of the disc is uniform in the radial direction and exponential in the vertical direction. The panels show the effects of different disc scale heights: $h_z = 4$~pc (left), 20~pc (middle) and 40~pc (right). Note that the complex velocity structure is considerably smeared away when $h_z = 20$~pc, and is almost completely gone when the thickness is 40~pc.}\label{f_discpeter}
\end{figure*}

We also note that there are observational indications that $v/\sigma$ may not always be a good indicator of disc thickness. For example, \citet{Neumayer2007} note the elongated contours in both flux and velocity dispersion maps of \H2\ emission in \n5128 (Centaurus~A), which they suggest is consistent with high-velocity-dispersion gas residing in a disc on scales of 0.4~arcsec in radius. However, the $v/\sigma$ ratio is below 0.5 in this region; if this is interpreted in the disc-thickness paradigm, then the inner region should be effectively spherical and the contours should be much rounder than they actually are.

More generally, if such circumnuclear discs are indeed thick in their innermost regions, the ellipticity of their isophotes should decrease towards the centre once the thickened region is approached. In our sample, this might be the case in \n3627, where the ellipticity of the \H2\ emission line distribution does decrease inside a radius of 1~arcsec, and also in \n4536, where the \H2\ distribution is almost circular inside the nuclear ring \citepa[$r < $1~arcsec, see][]{Mazzalay2013a}. On the other hand, the \H2\ emission in the inner 1~arcsec of \n3351 is highly elongated, down to our resolution limit, which may indicate that \n3351 does not posses a thick nuclear disc, at least on scales larger than the PSF of our observations.

Additional clues about the geometrical and optical thickness of gas discs can potentially be obtained from their velocity maps. If a clear twist in the zero-velocity line is observed in the central region, as in the case of \n4536, then the disc cannot be both geometrically thick and optically thin in this region, because the velocity gradient perpendicular to the LON would then be erased by geometrical projection effects. This is demonstrated in Fig.~\ref{f_discpeter}, where we show projections of an optically thin disc model with different scale heights. This model uses the model velocity field from our fit to the HR velocity field of \n4536, assuming cylindrical rotation, with \H2\ emissivity following a uniform radial profile and an exponential vertical profile. We incline it at 67\degr{} and perform LOS integrations through the model while assuming different scale heights $h_z$. A thin disc ($h_z = 4$~pc, left panel) shows the same complexity as the input velocity field. But when the disc has a scale height of 20~pc (middle panel), the twist in the zero-velocity line is strongly smeared, and for $h_z = 40$~pc, almost none of the original spatial complexity is preserved. We conclude that the inner disc in \n4536 is either thin, despite its high velocity dispersion, or that it is both geometrically \textit{and} optically thick.

\subsection{Black hole mass estimates from gas measurements}\label{s_bh}

The galaxies discussed in this paper were originally observed with SINFONI as part of a project to measure central SMBH masses \citep[e.g.][]{Nowak2010,Rusli2011,Rusli2013a}. Although the primary aim of that project has been to use the \textit{stellar} kinematics to determine the SMBH masses (Erwin et al. in preparation), it is interesting to investigate to what extent the gas kinematics might also be used for this purpose.

This is particularly relevant given that gas kinematics has been one of the main tools for measuring SMBH masses (see references in Sections~\ref{s_intro} and \ref{s_discdyn}). Under standard assumptions, using gas kinematics is considerably easier than using stellar kinematics, since the latter involves the iterative calculation of thousands or tens of thousands of stellar orbits in a model potential as part of the Schwarzschild modelling process, which is computationally very intensive. In contrast, using gas kinematics can (ideally) be as simple as matching a model circular-velocity curve to a gas rotation curve. 

However, the drawback of the gas-kinematics approach is that it usually requires to make certain simplifying assumptions about the gas: in particular, that the gas is in \textit{ordered, circular} motion -- ideally in a kinematically cold disc. How often are these assumptions valid for the galaxies in our sample?

\n4569 and \n4579 are strongly ruled out in this sense: the circumnuclear gas motions in both galaxies are severely disordered, with clear evidence for multiple components in the case of the latter galaxy. In three more galaxies -- \n3627, \n4501 and \n4536 -- we see gas which \textit{is} in ordered motion, but which is also clearly not in \textit{circular} motion. Two of these galaxies have evidence for rotating, oval flows, while in \n4501 the situation is complicated by the presence of a separate gas filament which does not participate in the general rotation of the gas, severely distorting the observed velocity field.

Only one of the six spirals in this study -- \n3351 -- shows a velocity field that can be described by simple circular rotation. However, it also has high velocity dispersion in the gas, exceeding the rotation velocity out to a radius of 30--50~pc. 
If this velocity dispersion can be ascribed to the local turbulence (see Section~\ref{s_3351}), then the observed gas rotation can be interpreted as the circular velocity of the potential and the dispersion would not affect the SMBH mass estimate. However, if the velocity dispersion is dynamical, then the observed rotation will be less than the circular velocity, and the dispersion must be included in the SMBH mass estimate (e.g. via asymmetric drift or a pressure-like term in Jeans equations, see Section~\ref{s_discdyn}), and the estimated mass has to be corrected upwards from its value in the circular-rotation case. Such a correction can be rather significant; in the case of M84, \citet{Walsh2010} showed that even when the velocity dispersion remained less than 0.6 of the rotation velocity value, the correction was of order 100 per cent. In a galaxy like \n3351, where the dispersion is \textit{larger} than the velocity in the innermost regions, this correction could be even higher.\footnote{One should note however, that \citet{Walsh2010} refer to the intrinsic values, while we report the observed values, where part of the intrinsic velocity is seen as velocity dispersion.} The reality may be that the observed gas velocity dispersion is partly non-dynamical and partly dynamical; unfortunately, as the discussion in Section~\ref{s_discdyn} makes clear, we are currently unable to determine what the relative proportions are or should be for these galaxies. 

Moreover, if the velocity dispersion is dynamical and isotropic, it not only affects the SMBH mass estimate through altering the azimuthal velocity, but also leads to the thickening of the disc. Then a proper estimate of SMBH mass should include projection effects and integration of emission through the thick disc along the LOS. Thus far, all SMBH mass estimates that take into account velocity dispersion in the disc assume that the disc is geometrically thin. We are not aware of SMBH mass estimates that take into account the thickness of the gaseous disc in the innermost parsecs.

\begin{figure}
\includegraphics[trim = 0cm 19cm 11cm 1.5cm, clip=true, width=\columnwidth]{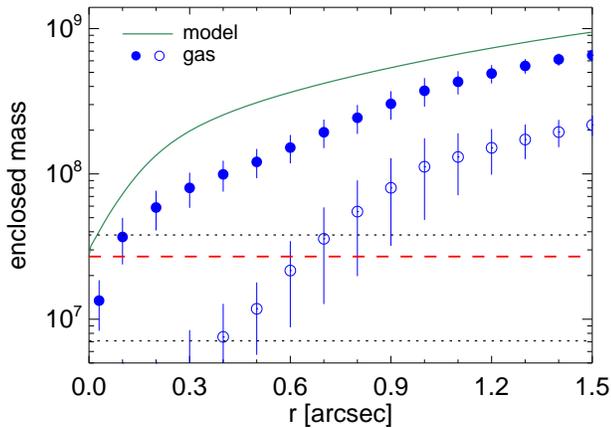}
\caption{Deprojected enclosed mass as a function of radius for \n4501 derived from the stellar dynamical modelling (green line, see Paper\,I and Erwin et al. in preparation) and from the \H2\ emission-line gas kinematics (blue circles, see text for details). The red dashed line indicates the SMBH mass derived from the stellar dynamical modelling and the black-dotted lines the upper limits given by \protect\citet{Beifiori2009}.}
\label{f_n4501enclosed}
\end{figure}

Of the galaxies in our sample, \n4501 is the only one so far for which the central SMBH mass has been estimated via stellar dynamical modelling (Erwin et al. in preparation). We can perform a simple exercise and compare the enclosed (dynamical) mass as a function of the radius given by that model (Paper\,I; Erwin et al. in preparation) with that expected from the gas kinematics under commonly made simplifying assumptions. Fig.~\ref{f_n4501enclosed} shows the deprojected enclosed mass as a function of the radius $r$ given by the stellar dynamical model (green line) and the enclosed mass derived from the \H2\ emission-line gas kinematics in two cases. In the first case, we discard any possible dynamical contribution from the velocity dispersion and assume that the \H2\ deprojected velocity derived from the \S\ data represents the circular velocity of the gravitational potential (blue open circles). In this case, $M(<r)=r\,v_{\rm dp}^2/G$, where $G$ is the gravitational constant. In the second case, we assume that the velocity dispersion is dynamically important and estimate the enclosed mass by $M(<r)=r\,(v_{\rm dp}^2+3\sigma^2)/G$ (blue filled circles). In both cases, we compute the \H2\ rotation curve while excluding the part of the velocity field which is clearly affected by the blue kinematic spiral observed in this galaxy (see Fig.~\ref{f_n4501_2D} and Section~\ref{s_4501} for details). Fig.~\ref{f_n4501enclosed} also shows the value of the SMBH mass derived from the stellar dynamical modelling, $M_{\rm BH}=2.7\times10^7$\Ms, and the two upper limits derived from the central ionized gas velocity dispersion by \citet{Beifiori2009}, $M_{\rm BH}=7.6\times10^6$ and $3.8\times10^7$\Ms\ (scaled to our adopted distance to the galaxy, Table~\ref{t_prop}). 

Fig.~\ref{f_n4501enclosed} shows that the mass derived from the \H2\ emission-line kinematics using these simple approaches significantly underpredicts the enclosed mass of the galaxy. When we assume that the velocity dispersion is dynamically unimportant, we obtain masses at least a factor of 5 lower than those derived from the stellar dynamical model. There is an improvement when we include the \H2\ velocity dispersion in the calculation, although the enclosed mass is still underpredicted by factors of between 1.5 and 3. 
Note that the innermost point of the enclosed mass estimates from gas dynamics including velocity dispersion in Fig.~\ref{f_n4501enclosed} indicates a mass which is still lower than that from the stellar-dynamical estimate, contrary to the common belief that mass estimates from gas velocity dispersion within a given aperture yield upper limits on mass. The latter is true when the observed velocity dispersion arises solely from circular motion in a cold disc within the aperture, but in our case the observed velocity dispersion is dominated by the intrinsic velocity dispersion of the gas, 
something which, if it is not continuously driven by an external source of energy, is being gradually lost via dissipation. 
A detailed comparison between SMBH masses measured via stellar dynamical modelling and gas for two of the other galaxies (\n3627 and \n4536) will be presented in Erwin et al. (in preparation).

\section{Summary and conclusions}\label{s_summary}

We have presented a kinematic analysis of the warm molecular and ionized gas in the inner parsecs ($r\lesssim 300$~pc) of a sample of six nearby spiral galaxies: \n3351, \n3627, \n4501, \n4536, \n4569 and \n4579. This is the second of two companion papers analysing the gas properties of these galaxies  based on new AO high-spatial-resolution $K$-band \S/VLT IFS observations, and is part of a series of papers reporting the results of a \S\ survey aimed at deriving accurate SMBH masses in the nucleus of galaxies from stellar dynamical modelling. 

The \S\ spectra of the sample galaxies display several \H2\ emission lines, and in three of the galaxies (\n3351, \n3627 and \n4536) we also observe emission lines from ionized gas. In \citeta{Mazzalay2013a} we analysed the morphology and physical properties of the gas,  as well as the mass content of the galaxies. 
Here, we focus on the kinematics of the \H2\ emission-line gas, complementing it with the stellar kinematics derived from the same dataset. A detailed kinematic analysis of each of the six galaxies is carried out, placed in the context of the large-scale structure of the galaxies. We scrutinize the suitability of assumptions commonly made such as circular motion and thin discs. 

From the analysis of the velocity fields we find different types of flows that can be ordered by increasing complexity. The simplest case is that of \n3351, where the velocity field is consistent with gas rotating in circular orbits in the plane of the galaxy. More complex velocity fields are shown by \n3627 and \n4536. In \n3627 we find signatures of shocks in the gas, possibly associated with two bright blobs of \H2\ emission that form part of the molecular bar in this galaxy, as well as a misalignment of the kinematic axis of the gas and stars in the inner $\sim 2$~arcsec radius. The latter feature can be explained as an oval gas flow, one aligned with the \H2\ distribution (but not with the stellar bar), implying a strong torque from the stellar bar on the \H2\ flow. \n4536 shows a strong twist in the zero-velocity line in the regions inside the nuclear star-forming ring. As in the case of \n3627, the departures from circular motions are well reproduced by an oval flow, possibly driven by an inner bar. In \n4501, the bulk of the \H2\ emission-line gas seems to be co-rotating with the stars in circular orbits in the plane of the galaxy. However, strong deviations are observed in the form of a blueshifted kinematic feature, resembling a (single) spiral arm. From our analysis we conclude that this feature is not a spiral density wave but rather a filament moving with respect to the rotating galaxy disc. Finally, \n4569 and \n4579 show complex, highly perturbed velocity fields, with no evidence for ordered circular motion.

In all the galaxies of our sample, the observed \H2\ velocity dispersion in the inner 1--2~arcsec ($\sim$50--160~pc) radius is high, in general higher than 50\kms, and increases towards the centre. The stellar velocity dispersion also increases towards the centre, except in the case of \n3351 and \n3627, where there is a drop in the stellar velocity dispersion close to the nucleus. 
In \n4536, there are regions of low velocity dispersion in both the stars and gas spatially coincident with the circumnuclear star-forming ring. 
Due to the large spatial extent of the high-velocity-dispersion gas in our galaxies, we can rule out PSF smearing of the rotation curve as a factor, so the high values are intrinsic to the emission-line gas. At this point is not possible to determine whether these high values of velocity dispersion are predominantly dynamical (which would mean they play an important role in the disc dynamics) or intrinsically non-dynamical (due to, e.g. turbulence).

We analysed the radial behaviour of $v_{\rm dp}/\sigma$ for the gas, which is sometimes used to estimate the disc thickness. We found that $v_{\rm dp}/\sigma$ is $< 1$ in these galaxies interior to a radius of 40--80~pc; the distance at which $v_{\rm dp}/\sigma$ is equal to unity is independent of the nuclear activity in the galaxies. We have also presented lines of evidence that argue against a thick circumnuclear gas disc interpretation for some galaxies (e.g. \n3351 and \n4536), suggesting that the $v_{\rm dp}/\sigma$ ratio may not always be a good indicator of the disc thickness. 

The complexity of the \H2\ emission-line kinematics displayed by the galaxies in our sample argues against the validity of simple assumptions commonly made to derive SMBH masses from gas kinematics, and highlights the importance of having full 2D information. Understanding the role of the gas velocity dispersion in the gas dynamics is key for deriving accurate SMBH masses from gas kinematics. We expect that future ALMA observations will provide the necessary information about the cold and dense molecular gas which, together with NIR IFS data, will help to shed light on the origin of the gas velocity dispersion and its role in the dynamics of the gas in galactic nuclei.

\section*{Acknowledgments}

We would like to thank Inma Mart\'inez-Valpuesta and Ric Davies for useful discussions. We thank the Paranal Observatory Team for support during the observations. PE was supported by the Deutsche Forschungsgemeinschaft through the Priority Programme 1177 'Galaxy Evolution'. SPR acknowledges support from the DFG Cluster of Excellence `Origin and Structure of the Universe'.
This research has made use of the NASA/IPAC Extragalactic Database (NED), which is operated by the Jet Propulsion Laboratory, California Institute of Technology, under contract with the National Aeronautics and Space Administration.

\newcommand{\noop}[1]{}

\appendix

\section{Oval flow model}\label{appendix}
In a few of the galaxies analysed in this paper, there is strong morphological (\n3627) or kinematic (\n3627, \n4536) evidence that the gas flow is highly eccentric. As pointed out by \citet{Spekkens2007}, such flows are not well described by simple departures from circular motion. Here we propose a simple model of an oval gas flow which can be used to fit and interpret the observed LOS velocity field.

\subsection{Location of the zero-velocity line for an oval flow}\label{s_app1}
\begin{figure}
\centering
\includegraphics[width=0.8\columnwidth]{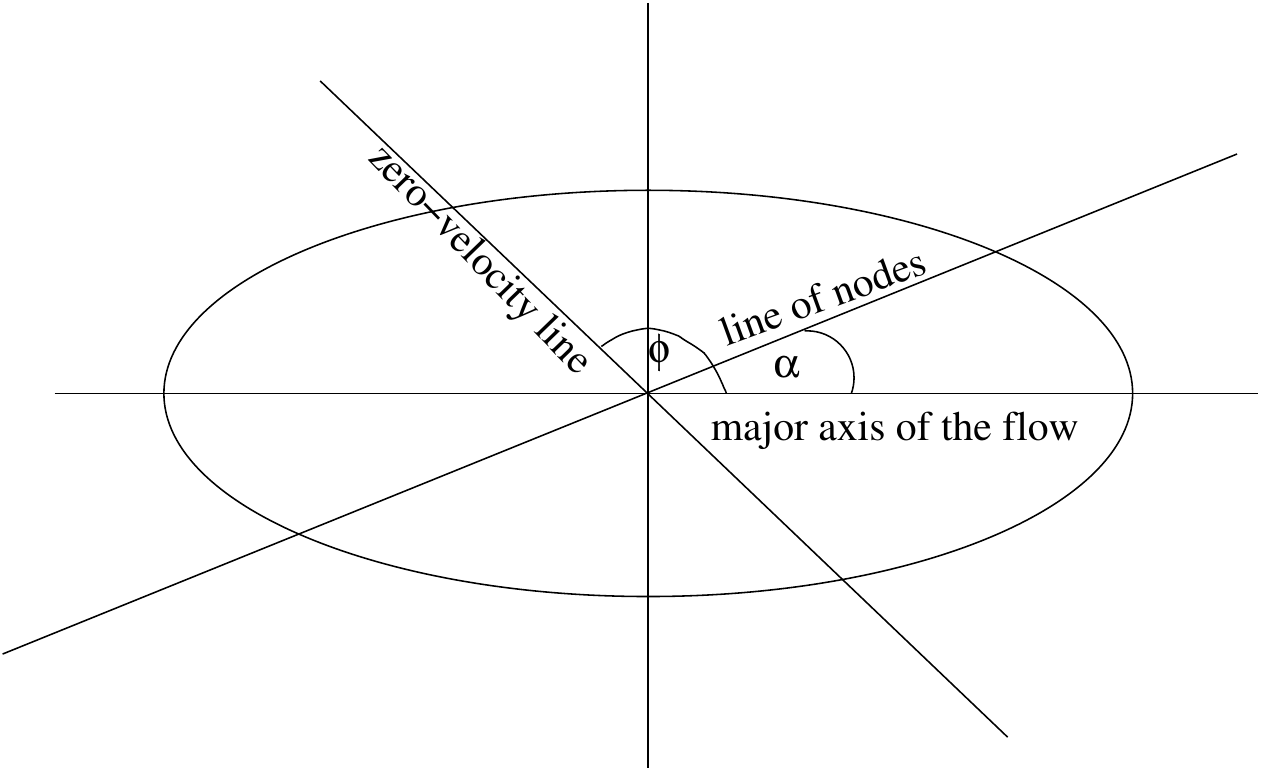}
\caption{Sketch of the geometric configuration of the oval flow in the galaxy plane.}
\label{f_ap1}
\end{figure}

\begin{figure}
\centering
\includegraphics[width=0.75\columnwidth]{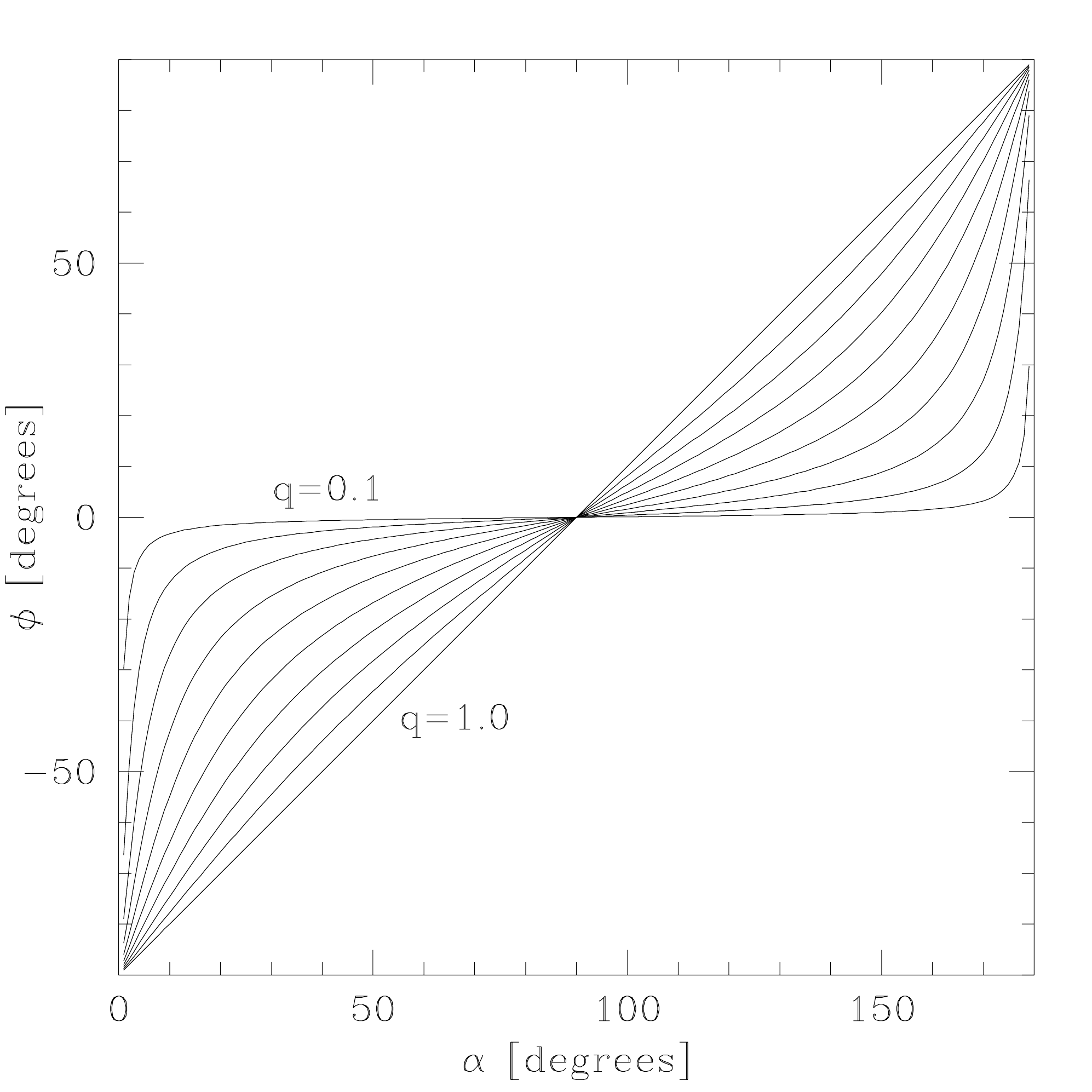}
\caption{Dependence of the angle between major axis of the flow and zero-velocity line ($\phi$) on the angle between major axis and the line of nodes ($\alpha$) for different values of axial ratio $q$.}
\label{f_ap2}
\end{figure}

For an oval flow, the observed maximum-velocity line (which corresponds to the LON in the case of circular flow) is not as robust an indicator of the flow as the zero-velocity line is. This is because the maximum-velocity line can change orientation depending on the distribution of speeds along the ellipse, while the zero-velocity line depends only on the geometry. In particular, if we have a non-rotating flow along concentric ellipses of constant axial ratio, $q=b/a$, then there is a simple relation connecting the angle between the major axis of the flow in the galaxy plane and the LON ($\alpha$) with the angle between that same major axis and the zero-velocity line ($\phi$):
\begin{equation}
\tan \phi \; \tan \alpha \; = \; -q^2.
\end{equation}
The orientation of these two angles is clarified in Fig.~\ref{f_ap1}, and the dependence of $\phi$ on $\alpha$ for various values of $q$ is plotted in Fig.~\ref{f_ap2}. As expected, for circular flows ($q = 1$) we have $\phi = \alpha - 90$, i.e. the zero-velocity line is perpendicular to the LON. However, when $q < 1$, $\phi$ assumes values closer to zero than in the case of circular flow. This means that the zero-velocity line shifts towards alignment with the major axis of the flow when the eccentricity of the flow increases.

\subsection{Formulae for the simple oval flow}\label{s_app2}
Our model of an oval flow is the same as the one already used by Maciejewski et al. 2012. Here we give the explicit formulae defining the flow. The oval flow is confined to the disc plane, and it is constructed by compressing the circular flow in one direction, as explained below. In order to allow for the possibility of an oval flow which itself \textit{rotates} as a pattern, the circular flow is compressed in a rotating frame, and then the compressed flow is transformed back to the inertial frame. 

\begin{figure}
\centering
\includegraphics[width=\columnwidth]{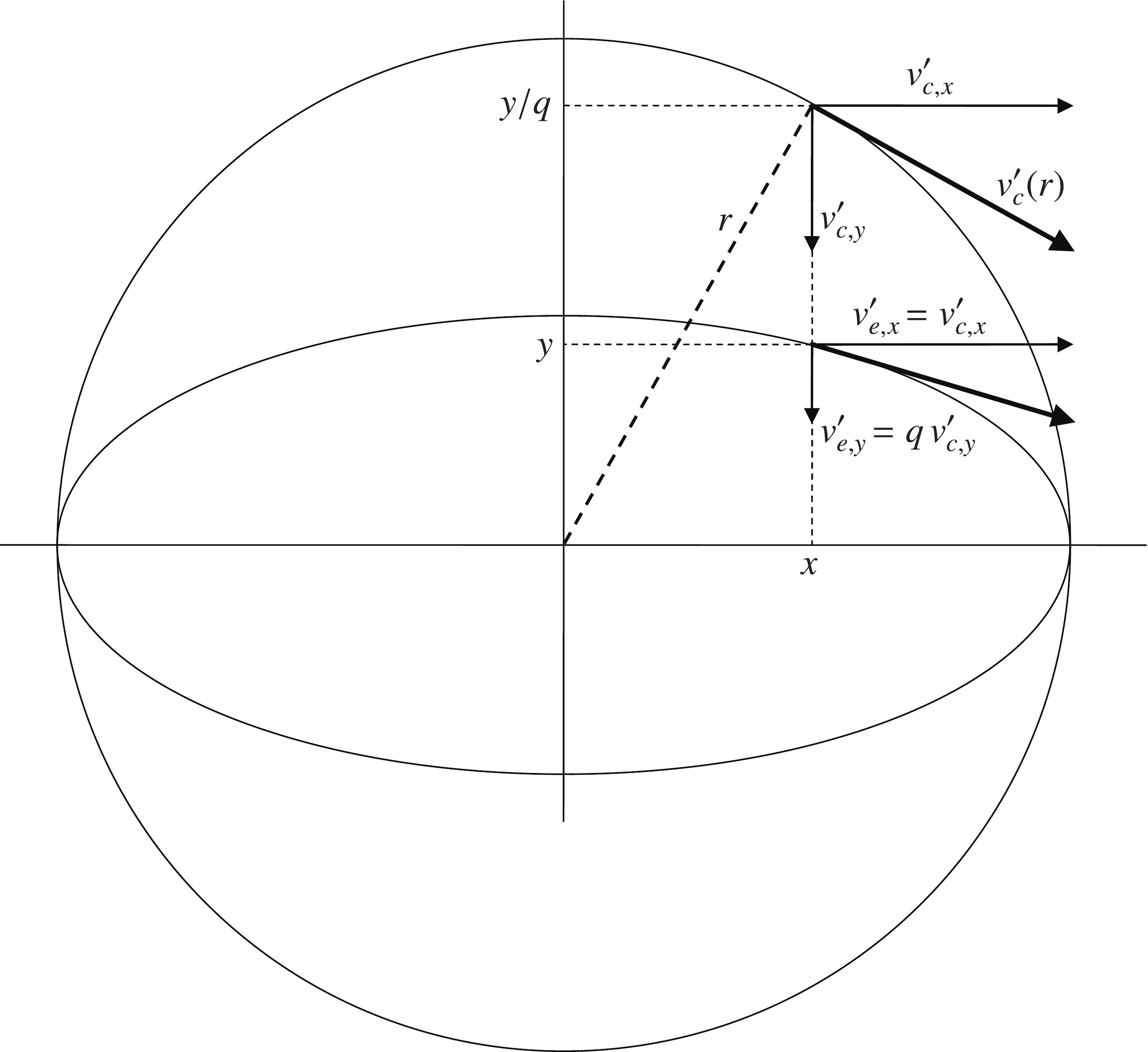}
\caption{ A graph illustrating how the oval flow is constructed by rescaling a circular flow in the rotating frame. Both the circle and the oval are in the plane of the galaxy disc. }
\label{f_ap3}
\end{figure}

Taking the $x-$axis in the disc plane as the major axis of the oval flow, compression of the circular flow means rescaling it along the $y-$axis by a factor of $q<1$, defined above. Thus the velocity of the oval flow in the disc plane at the $(x,y)$ position is dictated by the circular velocity $v_{\rm c}'$ at the position $(x,\frac{y}{q})$ (see Fig.~\ref{f_ap3} for an explanation; in the rotating frame, the velocities are indexed by a prime). Any circular velocity can be used, but we settled on a circular velocity that increases with the radius $r$ for $r < r_{\rm flat}$, and is flat outside this radius, with the velocity $v_{\rm c}(r) = v_{\rm flat}$. In  a reference frame rotating with the angular velocity $\Omega_P= v_{\rm c}(r_{\rm CR})/r_{\rm CR}$, where $r_{\rm CR}$ is the corotation radius, we have $v_{\rm c}'(r) = v_{\rm c}(r) - r \Omega_P$.

For the circular flow in the rotating frame, the components of the velocity vector at the position $(x,\frac{y}{q})$ are related to the velocity in the inertial frame, $v_{\rm c}$, by
\begin{eqnarray}
v_{{\rm c,}\,x}^{\prime}(x,\frac{y}{q}) & = & -\frac{y}{qr} \, v_{\rm c}(r) + \frac{y}{q} \Omega_P\\
v_{{\rm c,}\,y}^{\prime}(x,\frac{y}{q}) & = & \frac{x}{r} \, v_{\rm c}(r) - x \Omega_P,
\end{eqnarray}
where $r = \left[x^2 + (\frac{y}{q})^2\right]^{\frac{1}{2}}$.
After the rescaling along the $y$ direction, they become the velocity components of the oval flow at the position $(x,y)$:
\begin{eqnarray}
v_{{\rm e,}\,x}^{\prime}(x,y) & = & -\frac{y}{qr} \, v_{\rm c}(r) + \frac{y}{q} \Omega_P\\
v_{{\rm e,}\,y}^{\prime}(x,y) & = & \frac{xq}{r} \, v_{\rm c}(r) - xq \Omega_P.
\end{eqnarray}
Note that when we rescale the flow by a factor $q$ along the $y-$axis, so that the flow is now on ellipses of an axial ratio $q$, the $y$-component of the velocity is scaled by a factor $q$, while the $x$-component remains unchanged. This ensures that the velocity vector in the rotating frame remains tangential to the oval flow (see Fig.~\ref{f_ap3}). 

This oval flow is now placed back in the inertial frame by adding to the velocity components at $(x,y)$ a contribution from the uniform rotation:
\begin{eqnarray}
v_{{\rm e,}\,x}(x,y) & = & v_{{\rm e,}\,x}^{\prime}(x,y) - y \Omega_P \; = \; -\frac{y}{qr} v_{\rm c}(r) + y \frac{1-q}{q} \Omega_P\\
v_{{\rm e,}\,y}(x,y) & = & v_{{\rm e,}\,y}^{\prime}(x,y) + x \Omega_P \; = \; \frac{qx}{r} v_{\rm c}(r) + x (1-q) \Omega_P.
\end{eqnarray}
From these two velocity components one can calculate the LOS velocity, $v_{\rm LOS}$, as a function of the angle $\alpha$ between the major axis of the oval flow and the LON (Fig.~\ref{f_ap1}). Given the inclination $i$ of the galactic disc one gets
\begin{equation}
v_{\rm LOS} (x,y) = [-v_{{\rm e,}\,x}(x,y) \sin \alpha + v_{{\rm e,}\,y}(x,y) \cos \alpha] \sin i,
\end{equation}
where the coordinates $(x,y)$ in the plane of the galaxy disc are related to the sky coordinates $(x_s,y_s)$ by
\begin{eqnarray}
x & = & x_s \cos \alpha - \frac{y_s}{\cos i} \sin \alpha\\
y & = & x_s \sin \alpha + \frac{y_s}{\cos i} \cos \alpha,
\end{eqnarray}
where the $x_s$ coordinate is along the line of nodes. The angle $\alpha$ between the major axis of the oval flow and the LON in the plane of the galaxy disc is related to the angle $\alpha_s$ between the same two lines in the sky by
\begin{equation}
\tan \alpha_s = \tan \alpha \cos i
\end{equation}
In the models built in this paper, we used either a constant $q$, or a scenario where $q \sim r/r_{\rm CR}$ inside the corotation and $q= 1$ outside it, representing a flow which is circular at larger radii, but becomes increasingly eccentric towards the centre. In this second scenario, an explicit formula gives $q$ at the position $(x,y)$ in the plane of the galaxy disc
\begin{equation}
q(x,y) = \left(\frac{x^2 + (x^4 + 4 r_{\rm CR}^2 y^2)^{1/2}}{2 r_{\rm CR}^2} \right)^{1/2}.
\end{equation}

\end{document}